\documentclass[11pt]{article}         
\usepackage{epsfig}                   
\usepackage{a4p}                      
\usepackage{cite}                     
\begin{document}
\newcommand {\etal} {{\em et al.}}
\newcommand {\tb}[1]      {table~\ref{#1}}
\newcommand {\tbs}[1]     {tables~\ref{#1}}
\newcommand {\Tb}[1]      {Table~\ref{#1}}
\newcommand {\Tbs}[1]     {Table~\ref{#1}}
\newcommand {\fg}[1]      {figure~\ref{#1}}
\newcommand {\fgs}[1]     {figures~\ref{#1}}
\newcommand {\Fg}[1]      {Figure~\ref{#1}}
\newcommand {\Fgs}[1]     {Figures~\ref{#1}}
\newcommand {\SW}         {SiW}
\newcommand {\sctn}[1]      {section~\ref{#1}}
\newcommand {\sctns}[1]     {sections~\ref{#1}}
\newcommand {\Sctn}[1]      {Section~\ref{#1}}
\newcommand {\Sctns}[1]      {Sections~\ref{#1}}
\newcommand {\ec} {\epsilon}
\newcommand {\ecx} {\epsilon}
\newcommand {\ecy} {\epsilon}
\newcommand {\Zzero}   {{\mathrm Z}^0}
\newcommand {\Zprime}    {{\mathrm Z'}}
\newcommand {\MZ}      {m_{\mathrm Z}}
\newcommand {\mt}      {m_{\mathrm t}}
\newcommand {\mh}      {m_{\mathrm h}}
\newcommand {\shad}    {\sigma_{\mathrm {had}}^{\mathrm {pole}} }
\newcommand {\Ginv}    {\Gamma_{\mathrm {inv}}}
\newcommand {\Ghad}    {\Gamma_{\mathrm {had}}}
\newcommand {\Gll}     {\Gamma_{\ell^+\ell^-}}
\newcommand {\Gff}     {\Gamma_{ \mathrm{f \bar{f} } }}
\newcommand {\Gmu}     {\Gamma_{\mu^+\mu^-}}
\newcommand {\Gee}     {\Gamma_{\mathrm{e^+e^-}}}
\newcommand {\Gtau}    {\Gamma_{\tau^+\tau^-}}
\newcommand {\sff}     {\sigma_{ \mathrm{f\bar{f} } }^{\mathrm {pole}}}
\newcommand {\sll}     {\sigma_{\ell^+\ell^-}^{\mathrm {pole}}}
\newcommand {\sllpk}   {\sigma_{\ell^+\ell^-}^{\mathrm {peak}}}
\newcommand {\se}     {\sigma_{\mathrm{ee}}^{\mathrm {pole}}}
\newcommand {\smu}     {\sigma_{\mu \mu}^{\mathrm {pole}}}
\newcommand {\stau}     {\sigma_{\tau\tau}^{\mathrm {pole}}}
\newcommand {\GZ}      {\Gamma_{\mathrm Z}}
\newcommand {\nmm}     {\mathrm n_{\mu^+\mu^-}}
\newcommand {\x}      {{\mathrm X_0}}
\newcommand {\Rinv}    {R_{\mathrm {inv}}}
\newcommand {\RinvSM}    {R^{\mathrm{SM}}_{\mathrm {inv}}}
\newcommand {\Rhad}    {R_{\mathrm {had}}}
\newcommand {\Nr}      { N_{\mathrm R}}
\newcommand {\Nl}      { N_{\mathrm L}}
\newcommand {\lt}      { < }
\newcommand {\gt}      { > }
\newcommand {\ee }     {\mathrm{e^+e^-}}
\newcommand {\SwitA}   {\mbox{\sc SwitA}}
\newcommand {\SwitR}   {\mbox{\sc SwitR}}
\newcommand {\SwitL}   {\mbox{\sc SwitL}}
\newcommand {\SwitX}   {\mbox{\sc SwitX}}
\newcommand {\La}      {L_{\mathrm{A}}}
\newcommand {\Lrl}     {L_{\mathrm{RL}}}
\newcommand {\Dlr}     {D_{\mathrm{LR}}}
\newcommand {\rr}      {R_{\mathrm R}}
\newcommand {\rl}      {R_{\mathrm L}}
\newcommand {\ra}      {R_{\mathrm A}}
\newcommand {\phir}    {\phi_{\mathrm R}}
\newcommand {\phil}    {\phi_{\mathrm L}}
\newcommand {\er}      {E_{\mathrm R}}
\newcommand {\el}      {E_{\mathrm L}}
\newcommand {\ebeam}   {E_{\mathrm {beam}}}
\newcommand {\ecm}     {E_{\mathrm {cm}}}
\newcommand {\roots}   {\sqrt{s}}
\newcommand {\Nhad}     {N_{\mathrm {had}}}

\newcommand {\Bff}     {Br( \Zzero \rightarrow \mathrm{f \bar{f} })}
\newcommand {\Bll}     {Br( \Zzero \rightarrow \ell^+ \ell^-)}
\newcommand {\Bee}     {Br( \Zzero \rightarrow \mathrm{e^+ e^-})}
\newcommand {\Bhad}    {Br( \Zzero \rightarrow {\mathrm{had}})}
\newcommand {\Binv}    {Br( \Zzero \rightarrow {\mathrm{inv}} ) }
\newcommand {\Bnu}     {Br( \Zzero \rightarrow  \nu \bar{\nu} ) }
\newcommand {\Zg}      {{ {\mathrm Z} - \gamma }} 

\newcommand {\tech}     {\marginpar{$\bigcirc \!\!\!\!\! \dagger$ }}
\newcommand {\technote}[1] {\tech \footnote{#1} }
%
%
\newcommand {\hold }    {{hold}}
\newcommand {\SWSEG}    {{\sc swseg}}
\newcommand {\SWHILO}   {{\sc swhilo}}

\flushbottom
\begin{titlepage}
%
%
\begin{center}{\Large
EUROPEAN ORGANIZATION FOR PARTICLE PHYSICS
}\end{center}\bigskip
\begin{flushright}
       CERN-EP/99-136 \\ 28 Sep 1999
\end{flushright}
%
%
\bigskip
\boldmath
\begin{center}{\huge\bf
Precision Luminosity for $\mathbf {\Zzero}$ Lineshape Measurements
with a Silicon-Tungsten Calorimeter
}
\end{center}\unboldmath\bigskip
\vspace*{1cm}
\begin{center}{\LARGE The OPAL Collaboration
}\end{center}\bigskip\bigskip
%
%
\bigskip\bigskip
%
%
%

%
\vspace*{0.5cm}
%
%
\begin{center}{\large Abstract}\end{center}

The measurement of small--angle Bhabha scattering
is used to
determine the luminosity at the OPAL interaction point
for the LEP I data recorded between 1993
and 1995. 
The measurement is based on 
the OPAL Silicon-Tungsten Luminometer
which is composed of two calorimeters
encircling the LEP beam pipe, on opposite sides of the 
interaction point.
The luminometer detects electrons
from small--angle Bhabha scattering at angles between 25 and 58~mrad.  
At LEP center-of-mass energies around the $\rm{Z}^{0}$, about half
of all Bhabha electrons entering the detector fall within a $79$~nb
fiducial acceptance region.  
The electromagnetic showers generated in the stack of
1 radiation length tungsten absorber plates are 
sampled by 608 silicon detectors 
with $38,912$ radial pads of 2.5~mm width.
The fine segmentation of the detector,
combined with the precise knowledge of its physical dimensions, 
allows the
trajectories of incoming 45~GeV electrons or photons to be determined 
with a
total systematic error of less than 7~microns. We have quantified all
significant sources of systematic experimental error in the luminosity
determination by direct physical measurement.
All measured properites of the luminosity event sample are found
to be in agreement with current theoretical expectations.
The total systematic measurement uncertainty is
$3.4 \times 10^{-4}$, significantly below the theoretical error of 
$5.4 \times 10^{-4}$
currently assigned to the QED calculation of the Bhabha
acceptance, and contributes negligibly to the 
total uncertainty in the OPAL measurement
of $\Ginv/ \Gll $, a quantity of basic physical interest which depends
crucially on the luminosity measurement.

%
%
%
%
\bigskip\bigskip\bigskip\bigskip
\bigskip\bigskip
\begin{center}{\large
To be submitted to Eur. Phys. J. C}\end{center}
%
%
\end{titlepage}
\begin{center}{\Large        The OPAL Collaboration
}\end{center}\bigskip
\begin{center}{
G.\thinspace Abbiendi$^{  2}$,
K.\thinspace Ackerstaff$^{  8}$,
G.\thinspace Alexander$^{ 23}$,
J.\thinspace Allison$^{ 16}$,
K.J.\thinspace Anderson$^{  9}$,
S.\thinspace Anderson$^{ 12}$,
S.\thinspace Arcelli$^{ 17}$,
S.\thinspace Asai$^{ 24}$,
S.F.\thinspace Ashby$^{  1}$,
D.\thinspace Axen$^{ 29}$
% ...
}\end{center}\bigskip
\bigskip
$^{  1}$School of Physics and Astronomy, University of Birmingham,
Birmingham B15 2TT, UK
\newline
$^{  2}$Dipartimento di Fisica dell' Universit\`a di Bologna and INFN,
I-40126 Bologna, Italy
\newline
% ...
$^{  8}$CERN, European Organisation for Particle Physics,
CH-1211 Geneva 23, Switzerland
\newline
$^{  9}$Enrico Fermi Institute and Department of Physics,
University of Chicago, Chicago IL 60637, USA
\newline
% ...
$^{ 12}$Indiana University, Department of Physics,
Swain Hall West 117, Bloomington IN 47405, USA
\newline
% ...
$^{ 16}$Department of Physics, Schuster Laboratory, The University,
Manchester M13 9PL, UK
\newline
$^{ 17}$Department of Physics, University of Maryland,
College Park, MD 20742, USA
\newline
% ...
$^{ 23}$Department of Physics and Astronomy, Tel Aviv University,
Tel Aviv 69978, Israel
\newline
$^{ 24}$International Centre for Elementary Particle Physics and
Department of Physics, University of Tokyo, Tokyo 113-0033, and
Kobe University, Kobe 657-8501, Japan
\newline
% ...
$^{ 29}$University of British Columbia, Department of Physics,
Vancouver BC V6T 1Z1, Canada
\newline
% ...
$^{  e}$ and University of Mining and Metallurgy, Cracow
\newline
$^{  f}$ and Heisenberg Fellow
\newline
% ...
\newline
\newpage
\section{Introduction}
%
%
%

Precision measurements of the leptonic and hadronic 
production cross sections in 
electron-positron annihilations near the $\Zzero$ resonance are
a vital part of the LEP physics program~\cite{bib-old-opal,
bib-old-aleph,
bib-old-delphi,
bib-old-l3}.
The precision with which such cross-section
measurements can be made has often been
limited by the experimental systematic error on the luminosity
of the colliding beams.
For example, at PEP and PETRA where both wide-angle and small-angle
Bhabha scattering were used to determine the luminosity,
the experimental systematic errors were 2--5\% 
and much larger than the statistical error inherent in the number
of multihadron events produced~\cite{bib-old-petra}.

In LEP collider runs near the $\Zzero$ resonance, data samples
of order $5 \times 10^6$ $\Zzero$ decays 
have been delivered to each of the four experiments.
If these data are to be used efficiently, 
a very precise determination of the luminosity must be made.
In the initial running at the $\Zzero$ resonance,
OPAL was able to determine
the luminosity to significantly better than 1\%
but it was clear that the experimental systematic error could not be expected
to keep pace with the foreseen improvements in $\Zzero$
statistics.
In 1991 the decision was taken to build a second-generation luminosity
monitor.
The new detector, referred to as \SW,
employs calorimeters with tungsten absorber and
silicon sampling 
and was first operated in the 1993 LEP run.  
In this document we describe the determination of the OPAL
luminosity for the 1993, 1994 and 1995 LEP runs at the 
$\Zzero$.  The measurement has an experimental
systematic error of $3.4 \times 10^{-4}$,
an improvement of nearly
two orders of magnitude on the PEP/PETRA measurements, and
is substantially more precise than other luminosity measurements
at the $\Zzero$~\cite{bib-z-lumi}. 

%
%

\subsection{Physics motivation}
\label{sec:physics}
%
%
%
The measurement of the $\Zzero$ resonance parameters requires 
precise cross section measurements at several different energies.
From the shape of the resonance, the $\Zzero$ mass, $\MZ$, and 
total width, $\GZ$, are determined.  The overall normalization
of the resonance is given by the peak cross sections to
all accessible charged fermions, $\shad$,
$\se$, $\smu$ and $\stau$.  The measurement of these cross sections
requires an absolute determination of the luminosity. 
In principle, measurement of $\MZ$ and $\GZ$ requires only a
relative luminosity measurement.  In practice the situation
is more complicated. The ``off-peak'' data were collected in
energy scans in 1993 and 1995 with the bulk of the ``on-peak''
data taken in 1994.  To make the best use of the data for
determining $\GZ$, the
luminosity measurement must be stable between years, requiring
an analysis which is similar to the determination of
the absolute luminosity.

The importance of $\MZ$ as a fundamental input to the Standard
Model and the sensitivity of $\GZ$ to radiative corrections
involving the top mass, $\mt$, and the Higgs mass, $\mh$, 
are well known\cite{bib-yr-89}.
The physics interpretation
of the the total cross sections is somewhat more involved.
At tree level the cross sections for fermion-pair production
at the $\Zzero$ can be expressed as
\begin{equation}
\sff = \frac{12 \pi}{\MZ^2} \frac{\Gee \Gff}{\GZ^2},
\label{eq:sigff}
\end{equation}
where $\Gee$ is the partial decay width of the $\Zzero$ to
electrons, $\Gff$ the partial decay width to a given fermion pair,
and $\GZ$ is the total decay width of the $\Zzero$.
The physics of the absolute cross section can be seen most
clearly if equation~\ref{eq:sigff} is expressed in terms of
branching ratios:
\begin{equation}
\sff = \frac{12 \pi}{\MZ^2} \Bee \Bff.
\label{eq:sigbr}
\end{equation}
Thus the measurement of the pole cross section of the $\Zzero$
to fermion pairs, 
$\sff$, 
allows us to make measurements of absolute
branching ratios.

In particular, $\Binv$, the branching ratio of the $\Zzero$ to particles
such as neutrinos which are not visible in our detector,\footnote{
More precisely, the measured invisible width 
contains any contribution to the $\Zzero$ decay
width which is not accounted for by 
either the multihadron or
charged lepton pair analyses.}
is measureable
only by determining the absolute production cross sections of visible
final states.
An especially interesting quantity involving $\Binv$ is
\begin{equation}
\Rinv \equiv
             \frac{ \Binv }
                  { \Bll }
\label{eq:rinv}
\end{equation}
where $\Bll$ is the
leptonic width of a single generation.
Assuming charged lepton universality, 
the invisible branching ratio is
\begin{equation}
\Binv =  1 -  \Bhad - 3 \ \Bll
\label{eq:ginvt}
\end{equation}
where 
$\Bhad$
is the total branching ratio to the 5 quarks kinematically
accessible at the $\Zzero$.  
Using equation~\ref{eq:sigbr},  equation~\ref{eq:rinv}
can be rewritten as
\begin{equation}
\Rinv \ = \ 
\left(\frac{12 \pi} { \sll \MZ^2}  \right)^{\frac{1}{2}} \ - \  \Rhad \ - \ 3
\label{eq:ginve}
\end{equation}
where $\Rhad$ is the ratio of hadronic to leptonic $\Zzero$ decays
given by 
$\Rhad = \frac{ \Bhad} { \Bll }$.
Measurement of the quantities in equation~\ref{eq:ginve} 
allows the Standard Model
prediction for the coupling of the $\Zzero$ to neutrinos to be tested.
The Standard Model prediction for $\Rinv$,
\begin{equation}
\RinvSM \equiv   3 \frac{\Bnu}{\Bll}
\end{equation}
depends only weakly on
$\mt$ and $\mh$ since
the dependence of  $\Bnu$ and $\Bll$ on these quantities
nearly cancels in the ratio.
For example, using $\mt = 175 \pm 5.5$~GeV, 
$\mh = 300 ^{+700}_{-240}$~GeV and
$\alpha_s = 0.119 \pm 0.003$ gives the Standard Model
prediction of
$$ \RinvSM = 5.973 \pm 0.003. $$
The errors on the prediction of $\RinvSM$ 
from uncertainties on $\mt$ and 
$\mh$ are far smaller than the 
precision with which  $\Rinv$ can be measured. 
The contribution of the various experimental statistical
and systematic errors to the absolute uncertainty
in $\Rinv$ can be expressed as:
\begin{equation}
   \Delta\Rinv\; \approx\; 6 \;\frac{\Delta N_l}{N_l}\:\: \oplus \:\:
                21 \;\frac{\Delta N_{had}}{N_{had}}\:\: \oplus \:\:
                15 \;\frac{\Delta L}{L}
\label{eq:marcello}
\end{equation}
\noindent
where $N_l$ and $N_{had}$ are the numbers of leptonic and hadronic
decays observed, corrected for full acceptance,
$L$ is the integrated luminosity, and $\oplus$ indicates
summation in quadrature.  The large coefficients of the hadronic and
luminosity terms indicate their relative importance in the measurement.
It was equation~\ref{eq:marcello} which originally
motivated our interest in improving the OPAL luminosity measurement
to be commensurate with the achievable
precision in measuring hadronic events.
The relative uncertainties ultimately achieved are summarized in \tb{tab:rinv}.
The theoretical uncertainty on the photonic radiative corrections to $\sll$
affect the determination of $\Rinv$ in the same way
as the error on the luminosity.
Recent theoretical work~\cite{bib-pcp} has significantly reduced this
uncertainty to the point where it no longer makes a significant
contribution.
For the final OPAL 
LEP I data sample, the errors on the
measurements of the luminosity, the number of
lepton pair events and the number of multihadron
events contribute approximately equally to the
uncertainty on $\Rinv$, making a 0.4\% measurement
possible.
Because of the small theoretical error on
the prediction for $\Rinv$, 
a measurement of $\Rinv$ which differs from the Standard
Model would be a definite indication of 
new physics.

\begin{table}[htbp]
\begin{center}
\begin{tabular} {|l||c|c||c|}
\hline 
Quantity   & Relative         & Relative         & Relative (stat. + syst.) \\
           &statistical error & Systematic error & error on $\Rinv$   \\
           & ($\times 10^{-4}$) & ($\times 10^{-4}$)& ($\times 10^{-4}$) \\
\hline
\hline
Acceptance corrected hadrons   &  6     &    7    & 34 \\
Acceptance corrected leptons   & 17     &    13   & 21 \\
Luminosity (theoretical)       &  0     &    5.4  & 14 \\ 
Luminosity (experimental)      &  3     &    3.4  &  9 \\ 
Photonic correction to $\sll$  &  0     &      2   &  6 \\
\hline
\hline
Total                        &          &       & 44 \\
\hline
\end{tabular}
\end{center}
\caption[Contributions to error on $\Rinv$]
{ Approximate contribution to the relative 
error on $\Rinv$ from the various
quantities measured by OPAL.
Before the introduction of the SiW detector, the experimental
luminosity systematic error contributed $103\times10^{-4}$ to
the relative error on $\Rinv$, and represented the dominant
source of error in its measurement.
\label{tab:rinv}}
\end{table}

If the Standard Model couplings for neutrinos and charged
leptons are assumed, a measurement of
$\Rinv$ allows the parameter space of other models to
be explored.  For example, particles which are
candidates for cold dark matter and which can be produced
in $\Zzero$ decays would contribute to $\Binv$.  
These models are significantly constrained by
measurements of $\Rinv$\cite{kane}.

The values of $\shad$ and $\sll$ are also 
insensitive to the values of
$\mt$ and $\mh$.
This can be most clearly seen by
noting that the 
effects of $\mt$ and $\mh$ on
the partial widths are largely independent of the
fermion type\footnote{
Except for triangle vertex corrections to $\Gamma_{\mathrm{b}}$
which have an explicit $\mt$ dependence.
The resulting uncertainty is, however, negligable, given the known
constraints on $\mt$.
}
and therefore affect the numerator
and denominator of equation~\ref{eq:sigff} in the
same way.
In contrast,   
the final state QCD correction 
affects only the quarks, giving
the hadronic width a unique correction
which increases $\Ghad$ by a factor of 
approximately $1 + \frac{ \alpha_s}{\pi}$.
This correction gives $\sll$
sensitivity to
the value of $\alpha_s$ through 
the dominant contribution of $\Ghad$ to
the total width, $\GZ$.
Somewhat counter intuitively, $\shad$ exhibits less sensitivity
to $\alpha_s$ due to the partial cancellation of the
variation induced through $\Ghad$ and $\GZ$.
In \tb{tb:alphas} the statistical precision of measurements of
$\alpha_s$ using $\shad$, $\sll$, and $\Rhad$ are compared.
In the absence of systematic
errors, $\sll$ provides the most precise measurement of $\alpha_s$.
Measurements of $\alpha_s$ based on the
quantity $\Rhad$ do not depend on the luminosity measurements,
but are less precise than those from $\sll$.
%
%
%

In contrast to almost all other quantities used to measure
$\alpha_s$, uncertainties 
in higher order QCD radiative corrections to $\Ghad$
give rise to relatively small theoretical errors on 
$\alpha_s$\cite{bib-levan}.  Given the large 
statistics of the LEP I data sample, the measurement
of $\alpha_s$ from $\sll$ is one of the most precise.

Alternatively, if $\alpha_s$ is determined from
external measurements, 
the measurement of $\sll$ can also be used to constrain
anomalous $\Zzero$ decays which would be classified
as multihadrons or as an ``invisible'' decay.  

\begin{table}[htbp]
\begin{center}
\begin{tabular}{|c||c|c|}
\hline
Quantity   & Approximate fractional   &  Ideal statistical precision  \\
           & dependence on $\alpha_s$ &    of $\alpha_s$    \\
\hline    
\hline
$\sll$       & $1 - 1.4 \frac{\alpha_s}{\pi}$ & $\pi \sqrt{3.6/\Nhad}$ \\
$\shad$     & $1 - 0.4 \frac{\alpha_s}{\pi}$ & $\pi \sqrt{7.0/\Nhad}$ \\
$\Rhad$         & $1 + 1.0 \frac{\alpha_s}{\pi}$ & $\pi \sqrt{6.3/\Nhad}$  \\
\hline
\end{tabular}
\end{center}
\caption[Statistical precision of $\alpha_s$ determination]
{ The ideal 
statistical precision of $\alpha_s$ extracted from
various quantities expressed as a function of the total number of
{\it hadronic} events recorded, $\Nhad$.  In practice these precisions are
degraded by systematic errors and inefficiencies.
\label{tb:alphas} }
\end{table}
\subsection{Analysis strategy}
%
%

    The \SW\ luminometer and the associated luminosity selection 
    were optimized to exploit the characteristics of
    Bhabha scattering.
    The small--angle Bhabha cross section is dominated by the
    t-channel exchange of a photon, which is represented by the first term
    in the following expression for the Born--level differential
    scattering cross section:

\begin{equation}
\frac{d\sigma}{d\Omega} = \frac{\alpha^2}{2 s}
\left[ \frac{1 + \cos^4 \frac{\theta}{2}}{\sin^4 \frac{\theta}{2}}
  -2   \frac{    \cos^4 \frac{\theta}{2}}{\sin^2 \frac{\theta}{2}}
  +    \frac{1 + \cos^2 \theta}{2} \right] .
\label{eq:born}
\end{equation}

    This yields a $1/\theta^3$ angular
    spectrum, where $\theta$ is the angle of the out-going electron
    and positron with respect to the beam.  By instrumenting
    a region at sufficiently small angles, in our case 
    approximately 25 to 58~mrad, the counting rate
    from small--angle Bhabha events can be made larger than
    the rate of $\Zzero$ production, minimizing the contribution
    of the luminosity measurement to the statistical error of
    the measured cross sections.
 
    As far as experimental systematic errors are concerned,
    the forward-peaked $1/\theta^3$ Bhabha spectrum
    requires that the detector and 
    luminosity analysis 
    define the angle marking the inner edge of the acceptance with
    particularly  high precision.
    For example, to achieve a precision of 1/1000, the inner edge
    of the acceptance must be known to 10~$\mu$rad (see
    section~\ref{sec:radial_intro}),
    corresponding to $\sim$ 25~$\mu$m in the 
    radial coordinate of the 
    showers generated in the calorimeters located approximately
    2.5~m on either side of the interaction region.
Similarly, the $\sim$5~m distance between the two luminometers,
which is not easily accessible to direct measurement under operating
conditions, must be known to better than 2.5~mm.

The theoretical difficulty in ascribing meaning to the trajectory of a bare
Bhabha-scattered electron encouraged our fundamental decision to make a
calorimetric position measurement, which naturally includes the effects of
close-lying radiated photons. 
But practical limits on the spatial resolution which can be achieved by a
calorimeter are about 200$\mu$m. 
Similarly, the physical width of the LEP luminous region itself exceeds
100$\mu$m, and the angular divergence of the beams is about $100\mu$rad.
Furthermore, the movements of the beam spot with respect to the detector over
the course of data taking  are on the millimeter scale.
These effects are all much larger than the basic precision with which
the Bhabha scattering acceptance angle must be defined.
Only by ensuring
that they influence the detector acceptance in second-order is a luminosity
measurement surpassing a precision of 1/1000 possible.
%
%
%
%
%
    This is achieved by constructing the detector to be as 
    symmetrical and homogeneous as possible (section~\ref{sec:det}).
    The azimuthal symmetry of the detector, 
    coupled with the event selection
    described below in section~\ref{sec:sel}, 
    minimizes the effects of beam offset on
    the luminosity (section~\ref{sec:beam}).  
    Careful reconstruction of the radial coordinate 
    (section~\ref{sec:rad_meas})
    gives a resolution function which is nearly symmetric around the
    inner and outer radial cuts.
The goal here is not to attain the highest possible resolution, but rather to
minimize any possible bias. 
In the end our chief success is to demonstrate that the net bias in our
angular acceptance boundary is less than 4\% of our resolution.

    Apart from the determination of the 
    radial acceptance of the detector, 
    the most important measurement is that of
    energy (section~\ref{sec:en}).  
    As discussed in sections~\ref{sec:sel} and
    \ref{sec:back}, the background from
    off-momentum beam particles requires that rather tight energy
    cuts be applied.  
    Even in the absence of this background,
    energy cuts would be required to reduce the sensitivity
    of the selection to radiative Bhabha events
    with small values of t-channel momentum transfer.
    The systematic error arising from the imprecision
    with which we understand
    the low energy tails of the detector energy response 
    is reduced by imposing a cut on the acollinearity
    of the scattered electrons, 
    which causes our acceptance for radiative events to be
    determined by geometry rather than energy.

    Another consideration is the treatment of events
    with multiple showers in one (or both) of the calorimeters.  
    Here we exploit the fine granularity
    of our detector to reject spurious electromagnetic 
    showers from off-momentum
    beam particles, while retaining showers from final state radiation.  
    Using this approach (section~\ref{sec:ev_recon}), 
    the dependence of the luminosity measurement 
    on the theoretical description of final state radiation 
    and on the off-momentum background is minimized.

%
%
    The \SW\ detector and luminosity analysis were designed
    to provide sufficient redundancy and
    internal consistency checks 
    to allow the performance of the detector to be closely 
    monitored.
    Here we cite a few examples.  
    The fine radial and longitudinal granularity of the 
    detector are exploited to produce precise and
    continuous radial coordinates.
    This allows
    the acceptance of the luminosity selection to be
    probed by varying
    the inner and outer radial cuts
    (section~\ref{sec:anchor}).
    The continuous radial coordinate also
    allows us to exploit radiative and non-radiative
    Bhabha events to measure the energy response 
    and linearity of the detector
    during OPAL operation, as illustrated in section~\ref{sec:en}.  
    Finally, extensive monitoring of trigger
    signals allows us 
    to measure the background from random coincidences of
    off--momentum beam particles (section~\ref{sec:back})
    as well as to measure the efficiency of the trigger signals
    (section~\ref{sec:trig}).

%
%
    The luminometer acceptance calculation
    for Bhabha scattering was designed to be as 
    direct as possible.  
The simulation of electromagnetic showers in this calculation is based on
a parametrization of the detector response obtained from the data, rather
than the usual approach of utilizing a detailed Monte-Carlo simulation
of shower development.
This approach gives a much more reliable description of the tails
of the detector response functions, which are primarily due to extreme
fluctuations in shower development, than we could obtain using any existing
program which attempts to simulate the basic interactions of photons and
electrons in matter.
    The measured LEP beam size and divergence, as well as
    the measured offset and tilt of the  beam with respect to
    the calorimeters are also incorporated into this simulation. 
    This approach, described in section~\ref{sec:detsim},
    allows us to propagate the uncertainties
    in the measurements through to the final luminosity error
    in a transparent manner.
    As discussed in section~\ref{sec:anchor}, the radial coordinate
    bias is determined directly from data and explicitly accounted for.
    The Monte~Carlo, therefore, is only used to further 
    correct the acceptance
    for the effects of the detector 
    energy response, LEP beam parameters and
    coordinate resolution.

    The largest 
    acceptance correction generated by the differences between
    the raw four-vectors and the reconstructed detector
    response is of order $-5 \times 10^{-4}$ and due mostly to the finite
    resolution of the energy measurements in the two calorimeters.  
    A detailed parameterization of the energy 
    response of each calorimeter is required in order to 
    obtain a reliable calculation of this correction.
    This is discussed in section~\ref{sec:en}.
    The residual effect of the LEP beam parameters on the
    acceptance is of similar magnitude, but opposite sign. 
    In section~\ref{sec:beam} we demonstrate that 
    the effect of beam parameters on the
    acceptance has been understood and accounted for.

The overall acceptance correction is of order $\sim 2\times 10^{-4}$
and is largely independent of the details of theoretical input.
This will allow future improvements in the theoretical understanding
of Bhabha scattering to be incorporated into our results.
The interplay between the theoretical and experimental components
of our acceptance calculation is discussed in section~\ref{sec:theory}.

%
%

    In the remainder of this paper, the detector and luminosity analysis
    are systematically described.  We begin in 
    section~\ref{sec:det}  with the detector design and
    construction.  Also included in this
    section are the axial and radial metrology of the
    detector.  
    In sections~\ref{sec:ev_recon} and \ref{sec:ev_meas}
    event reconstruction and the measurement  of shower
    coordinates and energy are described.
    In sections \ref{sec:back}, \ref{sec:trig} and \ref{sec:beam}, 
    the sensitivity of the measurement to 
    background, trigger efficiency and 
    the LEP beam parameters is discussed. 
    This is followed by a discussion of the detector simulation
    in section \ref{sec:detsim}.
    In section~\ref{sec:acceptance}
    we discuss the calculation of the acceptance and summarize the experimental
    and theoretical systematic errors.
    Finally, in section~\ref{sec:ev_samp} we discuss the properties of the
    selected sample of Bhabha scattered events.

    One exception is made to this systematic description of the
    analysis.  We pause here to
    describe the final event selection.  
    This would logically follow
    the section on event measurement, but by introducing
    the event selection now we simplify and motivate 
    much of the following discussion.
\subsection{Bhabha event selection}
\label{sec:sel}

    The event selection criteria can 
    be classified into
    {\em isolation} cuts which isolate a sample
    of pure Bhabha scattering events from the off-momentum
    background, 
    and acceptance defining, or {\em definition} cuts.
    The isolation cuts are used to define a fiducial set of events 
    which lie within the good acceptance of both calorimeters and are 
    essentially background free.
    The definition cuts then select subsets of events from within the
    fiducial sample.
    As the nomenclature suggests, the characteristics of the 
    definition cuts largely determine the quality of the 
    measurement.
    Showers generated by incident electrons and photons are
    recognized as clusters
    in the calorimeters (section~\ref{sec:ev_recon})
    and their energies and 
    coordinates determined (section~\ref{sec:ev_meas}).
    The fine segmentation of the detectors allows incident
    particles with separations greater than 1~cm to be individually
    reconstructed with good efficiency.

The coordinate system used throughout this paper is cylindrical, with the
z-axis pointing along the direction of the electron beam,
passing through the centers of the two calorimeter bores.
The origin of the azimuthal coordinate, $\phi$, is in the horizontal
plane, towards the inside of the LEP ring.
    All radial coordinate measurements are projected to reference planes 
    at a distance of 
    $\pm 246.0225$~cm from the nominal intersection 
    point.
    These reference planes correspond to the nominal position
    of the silicon layers $7~\x$ deep in the two
    calorimeters.
 
    The {\em isolation} cuts consist of the following requirements, 
    imposed 
    on
    {($\rr$,$\phir$)} and {($\rl$,$\phil$)},
    the radial and azimuthal coordinates of the highest 
    energy cluster associated with the Bhabha event, 
    in each of the right and left  
    calorimeters, 
    and on $\er$ and $\el$,
    the total fiducial energy deposited by the Bhabha event
    in each of the two calorimeters, explicitly including the energy
    of radiated photons:
$$ 
\begin{array}{lrcl}
 
\bullet~ \mbox{Loose radial cut, right (left)} &
    6.7~{\rm cm} \  < \ \rr &<&13.7~{\rm cm} \\
&
    (6.7~{\rm cm}\  < \ \rl  &<&13.7~{\rm cm}) \\
~~~~~~~~~~~~~~~~~~~~~~~~~~~~~~~~~~~~~~~~~~~~~~~~~~~~~~
&  &&\\ 
\bullet~ \mbox{Acoplanarity cut} & 
    \left| \left|\phir - \phil \right| - \pi \right|&<&200~{\rm mrad} \\
&  &&\\ 
\bullet~ \mbox{Acollinearity cut} & 
    \left| \rr - \rl \right|&<&2.5~{\rm cm} \\
&  &&\\
\bullet~ \mbox{Minimum energy cut, right (left)} & 
    \er&>&0.5\cdot \ebeam \\
& 
   (\el&>&0.5\cdot \ebeam )\\
& &&  \\
\bullet~ \mbox{Average energy cut} & 
    \left( \er + \el \right)/{2} &>&0.75\cdot \ebeam \\
\end{array}
 $$
%
Note that by defining the energy cuts relative to the beam energy,
$\ebeam $, the selection efficiency is largely independent
of $\sqrt{s}$.
 
    The acollinearity cut (which corresponds to approximately
    10.4~mrad) is introduced in order to ensure that the
    acceptance for single radiative events is effectively determined
    geometrically and not by the explicit energy cuts.
 
    The {\em definition} cuts, based solely on the reconstructed radial
    positions $(\rr,\rl)$ of the two highest energy clusters, then distinguish
    two classes of events which we count to determine the luminosity.
$$
\begin{array}{lr}
\bullet~  \mathrm{\SwitR} &
    7.7~{\rm cm}
    ~<~
    \rr
    ~<~
    12.7~{\rm cm} \\
~~~~~~~~~~~~~~~~~~~~~~~~~~~~~~&
~~~~~~~~~~~~~~~~~~~~~~~~~~~~~~~~~~~~~~~~~~~~~~~~~~~~~~~~~~~~~\\ 
\bullet~ \mathrm{\SwitL} &
    7.7~{\rm cm}
    ~<~
    \rl
    ~<~
    12.7~{\rm cm} \\

\end{array} 
$$
 
    The \SwitR\ and \SwitL\footnote{
    The ``{\sc {it}}'' 
    in the ``$\SwitX$'' acronyms stands for in-time coincidences,
    to distinguish these from the corresponding selections
    used for the analysis
    of off momentum particle backgrounds.
    See section~\ref{sec:back}.} 
    {\em definition} cuts have been chosen so as to
    correspond closely (within $\approx 20\mu$m) to 
    radial pad boundaries (see section~\ref{sec:det}) 
    in the silicon layers at the reference planes
    7 $\x$ deep in the 
    calorimeters.
    Expressed in terms of polar angles, these cuts 
    correspond to 
    31.288 and 51.576 mrad.

    The luminosity
    is measured from the average of the $\SwitL$ and
    $\SwitR$ counters, i.e.,  
\begin{equation}
    \Lrl \equiv
    \frac{1} { 2 A_{\mathrm {RL}} }
    \left(
    { N_{\SwitR} } 
    ~+~
    { N_{\SwitL} }
    \right)
\label{eq:lrl}
\end{equation}
    where $N_{\SwitR}$ and $N_{\SwitL}$ are the numbers
    of events which satisfy each of the selections and
    the {\em acceptance},
    $A_{\mathrm{RL}}$, is the average of the theoretically calculated
    Bhabha cross sections accepted by the \SwitR\ and
    \SwitL\ selections, corrected for all 
    effects of detector response and
    LEP beam parameters.
    Because a detailed study of the bias in the $\SwitR$ and $\SwitL$
    definition cuts is possible, 
    $\Lrl$ is used for the primary luminosity measurement.

    A secondary luminosity measurement is defined 
    in terms of  the
    average  of the radial positions of the clusters
    at the reference planes of the 
    right and left calorimeters, $(\ra)$.  
    The counter, called \SwitA\ has a definition cut
    given by:
$$
    7.7~{\rm cm}
    ~<~
    \ra
    ~\equiv~
    \frac{\left( \rr + \rl \right)}{2}
    <~
    12.7~{\rm cm}
$$
    The corresponding luminosity measurement is defined
    as
\begin{equation}
    \La ~\equiv~\frac{N_{\SwitA}}{A_{\mathrm A}}
\label{eq:la}
\end{equation}
    where  the acceptance, ${A_{\mathrm A}}$, is the calculated
    Bhabha cross section  of the \SwitA\ selection,
    corrected for all effects of detector response and
    LEP beam parameters.
    The \SwitA\ selection has the advantage that the
    radial cut is effectively made in the beam-centered
    frame and therefore ${A_{\mathrm A}}$ is  largely 
    free of beam spot corrections.
    A second advantage of the \SwitA\ selection is that it
    defines a unique sample of left/right symmetric
    events, and is frequently used in the evaluation
    of systematic errors such as those
    related to energy reconstruction.
    Since the $\La$
    and $\Lrl$ selections
    have  
    largely complementary
    sources of systematic error with respect to the 
    acceptance--defining cuts,
    the comparison of the two selections can be
    used as a check on our understanding
    of coordinate bias and the effects of
    the LEP beam parameters on the luminosity measurement.
\subsection{Data samples}
\label{sec:data}
%
%
In this paper we consider data taken during the 1993, 1994 and 1995
runs of the LEP collider.  In 1993 and 1995, energy scans were
performed and substantial amounts of data were taken at
two energy points approximately 1.8 GeV above and below the 
$\Zzero$ peak.  In the following, we divide the data into
nine different samples which have slightly different characteristics.

In the 1993 and 1995 scans, each energy point has been analyzed
separately.  This allows us to account for any differences in the
selections at the different energies, including those induced
by changes in the LEP beam parameters (section~\ref{sec:beam}).

The 1994 data were all taken near the $\Zzero$ peak.  
These data have
been divided into three different samples.  
The first two
samples are necessary because of a large change 
which occurred in the 
average vertical offset of the LEP beam 
at the OPAL  intersection point (see section~\ref{sec:beam}).
The third sample corresponds to approximately the last 
month of running in 1994.  In this period LEP changed
from 8 bunch running to 4 bunch running in preparation
for operating with bunch-trains in 1995.  In addition, the 
OPAL Silicon-Microvertex Detector was removed for 
repairs, significantly reducing the amount of material
in front of the left calorimeter.  

A precise 
determination of the OPAL luminosity has not been undertaken
for data recorded
before stable LEP operation was achieved preceeding
the 1993 and 1995 energy scans.  
The 1993 ``prescan'' period was used to achieve stable
operation with the newly installed \SW\ luminosity monitor and the
1995 prescan period was used to commission the
readout necessary for understanding operation of 
the luminosity monitor with bunch trains.

The number of luminosity and multihadron events recorded
in each of the samples is given in \tb{tab:data}.

\begin{table}[htbp]
\begin{center}
\begin{tabular}{|l||r|r|r|r|r|}
\hline
Sample  &  Energy & Multihadrons & $\SwitR$ & $\SwitL$ & $\SwitA$  \\
        &  (GeV)  &              &          &          & \\
\hline
\hline
93  $-2$ &$89.45$&$  84710$&$ 697370$&$ 699041$&$ 702729$\\
93 pk    &$91.21$&$ 262735$&$ 689570$&$ 690487$&$ 694568$\\
93  $+2$ &$93.04$&$ 123895$&$ 682445$&$ 683642$&$ 687282$\\
\hline
94 a     &$91.24$&$ 267413$&$ 698423$&$ 701928$&$ 704489$\\
94 b     &$91.22$&$1238184$&$3238878$&$3252718$&$3267218$\\
94 c     &$91.43$&$  11135$&$  29066$&$  29211$&$  29293$\\
94 c'    &$91.22$&$  68324$&$ 177864$&$ 178154$&$ 179199$\\
\hline
95  $-2$ &$89.44$&$  83254$&$ 691317$&$ 690522$&$ 695735$\\
95 pk    &$91.28$&$ 139328$&$ 362725$&$ 362853$&$ 365130$\\
95  $+2$ &$92.97$&$ 126305$&$ 677469$&$ 675479$&$ 680982$\\
\hline
\hline
Total       &       &$2405283$&$7945127$&$7964035$&$8006625$\\
\hline
\end{tabular}
\end{center}
\caption[Data samples]{ The mean center-of-mass 
energy and  numbers of multihadron and luminosity
events recorded in the 1993, 1994 and 1995 data samples.
For the purposes of the luminosity analysis the c and c$^\prime$
data samples have been analyzed together.  They
are split in two when fitting  the lineshape because of
the large change in beam energy which occured in the 
middle of this sample.
\label{tab:data}
} 
\end{table}

\section{Detector}
\label{sec:det}
%
%
%
%
 
\subsection{Physical construction}

    The \SW\ Luminometer consists of two cylindrical
    small--angle calorimeters encircling the beam pipe at approximately 
    $\pm 2.5$~m from the interaction point.  
    Each calorimeter is a stack of 19 layers of silicon sampling wafers
    interleaved with 18 tungsten plates,  and 
    mounted as two 
    interlocking C-shaped modules around the LEP beam pipe.  
The first sampling layer lies in front of the first layer of tungsten
to provide improved presampling of the energy lost in upstream material.
    One such calorimeter is shown in \fg{fig:det_iso}.
    Electromagnetic 
    showers, 
    initiated by small--angle Bhabha scattered electrons, are almost 
    totally contained within the 140~mm sensitive depth of the detector,
    which 
    represents 22 radiation lengths $(\x)$.  
    The first 14 tungsten plates are each 1~$\x$ thick, 
    while the last 4 are each 2~$\x$ thick.
    The sensitive area of the calorimeter fully covers radii 
    between 62 and 142~mm from the beam axis.
 
    Mechanically each half calorimeter, or ``C'', consists of 
    19 half-layer 
    assemblies.  
    Each half-layer consists of a precisely machined tungsten 
    half-disk glued to the inner radius of a 2~mm thick 
    aluminum support plate.  
    The edges of the tungsten half disks that mate
    when the two C's close
    are bevelled 
    so that projective cracks in the absorber are
    eliminated.   
    The plate carries eight overlapping detector wedges 
    and a semicircular printed circuit board.  
    The detector wedges physically overlap, so that the active 
    areas of all adjacent
    wedges are contiguous.
    The four C's contain a total of 608 detector wedges.
    Each is a large, thick-film, ceramic hybrid carrying a 64-pad 
    silicon wafer diode,
    four AMPLEX readout chips and about 130 other electronic components.
    The pad layout of the silicon diodes is shown in 
    \fg{fig:det_wafer}.
    The pads have a radial pitch of 2.5~mm, and are arranged in an $r-\phi$
    geometry.
The diodes were fabricated by the Hammamatsu Corporation, and have proved
to be extremely reliable.
The average leakage current at $22^\circ$C and 80~V bias was 0.7~nA/pad
(1~nA/$\mathrm{cm}^2$) when the detector was installed.
In the entire sample of 785 delivered diodes (50,240 pads) the largest
leakage current for a single pad was 12~nA.
Radiation damage from catastrophic losses of the stored beam into the
calorimeters during seven years of operation at
LEP has increased the total leakage current by about a factor of ten, but
only three pads now exceed the AMPLEX bias current limit of
$\sim200\,\mathrm{nA}/\mathrm{pad}$.

   Every second layer is rotated by 11.25 degrees,
    (half the silicon wafer angle) 
    to 
    reduce any systematic effects which might be 
    introduced by the physical 
    overlapping of adjacent wafers in each layer.
    As each wafer contains two azimuthal
    segments, the azimuthal
    and radial structure
    of the pads in each layer is identical.  
    The projective
    structure of the azimuthal segmentation
    of the detector facilitates the
    cluster and coordinate reconstruction 
    described in sections~\ref{sec:ev_recon} and 
    \ref{sec:ev_meas}.  The radial segmentation
    of the detector is cylindrical,  but the approximate projective geometry
    with respect to the interaction point can
    still be exploited by the clustering algorithm.
    The lack of radial projectivity is in fact crucial to our ability
    to reconstruct a continuous radial coordinate with minimal
    systematic uncertainty (see section~\ref{sec:average_coord}).

    Two 15~mm precision dowels penetrate the entire stack,
    including the 
    aluminum front and rear plates, to hold the half-layers in 
    alignment (see \fg{fig:det_iso}).  
    Distortion of each half-layer out of a plane is limited by 
    spacers at the 
    inner and outer radii.  
    Spring loaded clamps on each of the sixteen 3.0~mm \
    cooling pipes which penetrate the 
    detector constrain the position of each layer 
    along the detector axis.
    These clamps also ensure adequate thermal 
    conductivity to the cooling water.
 
    The two C's slide together on two massive brass 
    dowels which ensure that 
    the silicon wafers, which overlap as the detector is closed, do not 
    violate their 
    $300~\mu$m clearances during assembly around the beam pipe.
    The relative alignment of the C's in the direction of
    closure is crucial in determining the radial 
    geometry, and is determined by clamping the C's
    together against massive and precisely machined
    lips on the upper and lower support blocks.
    This method of alignment is an order of magnitude
    more precise than the $\sim 10~\mu$m commonly achieved by 
    conventional techniques of doweling transversely to the
    closing direction.

    Sixteen 3~mm diameter water cooling pipes penetrate each 
    calorimeter 
    C at a radius of 220~mm to remove the 170 Watts generated by 
    the front 
    end electronics as close as possible to the source.  
    During operation 3~$\ell$/min  of 
    cooling water per C flows through the detector.
    The nominal temperature of the cooling water is
    16 $^\circ$C.
    Maximum temperature differences within the
    detector are measured to be 
    less than 2.0  $^{\circ}$C.
    During operation of the detector at LEP, 
    the temperature of each half ring is monitored with
    thermistors located either near the coolest or the warmest
    part of the half ring,  allowing
    the temperature gradients within the detector
    to also be monitored.
    Due to the slow accumulation of debris 
    in the filter of the  heat exchanger used 
    to cool the $16~^{\circ}$C water,
    the 
    average 
    temperature of the half-planes 
    slowly increased from 
    22.5~$^{\circ}$C, at the start of 1993 scan,
    to 24.5~$^{\circ}$C at the end of 1995 scan,
    when the reason for the temperature increase
    was understood and corrected.

    The effect of temperature on the radial metrology is discussed
    below in section~\ref{sec:rmetro}.

\begin{figure}[tbh!]
 \mbox{\epsfxsize17cm\epsffile{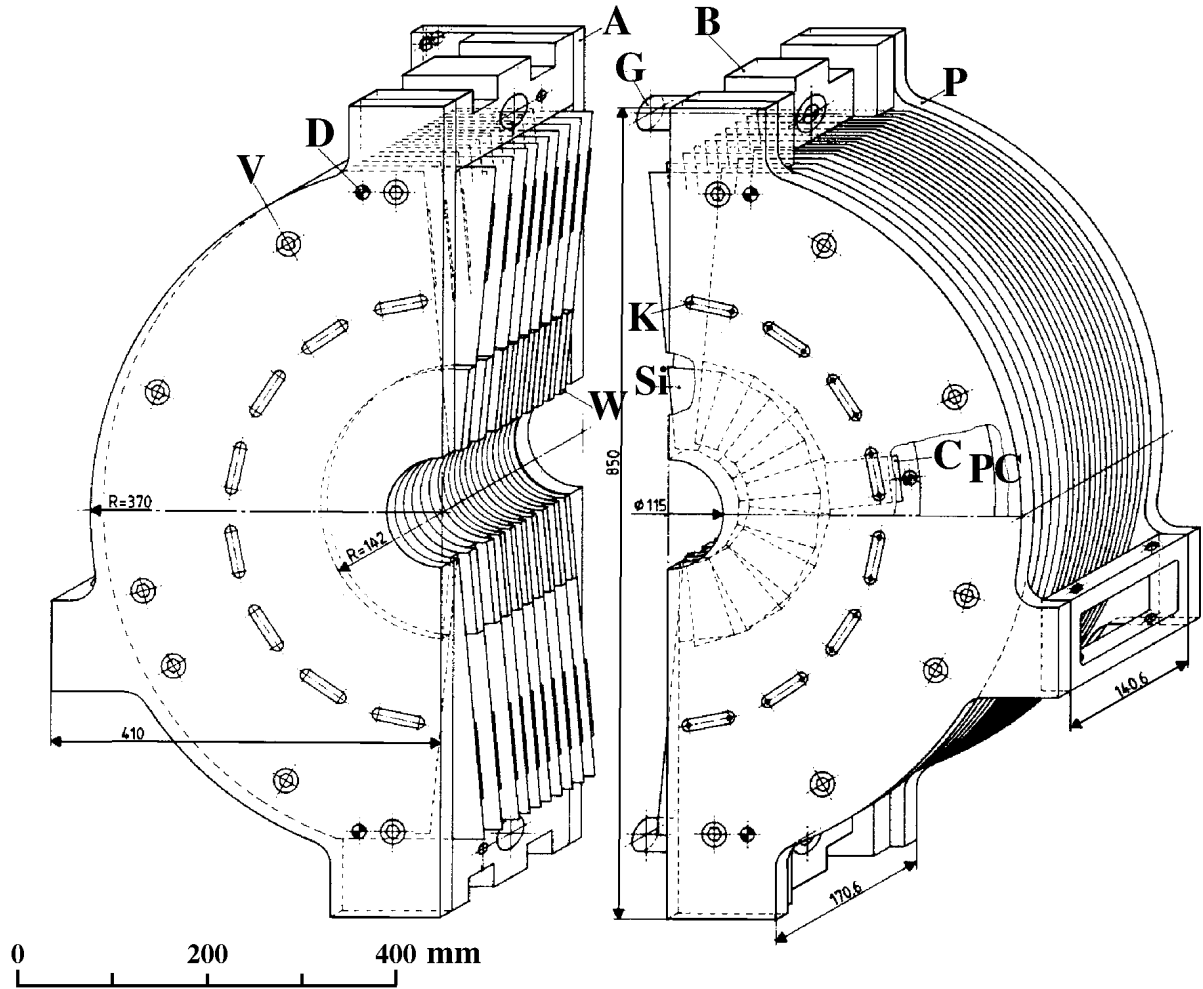}}
    \caption[\SW\ isometric view]{
     An isometric view of one of the \SW\ calorimeters, separated
    into two C's, {\bf A} and {\bf B}.  
    The 15~mm precision dowels which radially align
    the individual half layers are labeled as {\bf D}.
    The holes
    for the cooling pipes are labelled as {\bf K}.
    The clamps which fix the axial position of the
   half layers on the cooling pipes are not visible.
    The axial position of the half layers is also constrained
    by bolts and spacers at locations {\bf V}, as well as by
spacers at the inner bore.
    One of the brass dowels which align the two
    C's is labelled as {\bf G}. 
    On the left C, the beveled edges of
    the tungsten layers are visible and labeled as {\bf W}.  
    The silicon diode wedge structure in two successive
    layers is shown on the right C and labelled {\bf Si}.  Also shown
    on the right C 
    is a PC motherboard ({\bf PC}) and the ceramic hybrid ({\bf C}).  
    The dimensions shown on the figure are mm. 
    }
    \label{fig:det_iso}
\end{figure}

\begin{figure}[tbh!]
\begin{center}
 \mbox{\epsfxsize17cm\epsffile{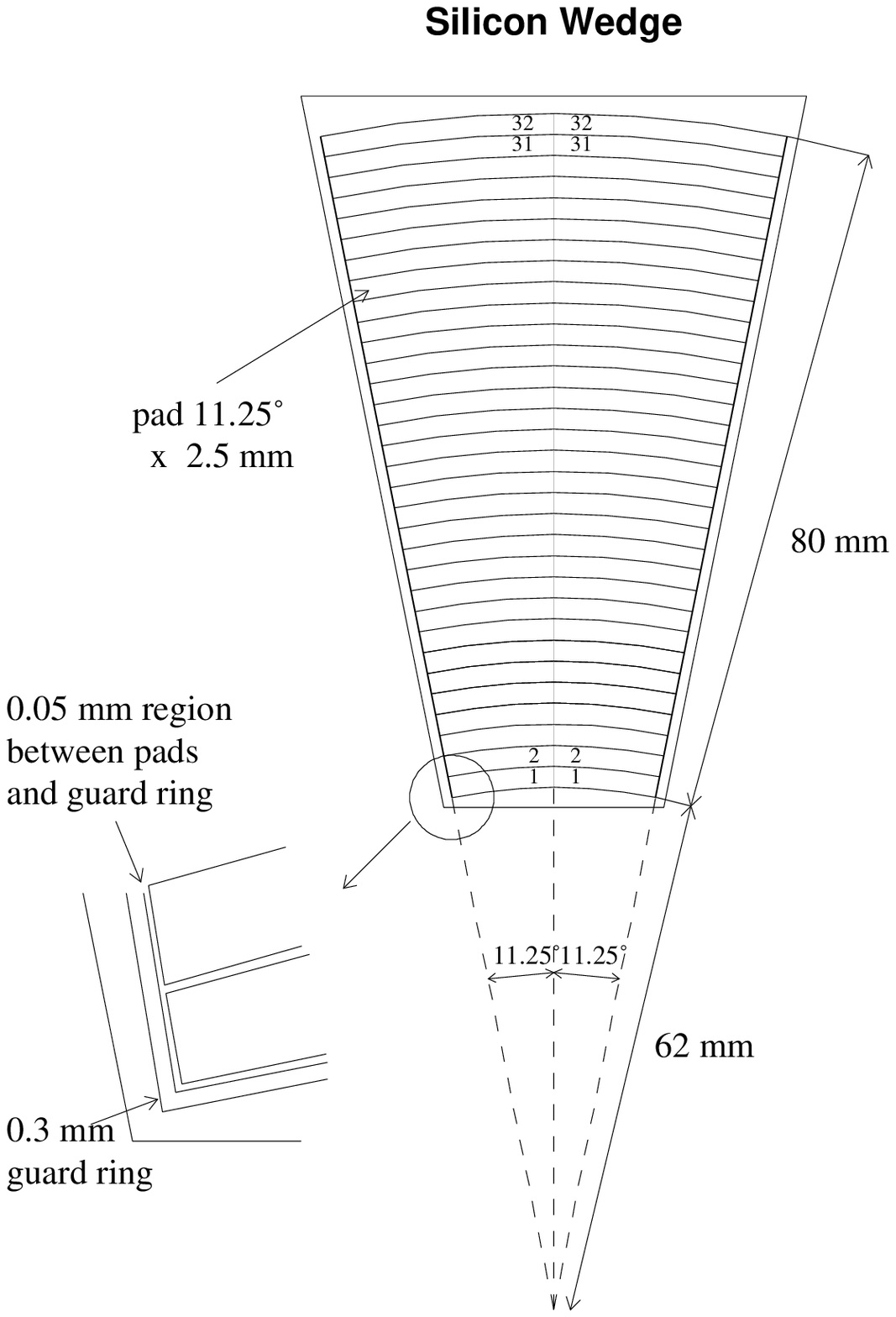}}
\end{center}
    \caption[Pad geometry]{
    A schematic illustration of the silicon diode pad geometry.
    The silicon detectors are divided into 2 rows of 32 pads, 
    each covering $11.25^{\circ}$ in azimuth.
    The radial pad segmentation is 2.5~mm.
    Silicon detectors within each layer of the calorimeter
    are physically overlapped, so that the azimuthal boundaries
    of their active regions coincide and there are no dead or
    ``double counted'' regions.
    }
    \label{fig:det_wafer}
\end{figure}

\subsection{Beam pipe and upstream material}

An important consideration in the design of the 
\SW\ calorimeters, over which we had little control,
was the location of material associated with the existing 
parts of the OPAL detector.
A flared beam pipe which
would have allowed particles to exit the beam pipe at normal
incidence was ruled out by the installation 
logistics of the  OPAL microvertex detector.
Instead, 
the material traversed by
particles originating at the interaction point 
was  reduced by extending the beryllium portion of the
cylindrical OPAL beam pipe and modifying its supports.
The distribution of material upstream of the calorimeters
is  shown in 
\fg{fig:det_mat}.  
Note that in the crucial region of the inner acceptance
cut the upstream material totals approximately 0.25 radiation lengths.
It was not possible to further reduce the shadow
cast in the middle of the detector's
radial acceptance by the microvertex detector cables
and by the flanges and support structures of
the OPAL pressure
pipe~\cite{bib-beam-pipe}.
Fortunately, the reconstruction of the shower position
remains largely unaffected by this additional material
(see section~\ref{sec:rad_meas}). Furthermore,
this region is not crucial for
the LEP I luminosity measurement.  
The effects of the degraded
energy resolution are important, but measurements
of the longitudinal development of the showers can be
used to correct for energy which is deposited in this dead
material (see section~\ref{sec:en}).
 
\begin{figure}[tbh!]
\begin{center}
 \mbox{\epsfxsize15cm\epsffile{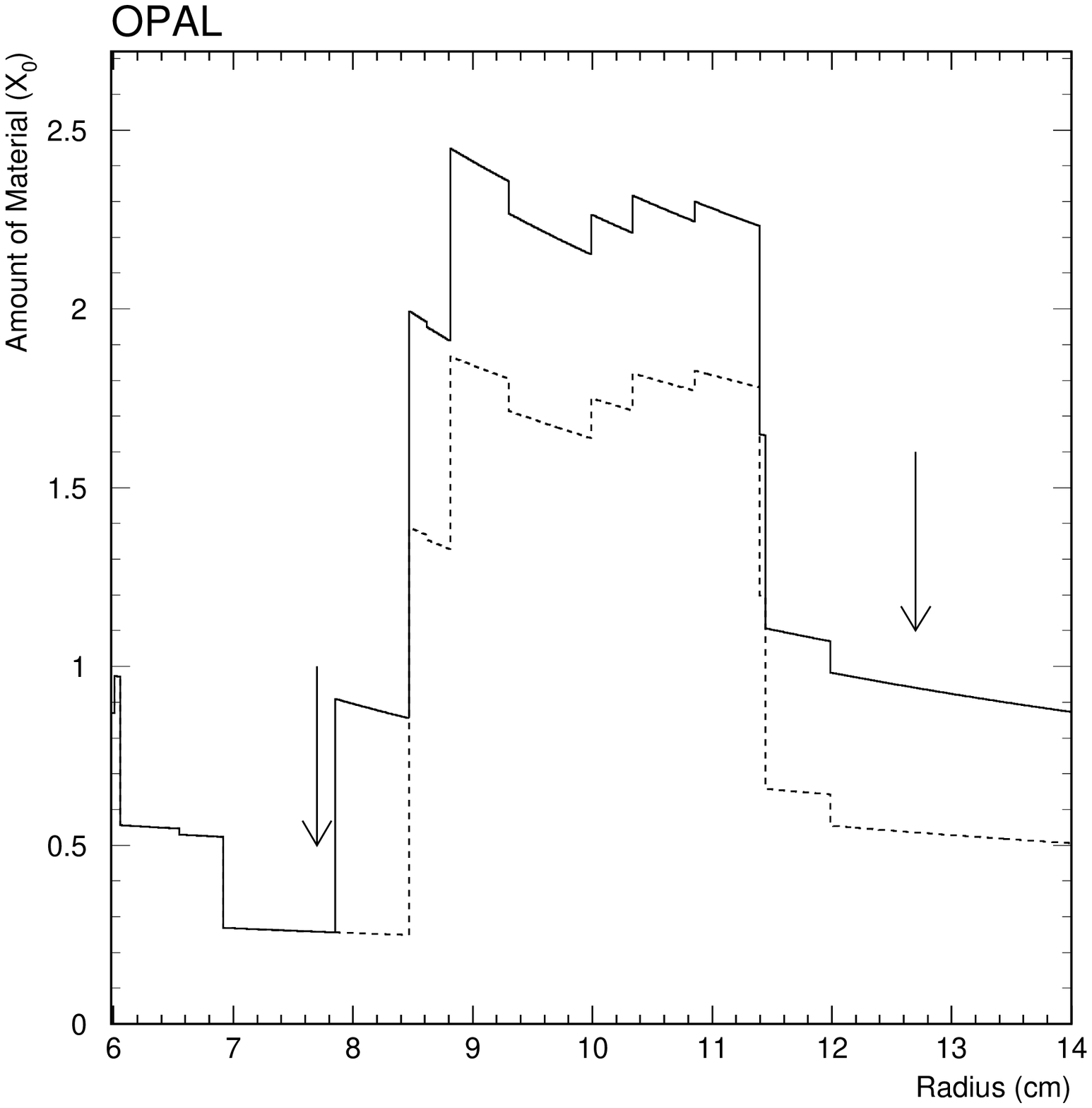}}  
\end{center}
    \caption[Upstream material]{
    The calculated material traversed by particles originating at
    the interaction point as a function of calorimeter
    radius, measured at the reference plane (246~cm) for the 1993--1994
    detector configuration.
    The solid curve corresponds to the left, the dotted curve to the right
    side. The larger amount of material on the left is
    due to the passage of cables from the OPAL microvertex
    detector.
    The arrows show the location of the 
    acceptance definition cuts on shower radius.
    }
    \label{fig:det_mat}
\end{figure}

\subsection{Radial metrology}
\label{sec:rmetro}

One of the most demanding aspects of the SiW luminosity measurement is
to establish the absolute radial dimensions of the detector.
The details of our techniques and measurements can be found 
in reference~\cite{bib-rmetro}.
Here we give a simplified description of the two essential steps of
the metrology: determining the geometry of the detector under
ideal conditions in the laboratory, and monitoring the
changes which occur when the detector is brought into operation
at LEP.

\subsubsection{Laboratory measurements}

The geometry of the Si wafers themselves can be defined very precisely.
The semi-conductor
industry routinely fabricates diode structures across the full width of
a 10~cm Si~wafer with sub-micron geometrical precision, and
electrostatic calculations confirm that the sensitive boundaries of the
detector pads correspond within a fraction of a micron with the midline
of the $50\mu$m inter-pad gap.
Sharply defined visible boundaries in the detector
aluminization are created above the diode implants in a subsequent 
photo-lithographic step, and
provide convenient visible references which are directly
related to the diode boundaries.

We used several microscopes mounted at fixed radii on stable
rotating arms to align and measure each of the 608 silicon wafers in the
luminometer using these visible references.
All wafers were individually measured during construction in 1993.
A sizable fraction were remeasured twice in 1994, and all of the wafers used in
coordinate determination were remeasured twice in 1995.
A smaller set was remeasured in 1996.

The absolute radii of the microscopes were calibrated and monitored against
a calibration plate, whose dimensions were determined with an interferometer,
and checked against the principal metrology instrument of the central CERN 
Metrology lab.
The absolute radius swept out by the primary microscope graticule
is
\begin{center}
   $R_{\mathrm{micro}} = 142.0305(19)$mm  at $T = 21.5\,^\circ$C
\end{center}
where the quoted error includes the effects of the absolute uncertainty in
the radius of the calibration plate ($0.7\mu$m), the allowance for possible
calibration plate distortions ($1\mu$m), and the uncertainty in the long term
stability of the microscope ($1.45\mu$m).

    The calorimeters each consist of two stacks of half rings,
    on which the silicon detectors are mounted.
    Within each stack, individual silicon detectors were verified
    to lie on the best--fit semicircle with an 
    RMS scatter of about $1.3\mu$m, as shown in~\fg{fig:half_ring_residuals}.
    When the two calorimeter halves are brought together, there is some
    residual misalignment of the two halves with respect to each other.
These misalignments are on the order of $10-20\mu$m, and are quite
reproducible over time.
    When integrated in azimuth, the higher order corrections to the
    detector acceptance introduced by the relative separation and shear
    between half-rings ($\Delta x_{c}$ ,$\Delta y_{c}$)  are less
    than $10^{-6}$ so that only the average radial position 
    of each layer 
    of silicon detectors needs to be taken into account.

\begin{figure}[tbh!]
  \begin{center}
    \mbox{\epsfxsize12cm\epsffile{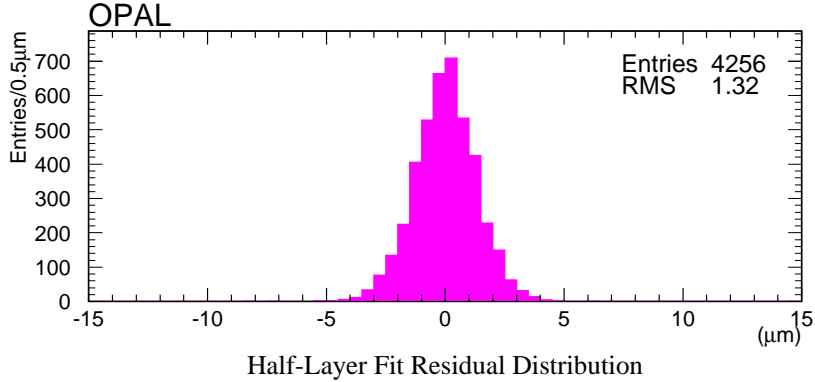}}

\caption[Half-ring layer residuals]{
The residuals of the individual detector radii from the fit half-circles
for all the standard metrology measurements.
The RMS of $1.32\mu$m includes the relative precision achieved in placing the
the individual detectors on each half-ring, as well as the resolution of
the subsequent measurements of their positions.
  \label{fig:half_ring_residuals} }
  \end{center}
\end{figure}

Variations in the individual half-ring separations in each layer as
observed each time the layer is stacked or unstacked
produce an RMS variation in layer radius of $1.9\mu$m.
The uniformity with which the separation between the two completed C's of
each calorimeter can be preserved when the detector is 
liberated from the stacking fixture produces
a further RMS variation in radius of $1.5\mu$m.

\begin{figure}[tbh!]
  \begin{center}
    \mbox{\epsfxsize12cm\epsffile{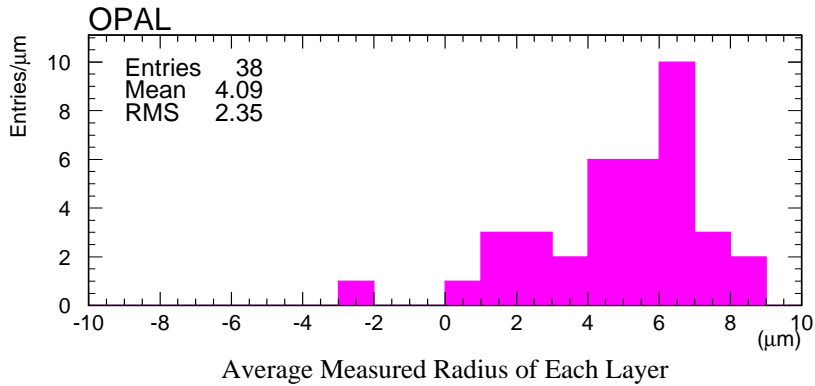}}

\caption[Average layer radii]{
The distribution of the average measured radii for each layer with
respect to $R_{\mathrm{micro}}$.
The RMS of $2.4\mu$m is due mostly to actual differences in the layer
radii, since the error on each entry is typically less than $1\mu$m.
The abscissa is in units of $\mu$m relative to the microscope graticule.
  \label{fig:full_layer_radii} }
  \end{center}
\end{figure}

Figure~\ref{fig:full_layer_radii} shows the distribution of the average
radius measured for each of the 38 layers.
The RMS of $2.4\mu$m is due mostly to actual differences in the layer
radii, since the error on each entry is typically less than $1\mu$m.

A list of the most interesting
parameters for layer 7 in the two calorimeters is shown
in \tb{tab:RADI}.
The entries are from the average of all the measurements
made on these layers.
The layer~7 radii, alone, determine the actual luminometer acceptance
boundary (see section~\ref{sec:rad_meas}).
The difference of each layer's radius with respect to layer~7 propagates
to the study of coordinate systematics
(the coordinate ``anchoring'' described in section~\ref{sec:anchor}).
Since the luminometer acceptance is determined by the {\em average} of the
right and left radii of the layer~7 pad boundaries, nominally at 77~mm,
we take

\begin{center}
$R_{\mathrm{std}} = 
77.000\mathrm{mm} + \frac{1}{2} ( 5.0 + 4.7 )\times 10^{-3}\mathrm{mm} =
77.0048 (32)\mathrm{mm}$
\end{center}

\noindent
as the effective acceptance radius of the luminometer in standard metrology
conditions.
The quoted error takes into account 
the effects of the absolute uncertainty in
items {\em a-f} of \tb{tab:radial_syst}.
The actual acceptance radius will differ from $R_{\mathrm{std}}$ due to dimensional
changes associated with bringing the detector from standard metrology
conditions into operation at LEP.

The small annual changes in the detector radius measured under standard
metrology conditions are well within the systematic errors we assign to the
stability of our absolute radial scale.
We found no evidence 
that the detector suffered any change in its standard
metrology radius as a function of time.

\subsubsection{Measurements in LEP operating conditions }
\label{sec:rmetrolep}

The largest single source of uncertainty in the detector radius arises
from the difficulty of measuring the dimensional changes which occur when
the calorimeters are removed from the laboratory and brought into actual
operation at LEP.
We have nevertheless made such measurements, and do not rely simply
on assumptions.

The first change required in bringing the detector into operation
is to rotate the calorimeter from horizontal to vertical, which
radically changes the mechanical loading placed on the main alignment dowels.
After rotating each calorimeter, the standard metrology measurements were 
carried out on the first layer.  The relative motion of survey marks
glued to the inner bore on deeper layers were also observed.

With each calorimeter in the vertical position, the flow of cooling
water was established, the front-end electronics were powered and the
detector allowed to come to thermal equilibrium.
The mean temperature of the inner and outer thermistors increased by 
0.5$^\circ$C and a thermal gradient of $\sim2.0^\circ$C was established.
The first layer metrology and the bore measurements were then repeated.

Some of the measured effects may be artifacts resulting from
the non-ideal metrology fixtures arranged for these tests.
The non-redundant nature of these measurements also precludes extracting
reliable estimates of systematic measurement errors from the observed scale
of fluctuations, and we have quoted errors representative of measurement
uncertainties achieved with the same equipment, but
in better conditions.
We then take a conservative approach by taking the scale of the
uncertainties from the differences
observed in the behavior of the two calorimeters, rather than from the
estimated measurement errors.
\Tb{tab:CHANGER} summarizes the observed changes.
Averaging over the two calorimeters, we estimate a radial change
of $-9\pm3\,\mu$m in bringing the luminometer from metrology conditions
into actual operation, where the error is half the observed difference
between the two calorimeters.

\begin{table}[tbh!]
 \begin{center}
  \begin{tabular}{|l||c|c|c|c|c|}
\hline
Calorimeter & Layer 7     &     1/2 Ring A        &  1/2 Ring B       &
                   $\Delta x_{c} = x_{c}(A)-x_{c}(B)$ &  
                   $\Delta y_{c} = y_{c}(A)-y_{c}(B)$ \\
  no.  & radius($\mu m$) &  radius($\mu m$) &  radius($\mu m$)  &
                           ($\mu m$)     &     ($\mu m$)        \\
\hline
\hline
Right & $   5.0\pm 0.9$ & $   1.2\pm 0.9$ & $   0.1\pm 0.7$ & $  13.0\pm 3.3$ & $   8.7\pm 2.1$ \\
Left  & $   4.7\pm 0.9$ & $   2.4\pm 0.5$ & $   1.1\pm 0.6$ & $   7.7\pm 3.4$ & $  18.7\pm 1.2$ \\
\hline
\hline
  \end{tabular}
    \caption[Average half-ring radii ]
             {
     Average results for fits to the metrology data taken for
     layer 7 of the two half rings.
     The two half-rings of each calorimeter are designated
     ``A'' and ``B''.  The coordinates  ($x_{c}$,$y_{c}$)
     give the centers of the half-circles fitted to the detectors of
     each half-ring.  The $x$-axis is parallel to the direction of closure.
     The errors are the RMS errors in the mean, evaluated from
     the spread observed in the fit results.
     The errors {\em do not} include the uncertainty in the absolute
     radial scale. All the radii are to be read as microns above 77\,mm.
     They are valid in standard metrology conditions, detector closed,
     at $T = 21.5\,^\circ$C. 
    \label{tab:RADI} }
 \end{center}
\end{table}

\begin{table}[tbh!]
 \begin{center}
  \begin{tabular}{|l|c|c|}
\hline
              &  $\Delta R_1$ ($\mu$m)   &  $\Delta R_2$ ($\mu$m) \\
\hline\hline
  Stack rotation (horizontal to vertical) 
&   $-1 \pm 1$  &  $-7 \pm 1$    \\
  Thermal loading          &   $-5 \pm 1$         &  $-5 \pm 1$    \\
\hline\hline
  Total measured radial change      &   $-6 \pm 1$         &  $-12 \pm 2$    \\
\hline
  \end{tabular}
    \caption[Changes between operating and standard conditions]
             {Summary of the different contributions to the radial change
             observed for the top layer radius of calorimeter 1 and 2
             ( right and left)
             in going from standard metrology to operating conditions.
             Stack rotation refers to moving the detector from
             the horizontal (standard metrology) to the vertical
             (operational) position.
    \label{tab:CHANGER} }
 \end{center}
\end{table}

The inner acceptance radius of the luminometer in normal operating
conditions is then
\begin{center}
    $R_{\mathrm{in}} = 76.9958(44)$\,mm at T = $22.0^\circ$C.    \\
\end{center}
where the quoted error takes into account
the effects of items {\em a-g} in \tb{tab:radial_syst}.

The temperature of the luminometer was continuously monitored during
operation, and the precision solid-state 
resistive temperature probes were carefully
cross-calibrated to the thermometer used when the absolute radius
of the calibration plate was determined.
The changes in the acceptance radii (77 and 127~mm) 
as a function of operating temperature
were measured to be

\begin{center}
$$
   \Delta R_{\mathrm{in}} = 3.9 [\mu\mathrm{m} /\,^\circ\mathrm{C}] \times
     ( T_{avg} - 22.0\,^\circ\mathrm{C})
$$
$$
   \Delta R_{\mathrm{out}} = 4.1 [\mu\mathrm{m} /\,^\circ\mathrm{C} ] \times
     ( T_{avg} - 22.0\,^\circ\mathrm{C} )
$$
\end{center}
\noindent
    with respect to normal operating conditions.  
The mean temperature of each data sample was determined and the
acceptance corrections (less than $3 \times 10^{-4}$, as shown in
table~\ref{tab:cor_all}) are calculated
on the basis of the relations given in section~\ref{sec:radial_intro}.
The systematic errors listed in table~\ref{tab:syssum_all} reflect a
$0.1\,^\circ\mathrm{C}$ uncertainty in the average operating temperature
and a 10\% uncertainty in the measured thermal expansion coefficient.
The smallest error (for 1993 pk) is taken as a component correlated among
all data samples.

The SiW luminosity acceptance calculation
assumes that the inner and outer acceptance radii
are exactly 77.000 and 127.000~mm.  Apart from the operating temperature
correction just discussed, it is therefore necessary to apply a base metrology
correction which reflects the $4.2\mu$m difference between the actual and
nominal dimensions, as well
as a small correction which accounts for the difference between
an ideal circular pad geometry and the actual polygonal pads implemented
in the detector fabrication masks, with chord
lengths of $\sim 1^\circ$.
This base radial metrology acceptance correction, $\Delta A / A$,
at $T_{avg} = 22.0\,^\circ$C is

\begin{center}
$$
\frac{\Delta A}{A} =  (+1.55\pm1.40)\times 10^{-4}
$$
\end{center}

\noindent
which is to be applied to all data taken between 1993 and 1995 (in
addition to the operating temperature corrections which are time dependent).

Diligent attention to the mechanical and thermal stability of the detector
has allowed us to reduce the uncertainty in the geometry of the
luminometers more than a factor of 5 beyond the level which would have
compromised our goal of a 0.1\% luminosity measurement.
The inner radius of the luminometer acceptance region is determined
with a precision of $4.4\,\mu$m
taking into account mechanical deformations,
temperature effects, and the precision of the optical apparatus used
for the survey.
This uncertainty in the acceptance radius contributes a systematic error
of 1.4$\times10^{-4}$ to the OPAL luminosity measurement
during 1993-1995.
\Tb{tab:radial_syst} gives the components of the total uncertainty
in the detector radius.

\begin{table}[tbh!]
 \begin{center}
  \begin{tabular}{|ll|c|}
\hline
Item &Systematic sources
            &  $\Delta R$  \\         
\hline\hline
{\em a} &   Calibration plate radius                &   $0.7\,\mu$m \\
{\em b} &   Calibration plate distortions           &   $1.0\,\mu$m \\
{\em c} &   Microscope stability                    &   $1.45\,\mu$m \\
{\em d} &   Half-ring separation stability          &   $1.9\,\mu$m \\
{\em e} &   Cover plate reproducibility             &   $1.5\,\mu$m \\
{\em f} &   Layer 7 measurement error               &   $0.6\,\mu$m \\
{\em g} &   Changes between metrology \& operation  &   $3.0\,\mu$m \\
{\em h} &   Operating temperature expansion         &   $0.4-0.8\,\mu$m \\
{\em i} &   Low detector polygon correction         &   $0.25\,\mu$m \\
\hline\hline
   & Total radial metrology systematic error       &   $4.4\,\mu$m \\
\hline\hline
   & Corresponding error in acceptance       & 1.4 $\times 10^{-4}$\\
\hline
  \end{tabular}
    \caption[Metrology systematic errors]
{Summary of the systematic errors for the radius measurements in
             standard operating conditions.
    \label{tab:radial_syst} }
 \end{center}
\end{table}

\subsection{Axial metrology}
\label{sec:zmetro}

The metrology which determines the separation of the detectors in $z$, 
along the beam axis, is described in detail in reference~\cite{bib-rmetro}.
Briefly, the measurement 
is based on interferometric and Johansson bar
measurements of the distance between flanges on the 
OPAL beam and pressure pipes~\cite{bib-beam-pipe}.  
These pipes are then used as rulers to measure the distance
between the calorimeters during operation.
The pressure pipe, which forms the primary length reference,
has a length of
$473.9858(62)$~cm at $22.3^{\circ}$C.  Optical grating position
monitors with 0.5~$\mu$m resolution are used to measure the position
of the calorimeters with respect to these pipes during detector
operation.
These monitors must track the relative motion of the calorimeters
as the expansion of the central detector pressure vessel during its
pressurization to 4~bar carries
them away from each other by about 10~mm.
The length of the pipe is corrected for thermal expansion
based
on measurements from probes located on the pipe.

Corrections to the nominal
detector half-separation of 246.0225~cm are made with a precision of
approximately 60~$\mu$m on the basis of these measurements,
corresponding to an 
uncertainty of about $5 \times 10^{-5}$ in the luminosity. 
The contributions to the systematic error from the axial
metrology are listed
in \tb{tab:det_zerrors}.

\begin{table} 
\begin{center}
\begin{tabular}{|l||c|c|}
\hline
Systematic sources                             &   1993--4    &   1995      \\
\hline
\hline
Position of layer 7 relative to calorimeter reference face    & $34\,\mu$m  & $60\,\mu$m  \\
Length of the pressure and beam pipes          & $31\,\mu$m  & $31\,\mu$m  \\
Position monitor stability                     & $5\,\mu$m  & $2\,\mu$m  \\
Reference pipe temperature during calibration  & $10\,\mu$m  & $0\,\mu$m  \\
Reference pipe temperature during operation    & $15\,\mu$m  & $4\,\mu$m  \\
\hline
\hline
Total axial metrology systematic error         & $50\,\mu$m  & $68\,\mu$m  \\
\hline
\hline
Corresponding error in acceptance &$0.41\times 10^{-4}$&$0.55\times 10^{-4}$\\
\hline
\end{tabular}
\caption[Axial metrology systematic error]
{Summary of errors in correcting the nominal 246.0225~cm 
half-distance between
the layer 7 reference planes of the two calorimeters}
\label{tab:det_zerrors}
\end{center}
\end{table}

%
%
\subsection{Front end electronics}
\label{sec:det_electronics}

%
%
The detector is read out using front-end electronics~\cite{bib-robert-report} based on
the AMPLEX~\cite{bib-AMPLEX} chip.  
Each AMPLEX contains 16 channels
which comprise a charge amplifier, a shaping amplifier and a
track and hold.  
Four AMPLEX chips are mounted on each ceramic allowing signals from the
64 pads on each silicon detector to be multiplexed to the common output.
The gain of the amplifier has been matched to 
signals expected from 45 GeV electromagnetic
showers~\cite{bib-ALEPH-AMPLEX}.
At shower maximum, a single layer of the calorimeter typically
records a signal equivalent to 300 to 400 minimum ionizing particles (mips)
or $\sim$ 1.0 to 1.3 pC for an incident 
electron of 45~GeV, and this is spread over a few pads 
(see section~\ref{sec:ev_recon}).  The signal seen by an
individual pad was well below the full scale
limit of the AMPLEX and the ADC used to digitize the signal,
which corresponds to more than 1000 mips.

Despite the large dynamic
range of the signals, the equivalent noise for each channel remained at
a level of 1500 to 2000 electrons for a 
typical detector capacitance of 20~pF,
giving better than 10:1 signal to noise for mips.  
While the detection
of single mips was not crucial in OPAL operation, the
good response of the detectors to muons was essential for the precision
alignment of the detectors in the test beam studies described in
section~\ref{sec:anchor}.  

The response of the AMPLEX to injected charge has been characterized
with both external and internal calibration signals,
as well as with signals from test beams and laboratory sources.  
The AMPLEX chips were individually tested with a custom--built
semi--automatic probe station, and sorted according
to gain.  The chips used on each ceramic were selected
to have equal gain, and a trimming resistor
was used to equalize the
gains of the ceramics giving an overall channel-to-channel
uniformity of approximately 1\%.
This degree of uniformity not only ensures optimum resolution
for the trigger threshold, but eliminates the need for a
database of calibration constants for off--line energy reconstruction.

The gain variations among the 16 channels of each AMPLEX, which can degrade
the uniformity of the radial position measurements, were appreciably
smaller.
Using the internal calibration system, the mean RMS gain variations within
each AMPLEX were measured to be $0.25\%$.
Real gain variations were likely to be smaller, since the observed width
could have been significantly broadened by variations in the calibration
system rather than in the AMPLEX gains themselves.

The AMPLEX gain and the cross talk among channels were
found to depend on the input capacitance
of each channel and of the \hold\  signal timing. 
Trimming capacitors, integrated into the
ceramic hybrid structure, were used to equalize the
total input
capacitance which otherwise would have
varied by
a factor of two from the inner to outer edge of the
detector.
We have optimized the \hold\ timing to provide a uniform response
to injected charge and to minimize cross talk.  
An empirical correction was made to the data to account 
for the residual $-2$\% cross talk of each channel to each of
the other 15 channels on the same AMPLEX.
This correction allowed us to accurately reconstruct
showers and to recognize close-lying secondary showers
which would have otherwise been masked by the cross talk
from the primary shower.  The effect of this cross talk
on the energy measurement is discussed in section~\ref{sec:en}.
The sensitivity of the cross talk to the \hold\
timing of the AMPLEX chip is further discussed
below in section~\ref{sec:bunchtrain}.

To minimize the 
effects of residual gain and cross talk variations that may be systematic
functions of the AMPLEX channel number, the AMPLEX channel ordering
was inverted between the two radial pad columns of each ceramic.

The reliability of the entire readout system has been excellent:
of the 608 installed detectors, only three are not currently functional.
For all of the LEP I running, the number of non-functional detectors
varied between one and three, none of them within three radiation lengths
of the shower maximum at layer $7\x$.
The signals from such detectors are ignored in event reconstruction,
and the missing energy interpolated from the depositions observed in
adjacent layers.
At the achieved level of reliability, such non-functional detectors cause
no measurable degradation in performance.

\subsection{Trigger electronics}

The AMPLEX chip also includes a trigger output.  When
used in trigger mode, the average of all 16 channels appears as a
single output.  
In the OPAL implementation, the signals from the
four AMPLEX chips which service each wafer are
further summed to provide a single analog trigger signal 
for each silicon wafer.  
Details of the OPAL 
design are given in reference~\cite{bib-robert-report}.

In trigger mode, the summed signals are further combined by
the trigger hardware, to provide whole calorimeter and wafer-tower sums.  
The tower signals allow the implementation of triggers
which require back-to-back showers in azimuth.  
Additional details for the trigger 
algorithm are given in section~\ref{sec:trig}.

On each LEP beam crossing, even if not triggered for
event readout, all of the signals used 
by the trigger are digitized and stored.  
This
trigger information is used for detailed monitoring of
the trigger (see section~\ref{sec:trig}). In addition, the trigger
information is used to determine independently  the
energy and azimuth of showers in the detector.  This capability
is exploited in background measurements 
(section~\ref{sec:back}) and 
to check for failures in the event readout and reconstruction
(section~\ref{sec:ev_recon}).

\subsection{Bunchlet triggering}
\label{sec:bunchtrain}

In normal operation LEP collides single bunches of electrons and positrons
at uniform intervals no shorter than $11\,\mu$s, but in 1995 LEP operated in
bunch-train mode, in which each normal bunch was replaced by a train of
up to four bunchlets separated by only a few hundred
nanoseconds~\cite{bib-bunchtrains}.
For the entire 1995 scan, LEP operated with 
3 bunchlets per train with an
inter-bunchlet 
spacing of 247~ns.  

Bunch-trains presented difficulties for the \SW\ 
detector, since the shaping time of the 
AMPLEX 
is comparable to the 1995 
inter-bunchlet 
spacing.
The AMPLEX operates in ``continuous time filtering mode'',
which means that the ionization charge injected by charged particles passing
through the silicon produces a shaped signal which peaks in about 270~ns, 
then decays away somewhat more slowly.
Readout of the detector requires that the \hold\  signal be asserted  at
a known time near the peak of the shaped signal, so that the instantaneous
analog signals are captured on storage capacitors.
In single-bunchlet LEP running the \hold\  can 
be issued at the proper time on each bunch crossing, but with bunch-trains, it 
can only be issued at the proper time with respect to a single bunchlet in 
each train.  

If we were to issue the \hold\  synchronous with the last bunchlet, we 
would measure only about half the energy of showers originating in the 
previous bunchlet, and only about 15\% of the energy of showers originating 
in the bunchlet preceding that.  Furthermore,  the inter-channel 
cross talk of 
the AMPLEX depends on the \hold\  timing.  
Even if we had been able to ensure 
adequate triggering on the degraded signals from the earlier bunchlets, it was 
not evident that our sophisticated clustering and coordinate reconstruction 
algorithms could have been modified to measure such stale showers precisely.

Another possible strategy
might have been to issue our \hold\  in time with one of the bunchlets chosen 
at random on each bunch crossing.  
Apart from the undesirable loss in statistics this would have entailed, 
it was not evident that the large number of stale signals from previous 
bunchlet 
events could have been distinguished from the low-energy tail of in-time 
radiative Bhabhas with sufficient precision.

We therefore decided to upgrade the luminometer with a ``wagon tagger''
to issue \hold\ in response to 
bunchlets
actually depositing energy in the calorimeters.
The wagon tagger is based on fast analog sums of the signals induced
by electromagnetic showers on the back-plane of the silicon wafers.
These sums
are locally discriminated and 
processed to provide a bunchlet synchronization   
signal.
The design of the wagon tagger was aided
by the analysis of test-beam data.
These data allowed us to ensure adequate sensitivity for both early
and late developing showers, with a minimum number 
of instrumented layers.
We included 6 layers (at 3, 7, 11 and 4, 8, 12 $\x$) in
the wagon tagger sums.

To achieve a low rate of false (or premature) tags, due to off-momentum 
electrons in bunchlets prior to the event of interest, while retaining 
sufficiently low thresholds to ensure full tagging efficiency, the wagon
tagger was operated in ``AND'' mode, 
which required coincident tags in the left 
and right calorimeters, in all but four runs 
during the 1995 scan.
In the absence of tags on earlier bunchlets the synchronization
signal was always issued for the last bunchlet in the train.
This allowed the efficiency of the wagon tagger to be precisely measured
on the last bunchlet, as discussed in section~\ref{sec:trig}.

%
\section{Event reconstruction}
\label{sec:ev_recon}
%
%
%
%
%
%

The detailed information available from the 38,912 individual pads of the
luminometers allows events to be reconstructed in a way
which is well matched to the strengths and weaknesses in the theoretical
treatment of the Bhabha scattering process.
The theoretical summation over increasing orders in perturbation theory
naturally delivers a more precise description of the higher-energy partons,
and becomes more approximate as their energy decreases.
The description of the angular distribution of radiated photons about the
primary axes of the event also becomes less precise as the event topology
becomes more complicated.
Both these features are mirrored in the natural characteristics of the \SW\
luminometer.

The lateral profile of electromagnetic showers in the dense medium of the
\SW\ calorimeters is characterized by a sharp central peak and broad tails,
as shown in \fg{fig:event_sh_shape}.
The sharpness of the shower cores and the fine granularity of the detector
allow us to reconstruct individual positions for incident electrons and
photons which are separated by about 1~cm, while
the inclusive nature of the calorimetric measurement of the energy in the
shower tails forms a natural summation over accompanying low-energy photons,
in a manner largely independent of their angular distributions.
We exploit the fine granularity of the detector and 
conservation of transverse momentum to exclude background clusters 
generated by accidental off-momentum electrons from the energy sum.

\begin{figure}[tbh!]
\begin{center}
 \mbox{\epsfysize21.7cm\epsffile{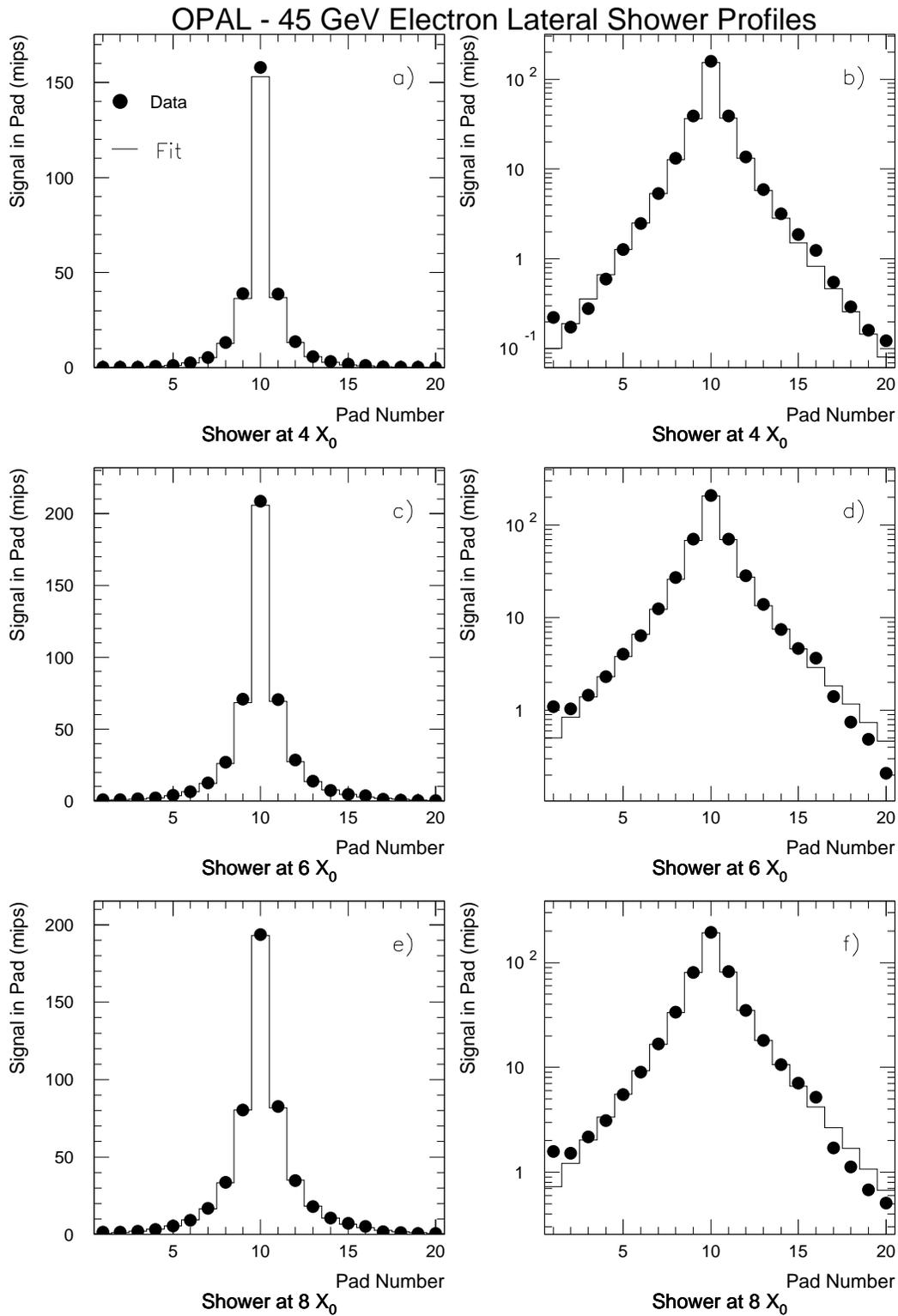}}
\end{center}
 \caption[Radial shower profiles]{
Average radial shower profiles at 4 $\x$ a) and b),
6 $\x$ c) and d), and 8 $\x$ (e) and (f).
The fit shows the parameterization of the electromagnetic
showers given in reference~\cite{bib-bernd}.
The left plots show the data on a linear scale, the
right plots on a logarithmic scale.
\label{fig:event_sh_shape}
}
\end{figure}

The final
luminosity measurement is not very sensitive to the parameters
used to distinguish individual shower cores
and to those used to 
determine which shower clusters belong to the Bhabha event. 


In this section we describe some details of
the cluster formation and describe how kinematic
information is used to select the clusters which
belong to an event.  
The first step of cluster finding is the peak-finding
algorithm which is used to find shower cores. 
Near the longitudinal shower maximum,
these shower cores are typically sharply 
peaked transversely with 
a FWHM of less than a single radial
pad width (0.25~cm). 
However, the tail of the energy distribution
extends laterally to
almost 10 pad widths (2.5~cm) as can
be seen in \fg{fig:event_sh_shape}.
This diffuse energy is associated with shower
cores to form ``clusters''.
The detector simulation described in section~\ref{sec:detsim} has been
tuned using actual data to reproduce the measured properties of this
clustering algorithm.

After all resolvable clusters have been assigned radial and azimuthal
coordinates, and their measured energies have been corrected for
showering losses in the dead material in front of the detector, as described
in section~\ref{sec:ev_meas}, the clusters belonging to a candidate Bhabha
are selected by a kinematic selection algorithm, described in
section~\ref{sec:kine}.  
This algorithm differentiates between showers from radiative Bhabha
processes and those due to background from the LEP machine by
exploiting the transverse momentum balance intrinsic in the set of
clusters belonging to a true Bhabha event. 
When the candidate Bhabha clusters are identified, the most energetic
cluster found on each side of the detector is used to define the event
coordinates, while the energy sum of all selected clusters on each
side is used in making the cuts on the minimum and average energies
required for fiducial Bhabha selection. 
By cutting on the total energy, rather than the energy of the most
energetic cluster, the  sensitivity of the luminosity measurement to
treatment of final state radiation in the BHLUMI Monte Carlo and in
the detector simulation is greatly reduced.

\subsection{Peak finding and cluster building}
\label{sec:peak}

Although the full 
three-dimensional information ($r$, $\phi$, $z$) of each
pad in the event is available,
for the purpose of identifying cores
of electromagnetic showers, a generalized
search in three dimensions is not
needed.  Instead, the nearly projective
geometry of the detector is exploited
to reduce the three-dimensional
pattern recognition problem
to a two-dimensional one.

The peak finding algorithm first
finds the 
energy deposited in pads in the 
three even and three odd layers 
nearest the estimated mean longitudinal
depth  
of each individual shower.
The energies in these two sets of
pads are then projected 
separately onto the $r-\phi$ plane.
The peak finder then searches for
coincident 
peaks in the $r-\phi$ projection
of the two sets of pads.
Peaks are identified by a threshold on the
second spatial derivative of pad charge, so that a 
sufficiently pronounced shoulder can be identified
as a secondary peak close to the primary.
The multiplicity spectrum of such identified peaks falls exponentially with
a factor of about 0.015 per additional peak.
Up to three peaks in a given calorimeter
are selected and pads in a
geometric region around each 
peak are associated to form clusters.
The collection of pads associated with
each cluster are then used to determine
the radial and azimuthal coordinates, as described in 
section~\ref{sec:ev_meas}.

In order to retain strangely shaped showers, a second
cluster algorithm is used to find
any remaining energy deposition.
This ``blob'' finder projects all of the unused energy
in the detector to the $r-\phi$ plane and looks for any simply
connected two-dimensional region of energy deposition.
Such blobs with more than
1~GeV of measured energy are subsequently treated in all respects
as normal ``peak'' clusters.
The azimuthal and  radial coordinates of
blobs are determined from the location of the energy
maximum in the $r-\phi$ plane.

The probability that a true shower is classified as a
blob is small.  By looking for events (with somewhat
looser selection requirements than those given in
section~\ref{sec:sel}) which have
a normal cluster on one side and only a blob on the other, the
rate of blobs is 
estimated to be less than 1~in~$10^4$~\cite{bib-ph-thesis}.
This small number shows that the main cluster finding succeeds
in correctly identifying almost all showers from 
Bhabha events.

We have also used the trigger system
to check the cluster finding efficiency. 
This is accomplished by requiring that one side has
a cluster with $E >0.85 \cdot \ebeam$ and a radial coordinate which
satisfies the definition cuts of section~\ref{sec:sel}.  
In addition, a back-to-back 
trigger (SWSEG, see section~\ref{sec:trig} ) is required
and the digitized trigger energy on each side
is required to exceed
$0.5 \cdot \ebeam$.  
We would expect that in all such events reconstructed
clusters would be found on both sides of the detector.  In the 
1993 and 1994 data, only one event without reconstructed
clusters on both sides is observed.  
In the 1995 data, 11 such events are observed in 
a sample of $1.7 \times 10^6$ events.
On the basis of these 11 events, a systematic
error of $0.1 \times 10^{-4}$ is assigned to the
cluster finding process.

The ability of the cluster finder to resolve two electromagnetic
showers has been studied by using events with 
a secondary cluster from final state
radiation.  
The two cluster separation efficiency
is measured by
overlaying the pad energies from the raw data
of the radiative cluster at a variable position near
the most energetic cluster. This is accomplished by
using events with a widely separated radiative cluster
and then rotating the pad signals of
one of the clusters in azimuth 
so that they overlap with the pad signals of the other cluster.
As the radial
distance between the two clusters varies, the efficiency
for the cluster finder to find both clusters is measured.
It is found that for clusters above 5~GeV, the cluster
finding efficiency is 50\% at a separation
of approximately 1.0~cm, corresponding to 4 pad widths,
as shown in \fg{fig:event_cluster_sep_eff}.
The measured values of the two cluster separation
efficiency have been incorporated into the 
detector simulation described in section~\ref{sec:detsim}.

\Fg{fig:event_cluster_pull} illustrates how the reconstructed
coordinate of the
primary cluster is disturbed by about 100~$\mu$m when
the secondary cluster is resolved, and up to 1.5~mm when the
cluster is fused.  There is no evidence of a net radial bias.

Although we have expended considerable effort to optimize the
cluster separation capabilities of the detector,
the acceptance of the Bhabha selection
is in fact insensitive to the details of cluster formation. 
For example, changing the cluster finding parameters so that
a large fraction ($\sim 20\%$) of true showers are split into
two clusters
reduces the acceptance by only $10^{-4}$.

\begin{figure}[tbh!]
  \mbox{\epsfxsize16cm\epsffile{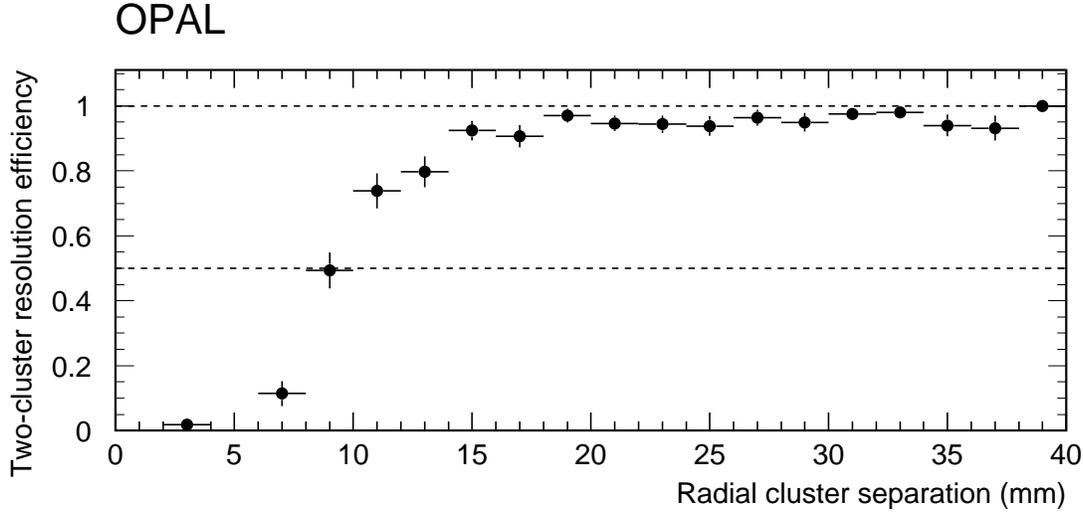}}%
\caption[Two Cluster Separation Efficiency]{ 
The measured probability of reconstructing the second cluster in a
radiative Bhabha event, where the second cluster energy lies between 5
and 10~GeV and it shares the same azimuth as the primary cluster,
plotted as a function of the radial separation between the two
clusters. 
\label{fig:event_cluster_sep_eff}
}
\end{figure}

\begin{figure}[tbh!]
  \mbox{\epsfxsize16cm\epsffile{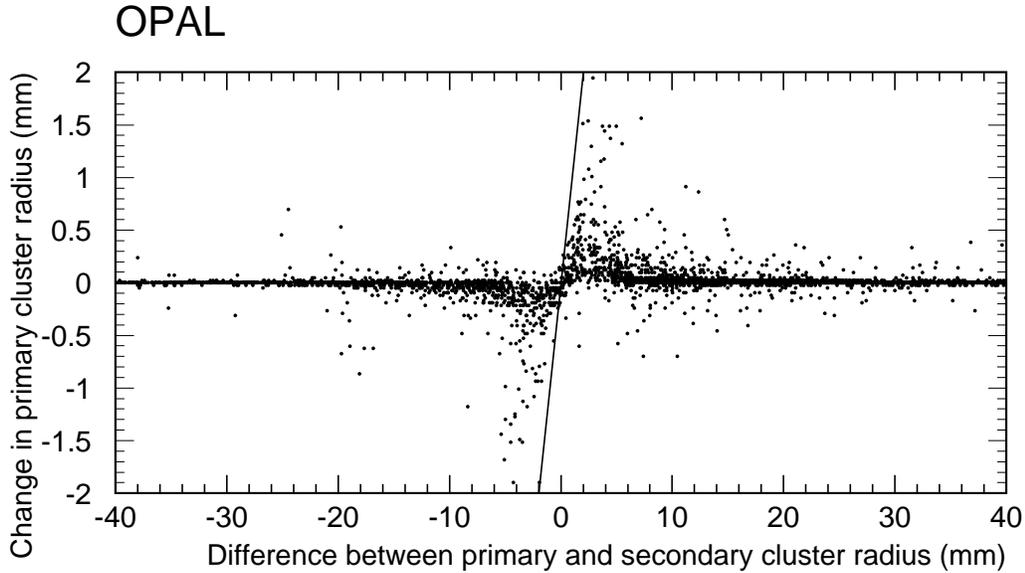}}
\caption[Two Cluster Radial Bias]{ 
The displacement of the radial coordinate of the highest
energy reconstructed cluster in the ``overlayed'' radiative event
sample, as a function of the radial separation between the primary
and secondary clusters.
Here the secondary cluster is selected to have an energy above 2~GeV\@.
The diagonal line shows the expected displacement if the secondary
cluster were simply mistaken for the primary.
\label{fig:event_cluster_pull}
}
\end{figure}

\subsection{Kinematic cluster selection}
\label{sec:kine}

\begin{figure}[tbh!]
\begin{center}
 \mbox{\epsfxsize16cm\epsffile{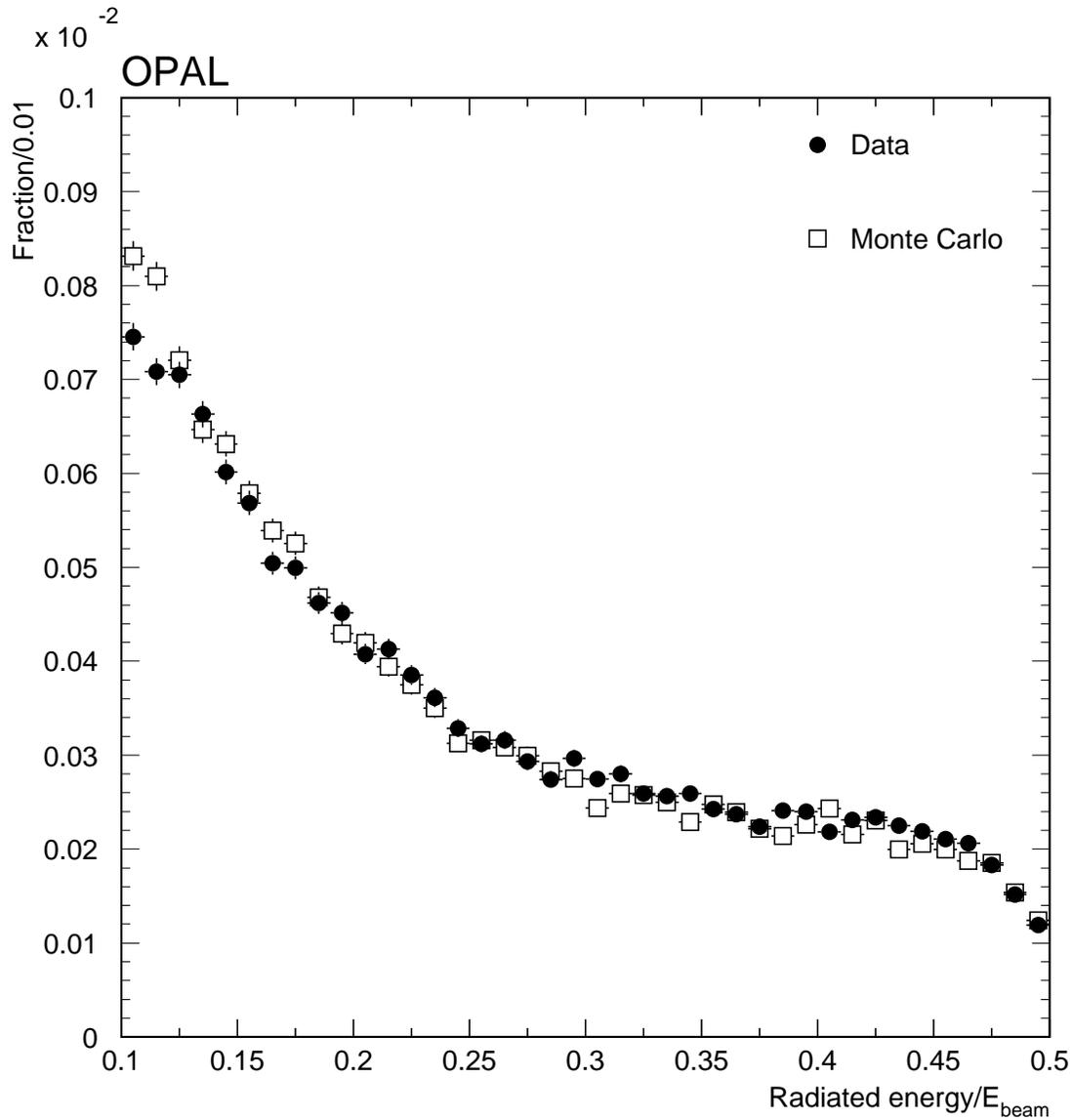}} 
\end{center}
 \caption[Energy of second cluster]{
The energy distribution of the least energetic cluster in $\SwitA$ events
where the kinematic cluster selection algorithm has selected exactly one
cluster on one side of the detector, and two on the other.
To ensure a sample of kinematically complete events, we furthermore  require
that the total measured energy in the event is
within $10\%$ of the LEP center-of-mass energy.
Only those clusters
separated by at least 2~cm in radius or 400~mrad in azimuth
are shown.  Data and Monte Carlo are independently normalized
to the total number of Bhabha events with a single  full--energy
cluster on one side of the detector.
\label{fig:sel_2clus}
}
\end{figure}

The kinematic cluster selection algorithm discriminates between radiative
clusters belonging to the Bhabha event and off-momentum machine background
by exploiting the conservation of transverse momentum in complete Bhabha events.
The algorithm first calculates the momentum vectors
associated with the clusters assuming they
originate from the measured beam spot.
The algorithm then evaluates each pairing of all possible
cluster combinations in each calorimeter to find the pair
of combinations which minimizes the total normalized transverse momentum
imbalance.
By minimizing transverse  momentum normalized to the
energy of the combination,
a possible bias towards selecting low energy combinations
is eliminated.
The probability that the kinematic selection algorithm finds a secondary
cluster belonging to a Bhabha event is about 1.3\% per calorimeter.
About 76\% of these secondary clusters exceed 10\% of the beam energy
and satisfy generous requirements for clear separation by lying either
more than 400~mrad in azimuth or more than 2~cm in radius from the
primary cluster.
The energy spectrum of such secondary clusters is shown in
figure~\ref{fig:sel_2clus} in section~\ref{sec:ev_samp}.
As explained above, when such radiative clusters are found, their
energy contributes to the fiducial energy of the event, but they are ignored
for the measurement of the radial and azimuthal coordinates.

The kinematic cluster selection algorithm has
been compared with other, simpler, algorithms and no significant
differences in acceptance have been observed.
The acceptance of the Bhabha selection is also quite insensitive
to the level of background clusters in the events, as discussed in
section~\ref{sec:back}.
%
%
\section{Measurement of shower position and energy}
\label{sec:ev_meas}
\subsection{Radial reconstruction}
\label{sec:rad_meas}
\label{sec:radial_intro}
%

The $1/\theta^{3}$ spectrum of the Bhabha scattering
process causes the luminosity measurement to be particularly sensitive to bias
in the cuts defining the angular acceptance.
Not only must the absolute physical dimensions of the detector itself be
carefully determined (as described in sections \ref{sec:rmetro} and
\ref{sec:zmetro}), but also the relationship between the reconstructed
shower positions and the physical detector elements must be established.  
Since this represents perhaps the most critical aspect of the luminosity 
measurement, we describe, in some detail, how we reconstruct the position
of showers from the pad signals, and especially how we determine the relation
of these reconstructed coordinates to the actual trajectories of incident
electrons in real space.
First, however, we describe quantitatively the sensitivity of the angular cuts.

The acceptance of the luminosity
measurement will be affected by any change, ${\Delta R}$,
in the inner and
outer edges of the acceptance as follows: 

\begin{equation}
    \frac{\Delta A}{A} 
    \approx -
    \frac{\Delta R_{\mathrm{in}}}{25 ~\mu \mathrm{m}}\times 10^{-3}
\label{eq:rinner}
\end{equation}
    and
\begin{equation}
    \frac{\Delta A}{A} 
    \approx +
    \frac{\Delta R_{\mathrm{out}}}{110~\mu \mathrm{m}}\times 10^{-3}.
\label{eq:router}
\end{equation}

Similarly, the acceptance will be affected by any change,
${\Delta Z}$, in the half-distance between the
effective planes of the radial measurements in the two calorimeters: 
\begin{equation}
    \frac{\Delta A}{A} 
    \approx +
    \frac{\Delta Z}{1.23 ~\mathrm{mm} }\times 10^{-3}.
\label{eq:zhalf}
\end{equation}
The coefficients in the expressions given above can be reproduced
    by simple analytic calculations, using only the
    $1/\theta^{3}$ Bhabha spectrum, the nominal
    half distance between the reference
    planes of the two detectors (2460.225~mm), and the inner and outer
    acceptance radii (77 and 127~mm).

Second order effects due to finite resolution also occur 
whenever a cut is imposed on a quantity with a steeply falling distribution.
An acceptance change is introduced due to the fact that more events
actually on the uphill side
of the cut will be measured to fall on the downhill side than vice--versa.
This {\em resolution flow} can be expressed as
\begin{equation}
    \frac{\Delta A}{A} =
    \frac{{\mathrm{d}}f}{{\mathrm{d}}x} \frac{\sigma_{x}^{2}}{2}
\label{eq:resflow}
\end{equation}
\noindent
where $f(x)$ is the intensity of events normalized to unity over the entire
acceptance, and $\sigma_{x}$ is  the resolution in the variable $x$ upon
which the cut is imposed.
Evaluating this expression for the $R^{-3}$ Bhabha distribution within our
acceptance, we find
\begin{equation}
    \frac{\Delta A}{A} 
    \approx +
    \left(\frac{\sigma_{\mathrm{Rin}}}{1.12 ~\mathrm{mm} }\right)^2 \times 10^{-3}
\label{eq:resrin}
\end{equation}
\noindent
and
\begin{equation}
    \frac{\Delta A}{A}
    \approx -
    \left(\frac{\sigma_{\mathrm{Rout}}}{3.04 ~\mathrm{mm} }\right)^2 \times 10^{-3}
\label{eq:resrout}
\end{equation}
\noindent
for radial resolutions of $\sigma_{\mathrm{Rin}}$ and $\sigma_{\mathrm{Rout}}$
at the inner and outer acceptance boundaries.

Similarly, a cut is imposed on $\Delta \mathrm{R}$, the difference of $\rr$
and $\rl$, in the tails of the acollinearity distribution (see \fg{fig:sel_dr},
section~\ref{sec:ev_samp}) where the density of events is low, but is
falling steeply.
This results in a resolution flow of
\begin{equation}
    \frac{\Delta A}{A}
    \approx -
    \left(\frac{\sigma_{\Delta R}}{3.54 ~\mathrm{mm} }\right)^2 \times 10^{-3}.
\label{eq:resacoll}
\end{equation}
\noindent
Any net bias, $\Delta R_{\mathrm{bias}}$, in the measured acollinearity also
causes a first--order change in the acceptance of
\begin{equation}
    \frac{\Delta A}{A}
    \approx 
    \frac{\Delta R_{\mathrm{bias}}}{416\,\mu\mathrm{m} }
    \times 10^{-3}
\label{eq:biasacoll}
\end{equation}
\noindent
due to the density of events near the acollinearity cut (see again 
\fg{fig:sel_dr}, section~\ref{sec:ev_samp}).
These acceptance changes, on the order of a few times $10^{-5}$ are
discussed in section~\ref{sec:detsim}.

%

The finite angular divergence of the LEP beams is in many respects equivalent 
to a further smearing of the detector resolution, and also creates a very small
change in acceptance.  This is discussed in section \ref{sec:beam}. 

One of the advantages of our highly-segmented calorimetric measurement
over a tracking measurement is the fact that valid coordinates can be
assigned to all clusters observed within the fiducial volume of the detector
with an inefficiency less than $10^{-5}$.
A typical 45~GeV electron shower deposits energy in $\sim 500$ pads in the
calorimeter.  From this large amount of information, it is possible to
reconstruct the radial position of the shower with a resolution of
130 to $170~\mu$m and a systematic bias everywhere less than $15~\mu$m.
At a large number of pad boundaries, distributed throughout the
volume of the detector, we can do even better, and demonstrate that
our total bias near these boundaries, including metrology errors, is less than
$6.5~\mu$m.

The shower-by-shower radial reconstruction begins with finding a radial
coordinate in each layer of the calorimeter (section~\ref{sec:layer_coord}).  
The better layer measurements are
then selected, projected onto the reference layer at 7~$\x$, 
and averaged (section~\ref{sec:average_coord}). The
average raw coordinate is finally smoothed to compensate for non-uniformities
in the resolution across the pad structure of the detector
(section~\ref{sec:smoothing}). This smoothed coordinate is then used
in making the primary cut defining the luminometer acceptance.
We then proceed to quantify the residual bias in the smoothed coordinate
and calculate a corresponding correction to our acceptance.
This procedure relies on a concept which we term the {\em pad boundary image},
which can be defined as the coordinate at which an
incident electron deposits, on average, equal energies on two adjacent pads.
The bias of the coordinate with respect to a large number of such
individual pad boundary images can be evaluated by a process we
term ``anchoring''.
The absolute offset of the pad boundary images themselves with respect to
the physical geometry of the detector is determined on the basis  of
test beam
measurements in which a fully instrumented \SW\ segment was exposed to a beam
of electrons and muons.  
Since the \SW\ pads are
sensitive to muons,
the muon tracks were used to determine the geometry of
the calorimeter pads with respect to a beam telescope.  
Measurements of the showering electrons
were then used to determine the small asymmetry in the energy deposited
in a pair of adjacent pads (which differ slightly in area due to the
$R-\phi$ geometry of the detector) when the electron trajectory lies
exactly on the boundary between the pads.
The evaluation of this bias is discussed in section~\ref{sec:anchor}.  We
now address the question of determining a radial shower coordinate from
the pad signals in the individual layers of the detector.

\subsubsection{Radial coordinate in each layer}
\label{sec:layer_coord}
%
%
%
%
%
%
%
%
%
%
%
%
%

    The radial coordinate reconstruction in SiW is based on the
    good approximation that pad boundaries and pad centers are symmetry
    points of the detector.
    In each layer,
    using the triplet of pads centered around the pad with the maximum
    energy deposition, we construct the function:
 
\begin{equation}
    D   \equiv \frac{E_{1}-E_{3}}{2\cdot E_{2} - E_{1} - E_{3}}
\label{eq:coord_D}
\end{equation}
 
\noindent where $E{_2}$ is the energy in the central pad of the triplet and 
    $E_{1}$ and $E_{3}$ are the energies in the adjacent
    pads.
    To within the approximation given above, the function $D$ so defined
    takes on the value of 0 for a shower hitting the center of the triplet of
    pads (in which case $E_{1}=E_{3}$), 
    and $\pm 1$ for showers hitting the outer or inner boundaries of the 
    central pad (in which case $E_{1}=E_{2}$ or $E_{3}=E_{2}$).
 
    These symmetry properties of the function $D$ would be perfect
    in a detector with Cartesian pad geometry.
    The extent to which this symmetry is altered
     by the $r-\phi$ configuration of the pads  depends on
    the scale of the lateral shower spread in the calorimeter compared to the 
    radius of curvature of the pads, as well as the pad size.
    The resulting bias is smaller at pad boundaries than it is at the pad
    center.
In section \ref{sec:anchor}, we discuss how we quantify the pad boundary bias
using
test beam  measurements.  This bias is, however, ignored in the process of 
shower coordinate reconstruction, which preserves these symmetries explicitly.

In the region between pad boundaries and pad centers, the variable $D$
is not a linear function of the true position of the incoming electron,
deviating from linearity by up to 20\% of the pad width, or $500\mu$m.
This nonlinearity can be conveniently parameterized using the easily
invertible functional form

\begin{equation}
\vert R_{\mathrm{p}} \vert = \frac{w_p}{2}
   \left[D_1 - \left(\frac{A}{\vert D \vert + D_0} \right)
   ^{\frac{1}{\beta}}\right]
\label{eq:coord_basic}
\end{equation}

\noindent where $R_{\mathrm{p}} \in [ -\frac{w_p}{2}, +\frac{w_p}{2} ]$
is the linearized coordinate relative to the pad center,
$w_p = 0.25$~cm is the pad width,
the quantity $\beta$ controls the nonlinearity
of the transformation, and $D_1 = 1.1$ is a constant.
The values of $D_0$ and $A$ are set by transferring the 
symmetry conditions
of $D$ to $R_{\mathrm{p}}$ during the optimization of $\beta$.
In particular, although not formally obvious, $R_{\mathrm{p}}$ is
constrained to be a function with odd symmetry about the pad center.
At a depth of $6\x$, the parameters take on the values: $\beta = 0.544,
D_0 = 0.372, A = 0.392$.

\begin{figure}[tbh]
  \mbox{\epsfxsize16cm\epsffile{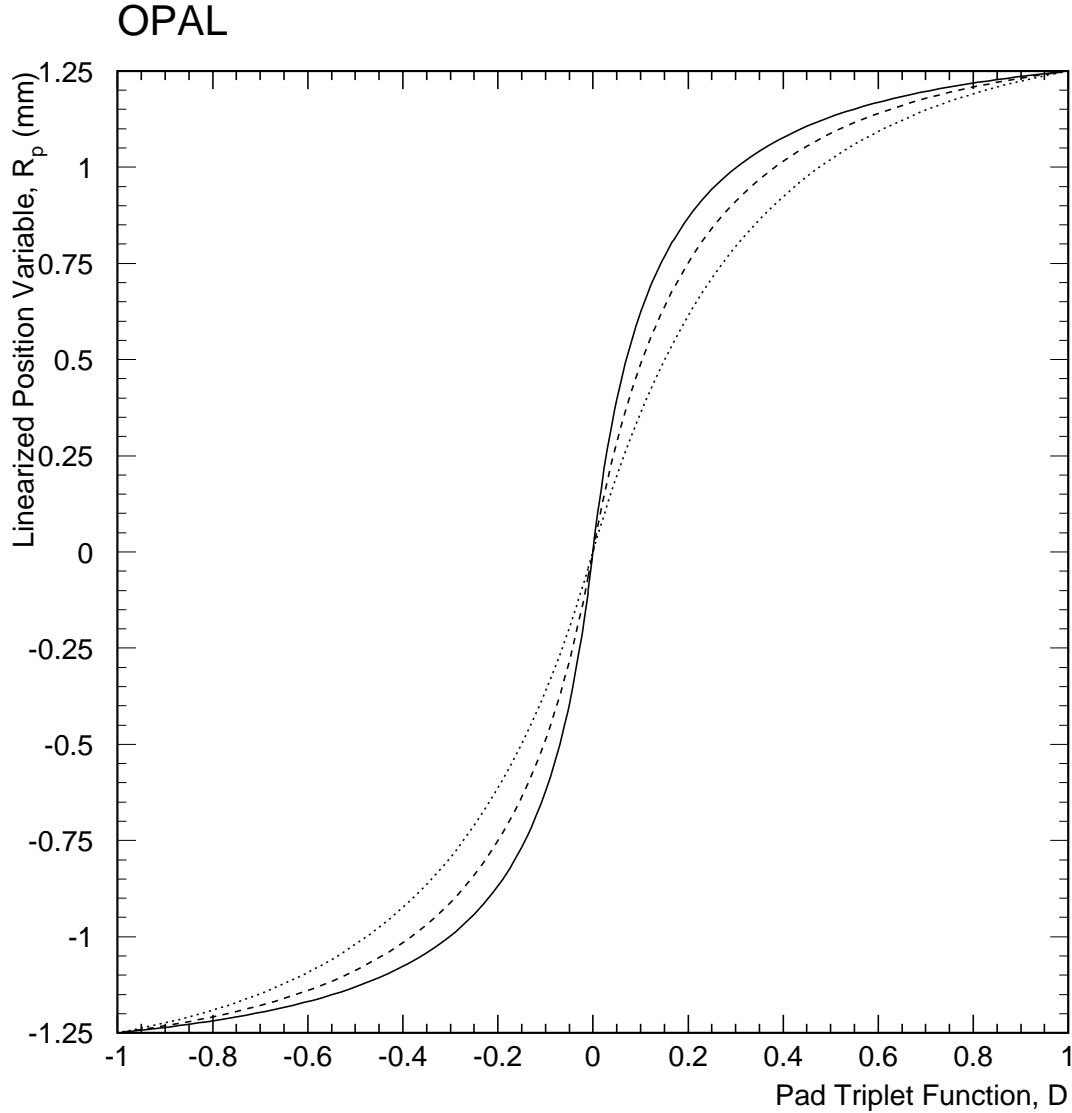}}
\caption[The layer radius versus D] { 
    The linearized position variable, $R_p$, in millimeters
    as a function of $D$, in units of half pad widths.
    This is shown for characteristic shower depths
    of 4 (solid line), 6 (dashed line) and 
    8~$X_{0}$ (dotted line).
    The center of the pad is at 0, and the boundaries at
    $\pm 1.25$~mm.
\label{fig:layer_coord_D}
}
\end{figure}

Figure~\ref{fig:layer_coord_D} shows $R_{\mathrm{p}}$ as a function of $D$
as parametrized according to equation~\ref{eq:coord_basic}
for shower widths characteristic of calorimeter layers near the depth
of maximum longitudinal shower development, which are more
critical in measuring the radial coordinate.
The extent of the evident nonlinearity depends sensitively ($\pm 100\mu$m)
on the scale of the lateral shower spread in the calorimeter compared
to the size of the pads.
We have optimized $\beta$ as a function of the locally
observed individual shower width while explictly preserving the desired
symmetries of the reconstruction function about the pad edges and centers,
using data taken both during normal running at LEP and in the test beam.
The result is a unique expression for $R_{\mathrm{p}}$ in terms of the charges
observed in the pad triplet.
Due to its explicit dependence on the apparent shower width,
the same functional form is valid, independent of layer.

The resolution of the radial coordinate in a single layer varies
strongly across the pad.
Near longitudinal shower maximum, it is about 300~$\mu$m
at the pad boundary, and 750~$\mu$m at the pad center.
This rapid variation in resolution makes a precise definition of the
residual bias in $R_{\mathrm{p}}$ rather subtle, since events falling in
regions of poor resolution are more likely to be measured as falling in
regions of good resolution than vice versa. 
We refer to $R_{\mathrm{p}}$ as being ``linearized'', in that it has a much
more linear relation to the true coordinate than the pad triplet quantity,
$D$. 
Precisely speaking, it is linearized in that the parameters
of equation~\ref{eq:coord_basic} have been chosen to minimize 
what might best be termed the integral bias throughout the pad
\begin{equation}
      {\rm Bias}_{\mathrm{int}}(R_{\mathrm{p}}) = 
     w_p \frac{N(R_{\mathrm{p}})}{N_{\mathrm{tot}}}
       -  (R_{\mathrm{p}} + \frac{w_p}{2})
\label{eq:int_bias}
\end{equation}

\noindent where $N(R_{\mathrm{p}})$ is the number of events
falling within the pad,
but below $R_{\mathrm{p}}$, in a clean sample of Bhahba events,
and $N_{tot}$ is the total number of events within the pad.
Note that we ignore in this optimization the expected $1/R^3$ fall-off in
the Bhabha intensity and attempt to approach a flat distribution
of events over the pad.
The odd symmetry of the linearization function~\ref{eq:coord_basic} about
the pad center, obvious from \fg{fig:layer_coord_D}, ensures
that no events cross from one half of the pad to the other, and prevents
this simplification from introducing a significant bias in the
optimization. 

Systematic errors in the layer coordinates derived from individual pad 
triplets can be induced by channel-to-channel variation in the AMPLEX
front-end chip gains.
All crucial coordinate measurements come from pad triplets which do not span 
AMPLEX boundaries.
For such triplets the gain variations have been measured to be
$\leq0.25\%$, with a precision limited by the uniformity of the
calibration system (see section~\ref{sec:det_electronics}).
The corresponding systematic fluctuations in $R_{\mathrm{p}}$ can be 
calculated from the slope of \fg{fig:layer_coord_D}, which varies strongly
across the pad.
At the pad boundary a $0.25\%$ gain fluctuation corresponds to
\begin{equation}
\label{eq:padgain}
\Delta R_{\mathrm{p}} = 2.5\,\mu{\mathrm{m}}.
\end{equation}
While at the pad center the same gain variation corresponds to
$\sim25\,\mu{\mathrm{m}}$.

The effect of such systematics will be reduced by the fact that many 
independent sets of pad triplets contribute to the measurement, as will be 
seen in the following sections.

\subsubsection{Average radial coordinate}
\label{sec:average_coord}
%
%
%
%
%
%
%
%
%
%
%
%
%
    As the next step in reconstructing a radial coordinate, 
    we project the positions measured in each layer
    using the variable $R_{\mathrm{p}}$
    onto the reference plane at 7~$\x$
    assuming the nominal beam spot position.
    The average,
    $R_{\mathrm{a}}$, is then determined from the projected coordinates.
The layers used in the average are selected from the 9 layers
at  depths\footnote{The range of 9 layers used is
shifted by up to $\pm~2$ layers in the case of
exceptionally early or late shower development.}
between 2 and  10~$\x$.
Only layers with ``good coordinates'', are used
but are not otherwise weighted in the average.
``Good coordinates'' are defined as coming from pad triplets with
\begin{itemize}
\item sensible observed shower width (characteristic of shower
depths from $1$ to $13 \x$ )
\item an energy sum exceeding a minimal threshold
 ($E_{\mathrm{triplet}} > 200$ MeV )
\item a useful estimated resolution ($\sigma_{\mathrm{r}} < 2.5$ mm)
\end{itemize}
The mean number of layers used in the average is
6.2 with an RMS of 1.5 for a 45~GeV shower within the fiducial acceptance of the
detector. The mean layer used in the average is 6.4 
with an RMS of 0.7 on the right,
and 6.1 with an RMS of  0.7 on the left.
The fact that the center of measurement is close to the reference plane limits 
the error introduced by any difference between the actual beam spot and the 
ideal beam geometry assumed in the projection.
Even for maximal beam spot offsets and highly abnormal longitudinal shower 
developments, the projection error is everywhere less than 10~$\mu$m, and
cancels to a fraction of a micron in the average over either azimuth
or shower development depth.

The process of averaging the layer coordinates also has the advantage of
reducing the possible systematic error due to pad gain variations.
Most significantly, any channel-to-channel gain variations which are
common to all AMPLEXes will be largely cancelled by the fact that the
radial direction of the AMPLEX channel ordering has been inverted between
the right and left pad columns of each detector hybrid.
This fact, in combination with the azimuthal staggering of each detector
layer means that the AMPLEX channel ordering alternates from radially inwards
to radially outwards for the pad triplets in each layer along the trajectory
of every measured shower.

Also, in averaging over the 6.2 layers which typically contribute to the average
coordinate for each shower, the effect of random channel-to-channel gain
variations will be sampled over most of a pad width and be statistically damped,
yielding a maximum systematic error in each tower of
\begin{equation}
    \Delta R_{\mathrm{gains}}^{\mathrm{a}} = \frac{1}{2}
    (25\,\mu{\mathrm{m}} + 2.5\,\mu{\mathrm{m}}) / \sqrt{6.2} =
    5.5\,\mu{\mathrm{m}}
\end{equation}
These random systematic errors will also be different in each of the 32
azimuthal towers in each calorimeter, yielding an expected error of only
$5.5\,\mu{\mathrm{m}}/\sqrt{32} = 1.0\,\mu{\mathrm{m}}$ when averaged
over $\phi$.

A typical $R_{\mathrm{a}}$ distribution is shown in \fg{fig:average_coord_raw}.
For this plot, the normal isolation cuts are imposed, except
for those involving the radial coordinate in either calorimeter.
This radial coordinate is no longer integral bias free, as evidenced by the
structure visible in the plot, which corresponds to a residual
nonlinear bias of up to $50\mu$m with a periodicity of 2.5~mm, given by the
radial pad pitch. 
In averaging $R_{\mathrm{p}}$ over many layers, we have improved its
resolution, but destroyed the original balance of resolution 
flow\footnote{Finite resolution, coupled with a sloping spectrum, 
causes events to flow across a boundary.  See equation~\ref{eq:resflow}.}
and inherent bias enjoyed by $R_{\mathrm{p}}$
in each layer. 

\begin{figure}[p]
\vskip -1.5cm
\begin{center}
  \mbox{\epsfxsize15cm\epsffile{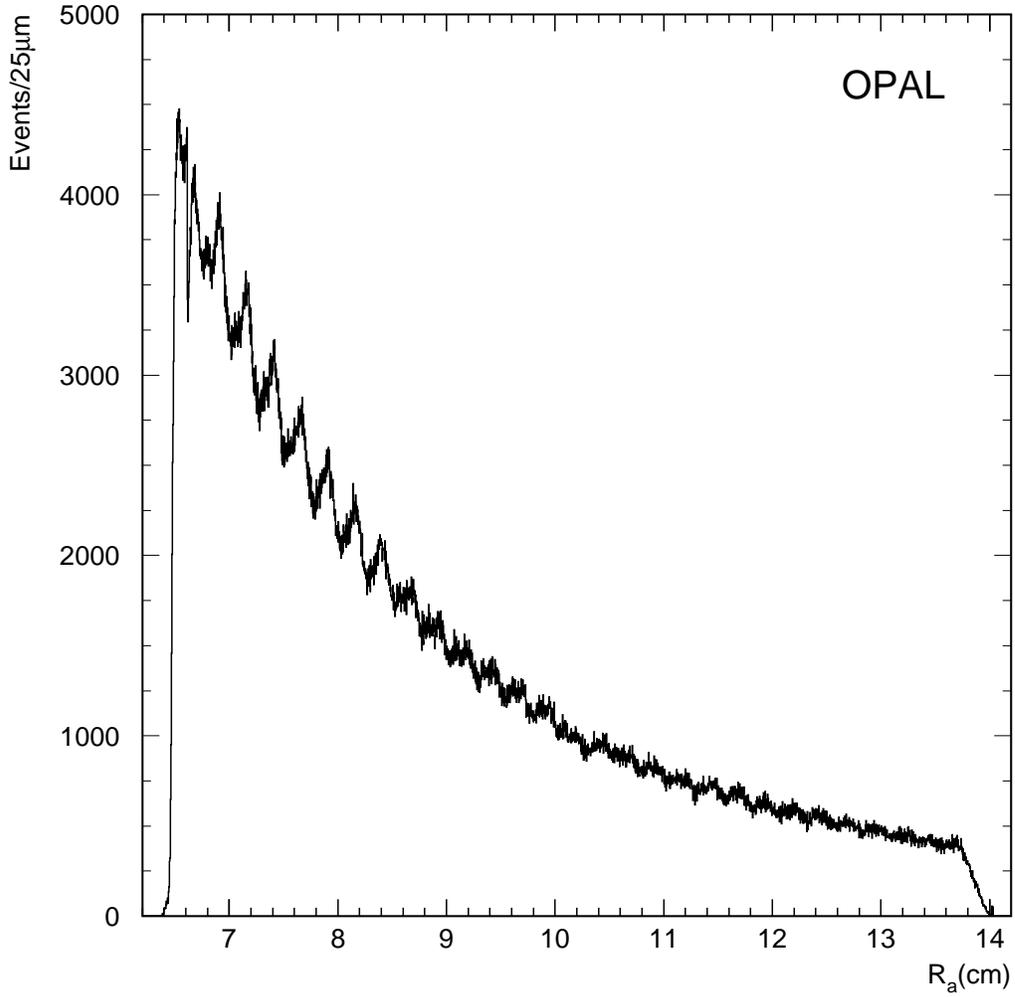}}
  \vskip -1.5cm
  \mbox{\epsfxsize15cm\epsffile{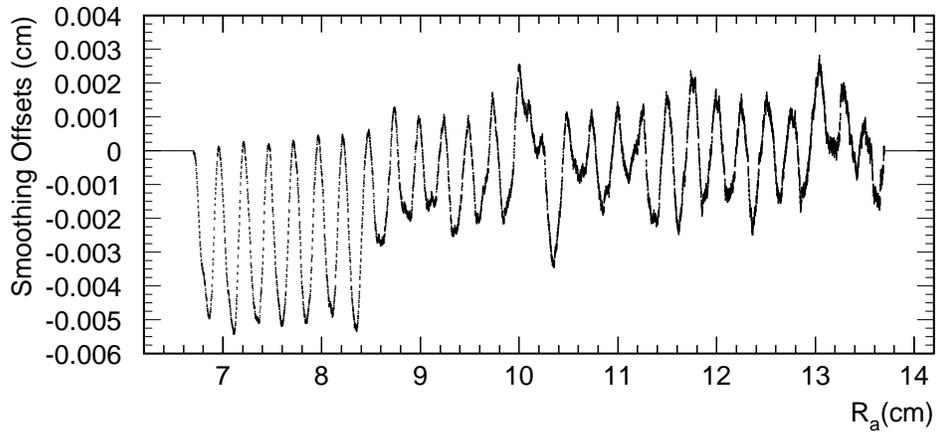}}
\end{center}
\caption[Raw radial coordinate averaged over layers]{ 
The distribution (upper plot)
of $R_{\mathrm{a}}$, the raw radial coordinate, averaged over layers, 
for the largest of the 9 luminosity analysis data sets (94b).  
Also shown (lower plot) 
is the bin-by-bin smoothing offsets, $S(R_a)$, determined by the smoothing
condition.  The quasiperiodic structure at the 2.5~mm period determined
by the pad width has a maximum amplitude of $\approx 50\mu$m.
The Bhabha 
isolation cuts have been applied, except for those involving the radial 
coordinate in either calorimeter.
\label{fig:average_coord_raw}
}
\end{figure}

The symmetry of the coordinate about the pad boundary has also been influenced 
to some extent by the averaging over layers, but we can expect it to be best 
preserved at layer 7, the reference layer, and center of measurement.
%
The fact that the pads possess a cylindrical, rather than projective
geometry, not only dilutes systematic effects, but even allows
the remaining systematic effects to be measured.
Near the inner edge of the acceptance, the projected coordinate lies on the 
pad boundary of one layer or another with a periodicity of about $200~\mu$m
in $R$.
This fact will be exploited extensively in section~\ref{sec:anchor}.
We now turn our attention to the problem of re-establishing a condition of 
minimum integral bias in the layer-averaged coordinate.
\subsubsection{Smoothing the radial coordinate}
\label{sec:smoothing}
%
%
%
%
%
%
%
%
%
%
%
%
%
%

If the bias, $B$, in the raw average coordinate, $R_a$, were known, then it
could be removed by direct subtraction, that is,

\begin{equation}
   R_{\mathrm{true}} = R_a - B(R_a) .
\label{eq:smooth_bias}
\end{equation}

\noindent
Although test beam measurements are able to give us some direct knowledge
of $B$, the interplay between the coordinates from each layer depends
strongly enough on the beam geometry and the material in front
of the detector that it was not optimal to directly apply
the bias measured in 
the test beam. 

We were able to arrive at a better estimate of $B$ under LEP conditions
empirically, by
studying the structure in the observed $R_a$ distribution, and extracting
a smoothing offset, $S \approx - B$, such that

\begin{equation}
   R_{\mathrm{smooth}} = R_a + S(R_a) .
\label{eq:smooth_disp}
\end{equation}

\noindent
The distribution of $R_{\mathrm{smooth}}$ is required to have minimal structure at
the 2.5~mm pad periodicity of the detector.  The lack of structure in
$R_{\mathrm{smooth}}$ is a requirement for the luminometer acceptance to be independent
of the position of the radial acceptance cuts with respect to the pad
boundaries.  This condition, although highly desirable for the 
\SwitR\ and \SwitL\
selections, is truly essential for the \SwitA\ selection, 
since the offset of the
beam spot with respect to the symmetry axis of the 
calorimeters means that
right and left shower coordinates individually lying well away from the
fiducial pad boundaries in each calorimeter will comprise the acceptance
boundary defined in terms of $R_A = \frac{1}{2}(R_R + R_L)$. 

Apart from requiring that the distribution of $R_{\mathrm{smooth}}$ be 
smooth, we also, and more importantly,
require the strict condition that the smoothing offsets cause no event to cross
any pad boundary in the reference layer at $7 X_0$.  This condition is
assured when $S(R_a) = 0$ at the pad boundaries.

We begin by parameterizing the $R_{\mathrm{smooth}}$ distribution as follows:

\begin{equation}
   \frac{dN}{dR_{\mathrm{smooth}}} =
   a_i R_{\mathrm{smooth}}^{-3}
 - \frac{s_i}{w_{\mathrm{bin}}} \cos \left( \alpha/2 \right)
 - \frac{d_i}{w_{\mathrm{bin}}} \cos \left( \alpha \right) 
\label{eq:smooth}
\end{equation}

\begin{equation}
   \alpha =  2 \pi \frac{R_{\mathrm{pad}}}{w_{\mathrm{pad}}} + \pi
\label{eq:smooth_alpha}
\end{equation}

\noindent where $R_{\mathrm{pad}} = R_a - R_i$, having the range
$[ -1.25, +1.25 ]$mm, is the radial coordinate
with respect to $R_i$, the center of the pad $(i)$ in which $R_a$ lies,
$w_{\mathrm{pad}} = 2.5$mm is the pad width, and $w_{\mathrm{bin}}=25\mu$m
is the radial histogram bin width.
The power-law constants $a_i$, $d_i$, and $s_i$ are determined as
follows, where $i=[ 1, 32 ]$ runs over all the radial pad rows.

The constants $a_i$ are fixed by the constraint that the smoothing
should not alter the number of events in any pad, giving

\begin{equation}
   a_i = \frac{2 N_{i}}
   {\left( R_i - w_{\mathrm{pad}}/2 \right)^{-2} -
    \left( R_i + w_{\mathrm{pad}}/2 \right)^{-2}} 
\label{eq:smooth_ai}
\end{equation}

\noindent where $N_i$ is the number of events in the sample to be smoothed
which fall within pad $i$ according to the raw radial coordinate, $R_a$.

Only if the radial distribution of events follows an $R^{-3}$ dependence
perfectly will  the $a_i$ terms determined above be continuous at the pad
boundaries. The sinusoidal terms in equation~\ref{eq:smooth} allow continuity
to be restored while imposing minimal structure on the shape of the smoothed
radial distribution.
The discontinuities of the $a_i$ terms at the pad boundaries are

\begin{equation}
   b_i = a_i \cdot w_{\mathrm{bin}} \cdot
   \left( R_i - w_{\mathrm{pad}}/2 \right)^{-3} 
   -  a_{i-1} \cdot w_{\mathrm{bin}} \cdot
   \left( R_{i-1} + w_{\mathrm{pad}}/2 \right)^{-3} .
\label{eq:smooth_bi}
\end{equation}

The sinusoidal terms in equation~\ref{eq:smooth} have zero integral over each
pad by construction, and therefore do not cause events to cross pad
boundaries. Their coefficients $d_i$ and $s_i$ are determined by the condition
that the right-hand side of equation~\ref{eq:smooth} pass through the
mid-points of the discontinuities, $b_i$, at each bin boundary.

The coefficients of the half-cycle sinusoids, $s_i$, can be called ``slope
modifier'' terms, since they tend to move events in a single direction across
the width of a pad. They compensate for any long-range deviation from the
assumed $R^{-3}$ dependence in the radial event distribution.
Such a deviation would cause the
$b_i$'s to remain of like sign from pad to pad.

The coefficients of the full-cycle sinusoids, $d_i$, 
can be called ``dust under
the carpet'' terms, since they tend to 
move events from both edges of the pad
toward the center, or vice versa.  
They compensate for pads having locally
either more or fewer events than could
be extrapolated from their two nearest
neighbors, leading to $b_i$'s which change sign from pad to pad.

The $d_i$ and $s_i$ coefficients are determined very simply from the
$b_i$'s as follows:

\begin{equation}
   d_i = + \frac{1}{4} \left( b_{i+1} - b_{i} \right)
\label{eq:smooth_di}
\end{equation}

\begin{equation}
   s_i = - \frac{1}{4} \left( b_{i+1} + b_{i} \right)
\label{eq:smooth_si}
\end{equation}

\Fg{fig:smooth_tech} shows the observed distribution of the raw radial
coordinate, $R_a$, and the target function for $dN/dR_{\mathrm{smooth}}$
as determined by equation~\ref{eq:smooth}
in the region near the inner acceptance boundary.
The departure of the $R_a$ 
distribution from $R^{-3}$ can be seen in the
steps between the virtually straight power-law 
segments given by the $a_i$.
In the inset, the final target
function can be seen to pass smoothly through the mid-points of these
discontinuities at the bin boundaries.

\begin{figure}[tbh!]
  \mbox{\epsfxsize16cm\epsffile{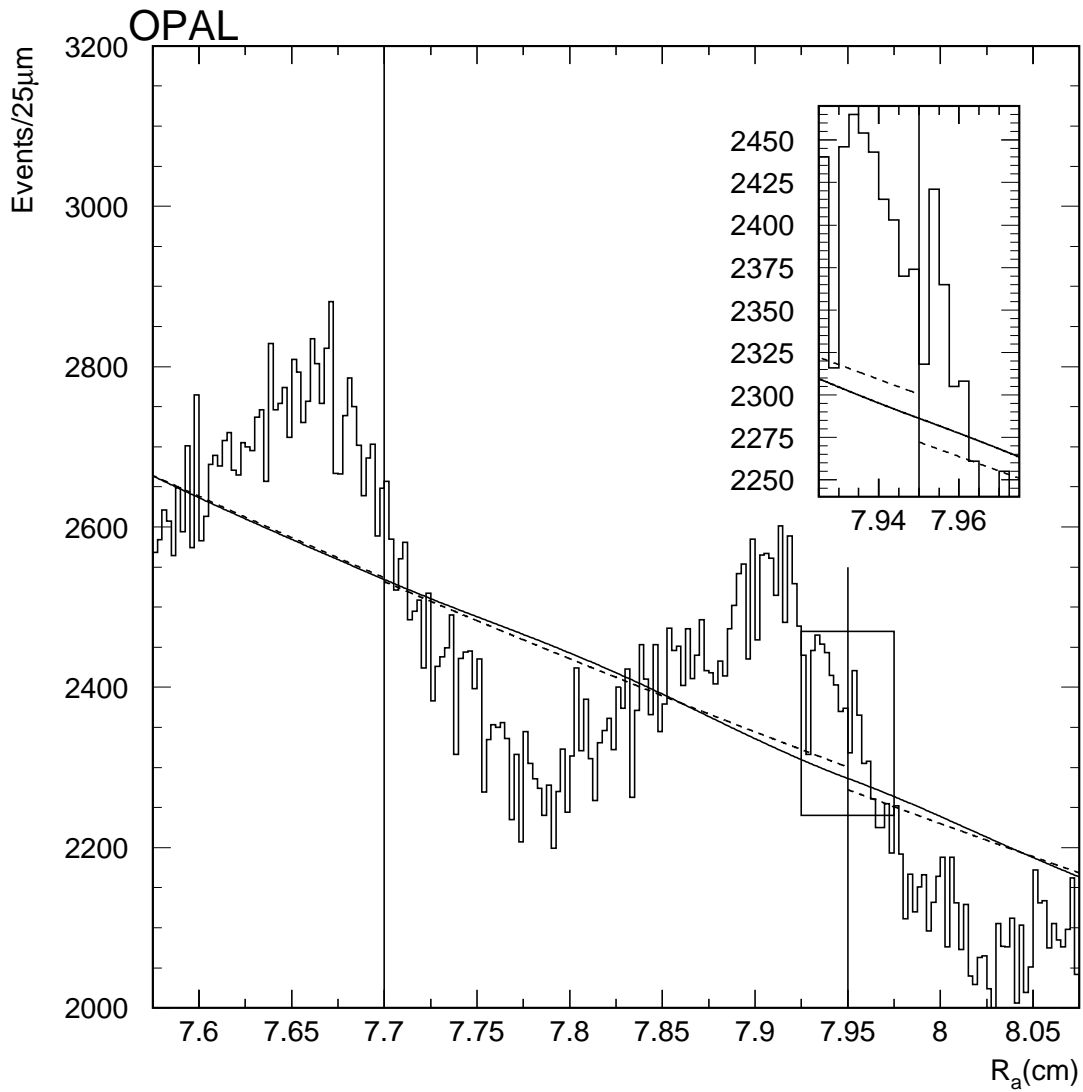}}
\caption[Radial distribution before smoothing]{ 
The observed distribution of the raw radial
coordinate, $R_a$, and the target function for $dN/dR_{\mathrm{smooth}}$
in the region near the inner acceptance boundary.
The vertical lines indicate the position of the pad boundaries in the
reference layer of the calorimeter.
The departure of the pad by pad
$R_a$ distribution from $R^{-3}$ causes the
steps between the virtually straight power-law segments (dashed lines)
given by the $a_i$.
The final target function is indicated by the solid curve.
The inset shows how the final target function  passes smoothly through the
mid-points of the power-law discontinuities at the pad boundaries.
\label{fig:smooth_tech}
}
\end{figure}

Having determined the desired shape of the radial event distribution in a 
way which ensures that no event will be allowed to cross a pad boundary at the 
reference layer in the detector, and which minimizes the structure in the 
radial distribution within the pad while guaranteeing continuity at the pad 
boundaries, we proceed to generate the smoothing offset of 
equation~\ref{eq:smooth_disp}, which when added to
$R_a$ will produce the desired shape of the smoothed radial distribution.

\noindent
The offset function $S(R_a)$ is determined by the condition

\begin{equation}
\int^{R} \frac{dN}{dR_a}\left( R_a + S(R_a) \right) dR_a  =
\int^{R} \frac{dN}{dR_{\mathrm{smooth}}} dR_{\mathrm{smooth}}
\label{eq:smooth_eqn}
\end{equation}

\noindent
where $dN/dR_{\mathrm{smooth}}$ is given by the right-hand side of
equation~\ref{eq:smooth}.
The smoothing offsets as found bin-by-bin from the $R_a$ distribution
are shown in \fg{fig:average_coord_raw}.
The offsets exhibit a quasiperiodic variation following the pad pitch, with
a maximum amplitude of $50\mu$m and an rms of $\sim 20\mu$m.
We then parameterize $S(R_a)$ with a Fourier decomposition
which represents
the lower portion of \fg{fig:average_coord_raw} 
everywhere within $1\mu$m, giving

\begin{equation}
   S_i(R_{\mathrm{pad}}) = \sum_{j=1}^7 {c_{ij} \cdot \sin( j \cdot \alpha)} .
\label{eq:smooth_fourier}
\end{equation}

\noindent
The restriction to sine terms only in the Fourier series guarantees that
$S(R_a)$ is identically zero at the pad boundaries, and hence that the
smoothing will not change the pad of any event.
An independent smoothing transformation is made for each of the 9 separate
luminosity analysis data samples in order to track any subtle influence of
beam spot motion on the reconstructed radial coordinate.
Noticeable differences between the data samples are observed,
but none of great significance.

\Fg{fig:smooth_Rsmooth} shows the distribution of the smoothed radial
coordinates near the inner cut boundary for the
same data sample as the raw radial distributions in
\fgs{fig:average_coord_raw} and~\ref{fig:smooth_tech}.
No structure is apparent beyond statistical fluctuations, and the measured
luminosity is independent of the radius chosen to define the inner edge
of the acceptance, as demonstrated in section~\ref{sec:anchor}.
Since the smoothing is constrained to make no change in the measured
coordinates at the pad boundaries, the primary
luminosity acceptances, $\SwitR$ and $\SwitL$,
which are defined by radial cuts corresponding to the pad boundaries between
pad rows 6 and 7, are insensitive to radical alterations, let
alone any residual imperfections, in the smoothing.

\begin{figure}[htb!]
  \mbox{\epsfxsize16cm\epsffile{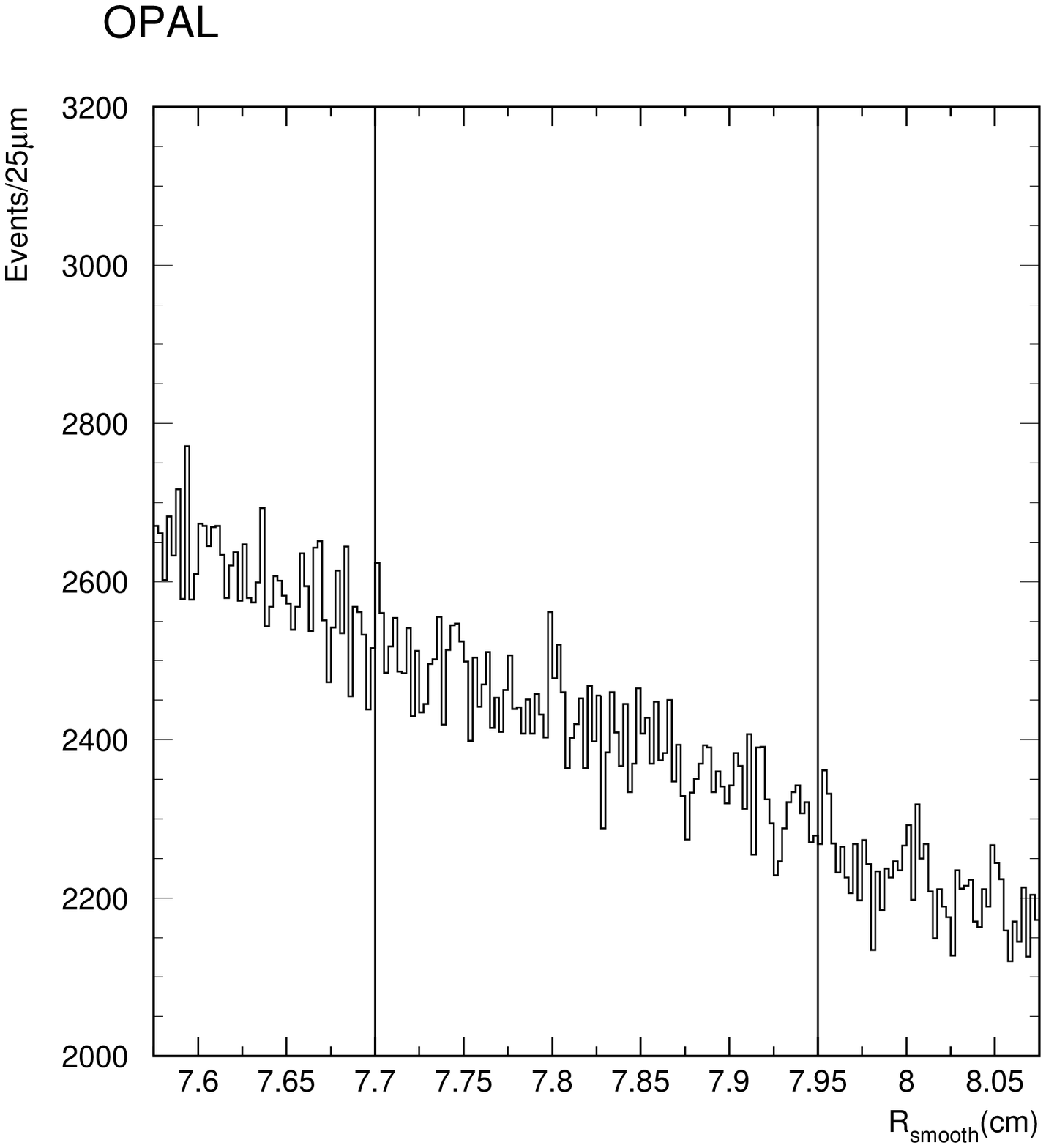}}
\caption[Smoothed Coordiante] { 
The distribution of the smoothed radial
coordinates near the inner cut boundary in the right side calorimeter for the
94b luminosity data sample, with all Bhabha isolation cuts applied,
except for any cut at all on the radial coordinate in either calorimeter.
The vertical lines indicate the position of the pad boundaries in the
reference layer of the calorimeter.
No structure is apparent beyond statistical fluctuations, and the measured
luminosity is independent of the radius chosen to define the inner edge
of the acceptance.
\label{fig:smooth_Rsmooth}
}
\end{figure}

It should be noted that the lack of integral bias provided by the smoothed
radial coordinates on each side of the luminometer does not guarantee the
lack of integral bias in the {\em average} radial coordinate, $R_A$,
taken between the right and left calorimeters.
The resolution of the average radius is improved by 
a factor of $\sqrt{2}$, and the
balance between resolution flow and inherent bias is again disturbed.
Test beam measurements provide a direct measurement of the variation of
resolution of $R_a$ (and hence $R_{\mathrm{smooth}}$) across the pads.
The resolution
varies from $130\mu$m near the pad boundary to $170\mu$m near the pad center.
We calculate that this level of variation induces $\sim 15\mu$m of
inherent, but compensated, bias in the smoothed coordinate, so that
the uncompensated
integral bias in the smoothed coordinates averaged between the right and
left sides, $R_A$, should be $\sim 10\mu$m.

Since the typical beam offset causes the radius of back-to-back showers to
differ by up to a full pad width, no structure is visible, or would be visible,
even if it were present, in the
distribution of the averaged coordinate, $R_A$. The mixture
of inherent biases is sampled fully as a function of azimuth, and any initial
structure will be washed out. In fact most of the residual integral bias will
also be washed out in the average over azimuth.  We estimate
that approximately $5\mu$m bias will
remain in $R_A$ at any particular radius in the clean region of acceptance
($R<8.5$~cm), after integrating over $\phi$.
\footnote{Behind greater amounts of preshowering material, the quality
of the measured coordinate is somewhat degraded, having consequences for
the systematics of the acollinearity cut (see section~\ref{sec:detsim}).}
The \SwitA\ luminosity counter therefore suffers an inherent
additional systematic error of $2 \times 10^{-4}$ due to this
bias uncertainty, and for this reason we use the average of the \SwitR\
and \SwitL\ counters as our primary measure of luminosity.  
The determination of the residual bias in the left and right
coordinates individually forms the topic of the following 
section.
\subsubsection{Anchoring}
\label{sec:anchor}
%
%
%
%
%

This section describes the method used to measure any
remaining biases
in the smoothed 
radial coordinate of the reconstructed electromagnetic
showers. 
In determining the radial shower coordinate, we have made every effort to
explicitly preserve the condition that a shower which deposits equal energies
on two adjacent pads in the reference layer at 7$\x$ will be reconstructed,
in the mean, to lie exactly at the boundary between the two pads.
Due to the $r-\phi$ geometry of the pads, however, the pad area increases by 
about 3\% per pad at the inner edge of the acceptance.
In the mean, therefore, the true position of such equal-pad showers will lie 
at a smaller radius than the physical pad boundary, and this bias should be 
reflected in our smoothed radial shower coordinates.

This pad boundary bias depends sensitively on the lateral shower profile, 
which we found was not consistently modeled by existing detailed Monte-Carlo 
shower simulations.
We therefore chose to measure this bias experimentally in a test beam.

The test beam utilized a SiW calorimeter module of 3 azimuthal wedges
fully equipped in depth, and a 
four-plane, double-sided Si micro-strip telescope~\cite{bib-bonn} 
with a resolution of
better than $3\mu$m for individual tracks.
The geometry of the calorimeter pads with respect to the telescope was 
determined using a beam of 100~GeV muons which could be tracked through
the individual pads of the calorimeter, since
the SiW electronics is sensitive to individual mips.
The muon beam was alternated with one of 45~GeV electrons, and measurements
made to determine the pad boundary bias.

It might seem
most straightforward to measure directly the radial bias
of the reconstructed smoothed 
coordinate in the test beam.
However, studies
of the reconstructed coordinate
(section~\ref{sec:smoothing})
show that it is sensitive
to the distribution and type of material in front of
the calorimeter as well as to the incidence
angle of the incoming particles, which at LEP changes continuously over
the face of the detector.
Because the divergence of the test beam was small,
the angle of the incident particles was appropriate to mimic
particles emanating from the beamspot at LEP
for only a small fraction of the test beam events.
This small sample of events was centered slightly 
above the region of the inner radial cut.
Since it was not possible to reproduce exactly
the OPAL beam geometry and the 
material in front of the calorimeters in the test beam 
studies, a more indirect approach, largely free of material and
incidence--angle dependence, is used for determining the bias
in the radial coordinate at LEP. 

In this approach, called {\em anchoring}, we take
the position of the pad with the maximum 
amplitude in a given longitudinal layer 
(the pad--maximum)
as our fundamental tool.
As the radial position of 
the incoming particles  
crosses a radial pad boundary in a single layer,
the average pad--maximum moves rapidly
from one pad to the next, giving an image of the pad boundary,
as shown in \fg{fig:anc_pmax}.

The anchoring procedure uses the pad boundary image
as the bridge between the OPAL data
and the test beam data.  
In the test beam measurements,
the pad boundary bias due to the finite shower size and
the curved pad boundaries is measured with respect to the external beam
telescope.  
In the OPAL data, the apparent pad boundaries are then used to measure 
any additional bias in the smoothed radial coordinate.

We now describe the anchoring procedure in somewhat more detail.
The curve fitted to the test beam measurements of the
pad boundary images in \fg{fig:anc_pmax}
is an error function (a Gaussian convoluted
with a step function) whose width,~$w$,
we term the {\em pad boundary transition width}.
We define the point at which the error function crosses the $50\%$
level as the {\em apparent pad boundary}.
We then take the difference between the apparent pad boundaries for
electrons and muons as the {\em pad boundary bias}.
We have measured the pad boundary bias
at a great number of pad boundaries
at different layers within the 
test beam module, with and without additional preshowering material.
We find that over a large range of conditions, the results
can be usefully described as a function of 
$w$, as shown in \fg{fig:anc_pbiass}.  
The parameterization of the
pad boundary 
bias as a function of $w$ was determined from a
linear fit to the test-beam data with the bare calorimeter.  
An error of 2.0~$\mu$m is assigned to the parameterization
which safely covers other possible parameterizations such as
the power law form also shown in \fg{fig:anc_pbiass}.
Data taken with additional
material in front of the calorimeter is found to be adequately 
described by the parameterization determined from the bare 
calorimeter.

\begin{figure}[tbh!]
\begin{center}
\mbox{\epsfxsize16cm\epsffile{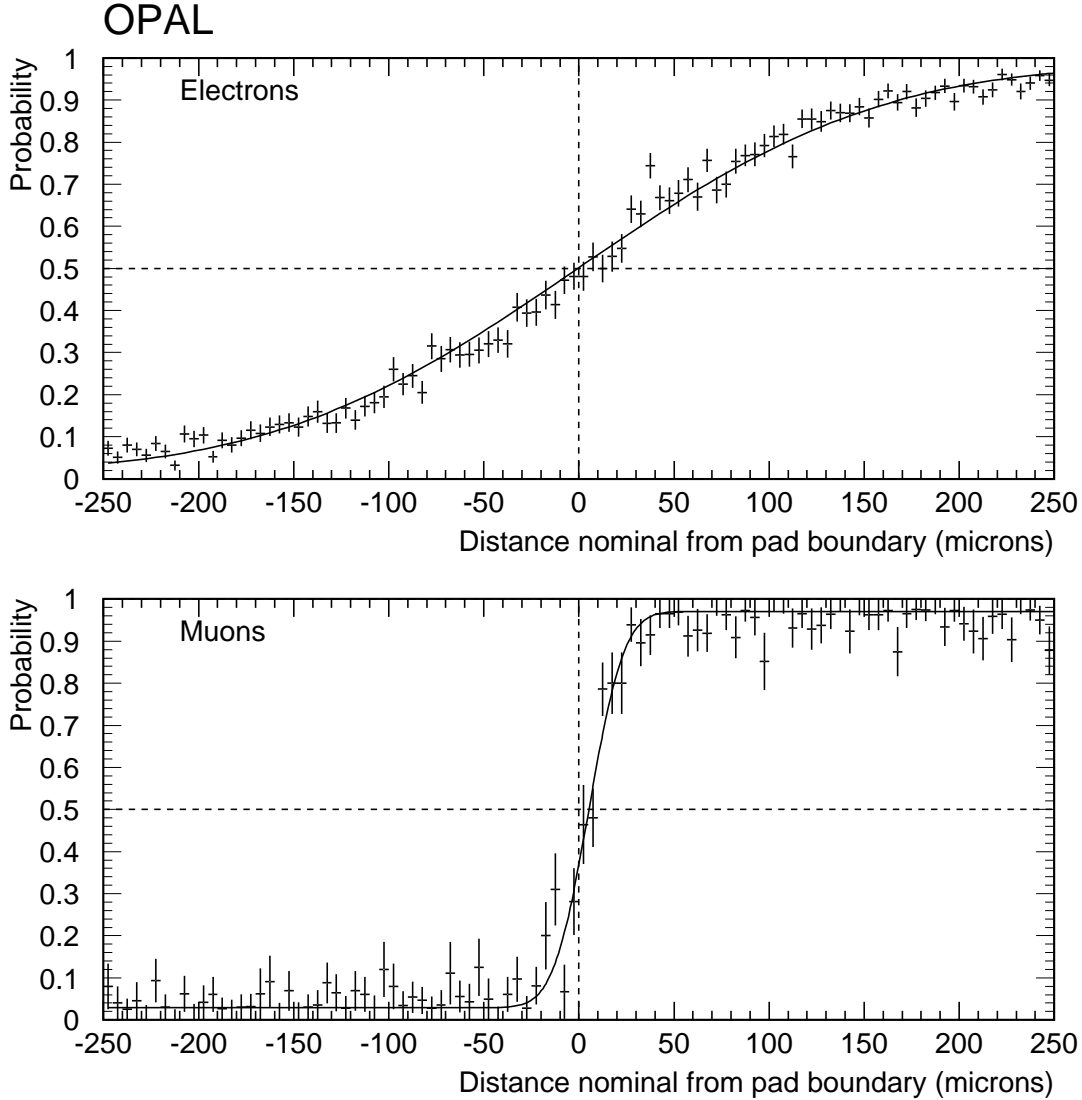}} 
\end{center}
\caption[Pad boundary image]
{ The pad boundary images for test beam electrons and muons
at layer 7. Each image is formed by plotting the
probability that the pad with maximum signal lies above
a given pad boundary as a function of the incident particle
trajectory.
The measured pad boundary bias is the {\em difference} between the intersection
of the fitted curves for electrons and muons with the 0.5 probability level.
The horizontal axis is in micro-strip coordinates only nominally
centered on the pad boundary.
\label{fig:anc_pmax}}
\end{figure}

\begin{figure}[tbh!]
\begin{center}
\mbox{\epsfxsize16cm\epsffile{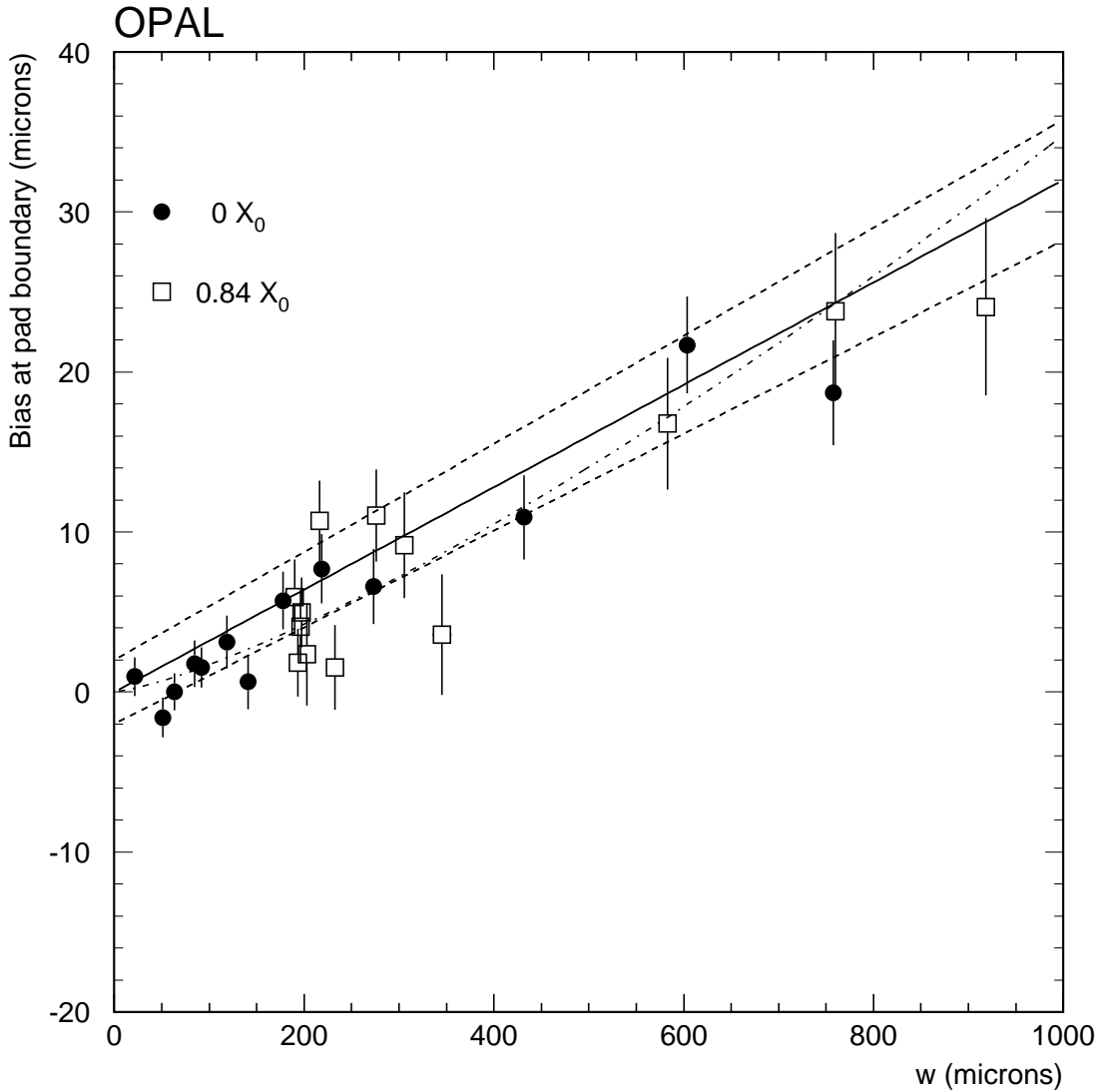}} 
\end{center}
\caption[Pad boundary bias]{ 
The pad boundary bias as a function of the measured
pad boundary transition width ($w$).  The data points
correspond to measurements at different depths in the
calorimeter.
The solid circles
show the bias for the bare calorimeter 
(0 $\x$) and the open boxes the calorimeter behind
additional dead material (0.84 $\x$).
The line was fitted to the 0 $\x$ data.
The dotted lines show the error assigned to the parameterization.
The dash-dotted lines shows an alternative two parameter fit
of the form  $a x^{b}$.
\label{fig:anc_pbiass}
}
\end{figure}

In OPAL running, we can not directly measure $w$
since we lack a micro-strip telescope to determine the
incident electron trajectories.
However, we can measure the apparent 
width, $w_a$, of the
pad boundary image in terms of the
smoothed radial coordinate.  
Since both
$w$ and $w_a$ can be measured in the test beam,
the test beam measurements can be used
to estimate $w$ from the
measured value of $w_a$ as shown in \fg{fig:anc_sigs}. 
The error bars shown on the points for $w$ indicate the uncertainty in
the conversion from $w_a$ to $w$.
This uncertainty is estimated from the
difference between the
test beam data with no additional material 
and with 0.84~$\x$ of material in front of the detector.
At the reference layer (7~$\x$) the $w_a$ to $w$ conversion contributes
about $2\,\mu$m to the uncertainty in the net coordinate bias.

\begin{figure}[tbh!]
\begin{center}
 \mbox{\epsfxsize16cm\epsffile{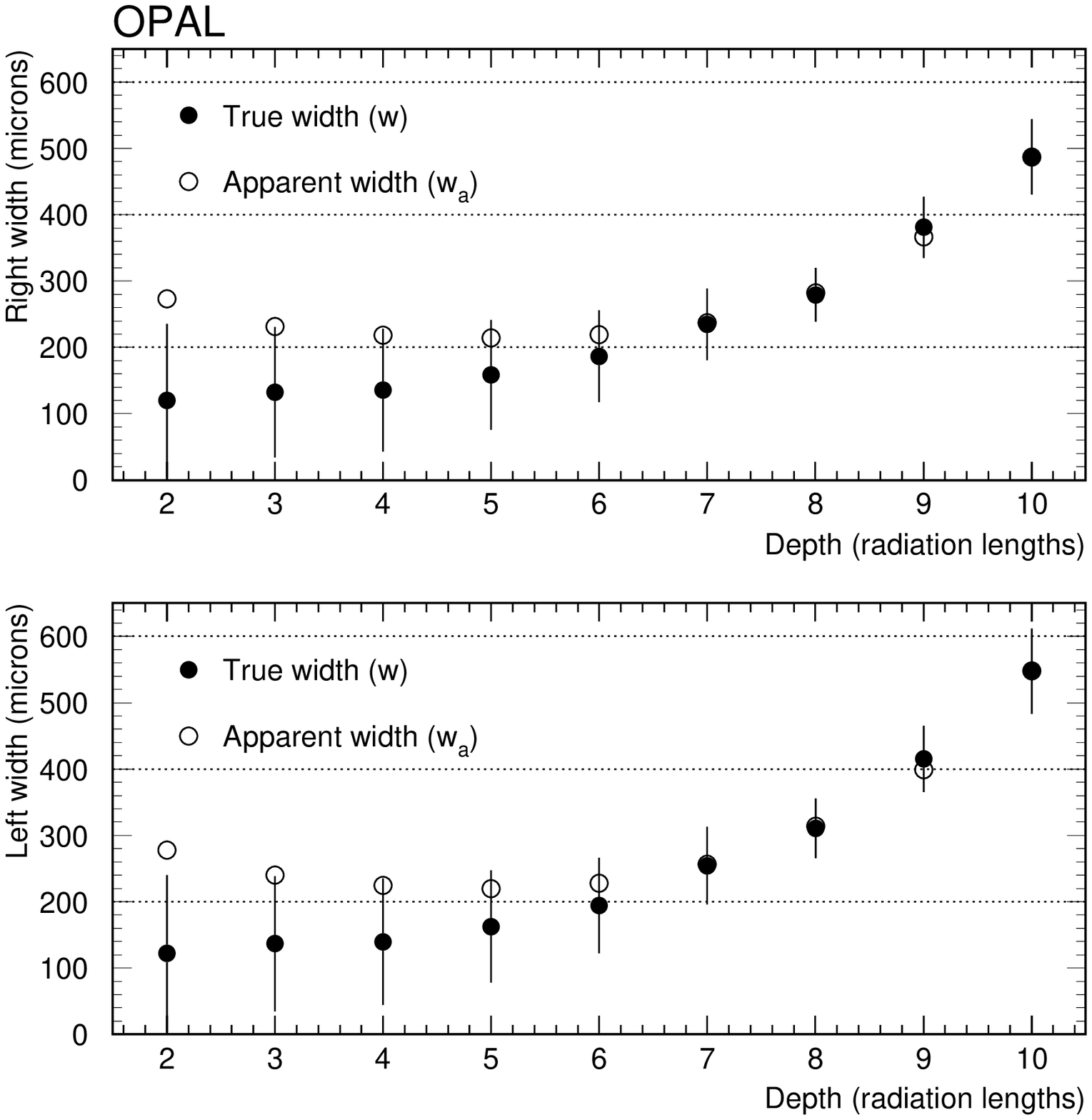}} 
\end{center}
 \caption[Pad boundary image width, $w$]
{ Apparent and estimated true values of the pad-boundary transition
width, $w$,
at the boundary between pad rows 6 and 7 (7.7~cm).
The upper plot
shows the right side, the lower plot the left side. 
The errors
on the estimated true width reflect the uncertainty in the
conversion of $w_a$ to $w$.
\label{fig:anc_sigs}
}
\end{figure}

To apply the anchoring method to the LEP data, events are selected
by applying all the \SwitR\ and \SwitL\ isolation cuts.
This eliminates background from the sample.
The coordinates on the right and left sides are then separately
anchored.
The first step of the
anchoring is to scan through the pads in each
layer to find the most energetic pad associated
with the cluster, $i_{\mathrm{max}}$, which is not more than 2 pad
widths (0.5~cm) from the average radial coordinate
of the shower.
This requirement is designed to eliminate
spurious pad--maxima, interference
from background, and multi-cluster events.

The image of the pad boundary is then formed by
plotting the fraction of events 
(in a given layer) 
with a pad--maximum above a given boundary (i.e., $i_{\rm max} \geq i$)
as a function of the radial calorimeter coordinate
${R_{\mathrm{smooth}}}$, as shown in \fg{fig:anc_l8}.  
As for the test beam measurements,
we model the pad boundary image with an error function
\begin{equation}
f(R;w_a,R_{\rm off} ) = \int g(t;R,w_a) \theta(t - R_{\rm off}) dt
\end{equation}
where $R$ is the distance from the nominal pad boundary, 
$g(t;R,w_a)$ is a Gaussian of width $w_a$ and mean $R$, 
and $\theta(t-R_{\rm off})$ is a step function with offset, $R_{\rm off}$, 
from the nominal pad boundary.  

\Fg{fig:anc_l8} shows that
a Gaussian resolution does not perfectly describe the tails of
the distribution, even at the 
center of measurement.
To the extent that the pad boundary image maintains an odd symmetry
about the apparent pad boundary, its non--Gaussian behavior
does not affect the determination
of $R_{\rm off}$, as can be seen from the close agreement of the data 
points and the fitted curve near the pad boundary.
The effect of non-Gaussian resolution becomes more important 
deeper in the calorimeter as the showers broaden and traces of asymmetry
appear.
We have also considered a model in which the apparent pad boundary is
taken as the median of the observed resolution function.
Over a reasonable range of depths, the differences observed with respect
to the Gaussian method set the scale for the $1.5\,\mu$m error we assign
to our method of fitting the pad boundary image.

%
%
\begin{figure}[tbh!]
\begin{center}
 \mbox{\epsfxsize16cm\epsffile{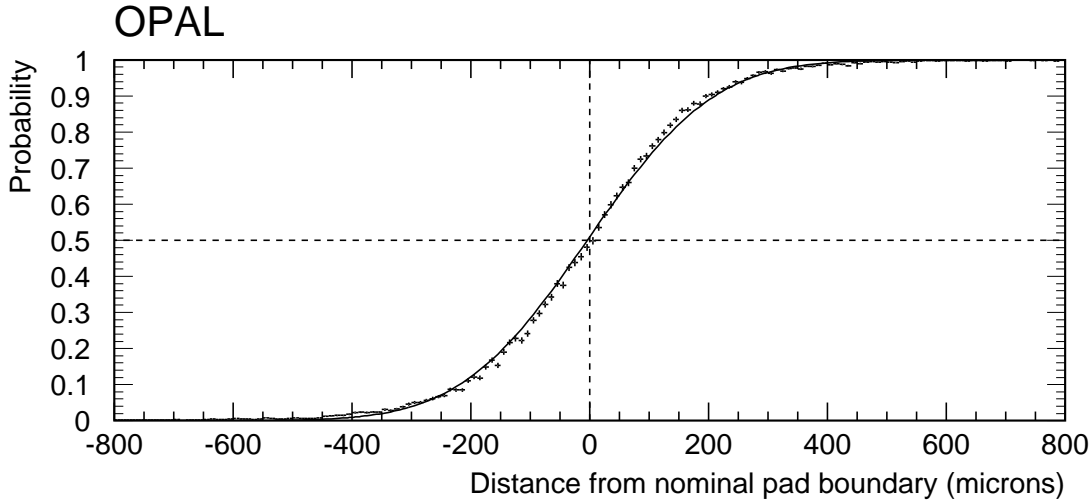}} 
\end{center}
 \caption[Pad boundary image (data)]{ 
The pad boundary image at LEP.
The fraction of events with maximum in pad row 6 or beyond, as
a function of distance from the pad boundary between rows 5 and 6, for
the layer located after 7~$\x$.  The solid curve shows the fitted
function.
  \label{fig:anc_l8}
}
\end{figure}

Using the measured values of $w_a$, the appropriate pad boundary bias measured
in the test beam can be added to $R_{\rm off}$ to
find the net bias in the smoothed coordinates for each of
the luminosity samples.
The results for our {\em primary anchor} at $R=7.7$~cm and a depth of
$7\x$ are shown in
\tbs{tab:final_anc_r} and \ref{tab:final_anc_l}.
Note that, as expected, the values of $R_{\rm off}$ are small, and
the inherent pad boundary bias dominates the net coordinate bias, so that
the effect of the finite shower size is to cause the coordinates of the
pad boundaries to
be reconstructed at smaller radii than their true position
(the net bias is positive).
Additional small corrections to the anchors, due to resolution 
effects, have been applied using
equations~\ref{eq:rinner} and \ref{eq:resrin}.  These
corrections account for
the difference between the resolution flows of the radial coordinate
and the apparent pad boundary.

Studies of $R_{\rm off}$ as a function of the 
32~azimuthal divisions of the calorimeters, taking into account
the detailed
radial and longitudinal metrology of the pads as well as beam offsets, 
find that the residuals are
approximately a factor of two larger than
would be expected
on the basis of statistical errors alone.  
The size of these extra azimuthal variations ($\sim7.5\,\mu$m)
is only slightly larger than the azimuthal variations
in the average radial coordinate which would be generated by the largest
pad gain fluctuations still compatible with the $\leq 0.25\%$
uniformity observed
in our electronic calibrations ($\sim5.5\,\mu$m, see
section~\ref{sec:average_coord}).
Such gain fluctuations affect the apparent coordinates much more than the
apparent pad boundaries.
Since the anchoring procedure aligns the coordinates to the pad boundaries,
it would be appropriate to use,
as a systematic error in the net coordinate bias,
the smaller gain-induced variations we could expect in the pad boundaries,
$\Delta R = 2.5\, \mu\mathrm{m} / \sqrt{32} = 0.6 \mu\mathrm{m}$
(see section~\ref{sec:layer_coord}).

We choose, however, to be more conservative,
and ascribe all the unexplained azimuthal variations to the anchors rather
than the coordinates,  and take the corresponding
$$
\Delta R = 7.5\ \mu\mathrm{m} / \sqrt{32} = 1.5 \mu\mathrm{m}
$$
as the systematic error in the net coordinate bias due to pad gain variations.

As expected,
the anchors in alternating azimuthal towers, which correspond to
the pattern in which the AMPLEX channel ordering is reversed
in the anchoring layer, do not show any
structure beyond the limit of statistical sensitivity of about
$2 \mu\mathrm{m}$.


The errors in \tbs{tab:final_anc_r} and \ref{tab:final_anc_l} are
dominated by systematic effects which we take to be common to
all of the data analysis periods.  The sources of these errors
are summarized in \tb{tab:anc_syse}.
 
As expected, the net coordinate bias does not
vary significantly as a function of data set, except for the
1995 data on the right side, where the amount of material in front
of the detector changed considerably due to the installation
of additional cables for the micro-vertex detector.  We therefore
average the 93-94 and 95 data separately, as shown in
\tbs{tab:final_anc_r} and \ref{tab:final_anc_l}. 
We then use these average net coordinate biases, as measured at the inner
fiducial boundary in layer 7, to calculate the acceptance correction
shown in \tbs{tab:final_anc_r} and \ref{tab:final_anc_l}.
We apply the average of the right and left inner anchor corrections
to the $(\SwitR + \SwitL)/2$ acceptance, as shown in
section~\ref{sec:acceptance}, table~\ref{tab:cor_all}.
In propagating the inner anchor error to table~\ref{tab:syssum_all},
we also assume that all the errors are correlated between the left and right
anchors, except for the measurement error in $R_{\mathrm{off}}$ and the
error allowed for azimuthal variations.

The reconstructed radial coordinate can be studied by 
simultaneously varying the value of the radial cut in the
data and in the Monte Carlo.  
The Monte Carlo assumes that the radial coordinate 
is reconstructed without bias.  
Thus any difference
in the acceptance of the data and Monte Carlo as
the inner cut is varied,
beyond that expected from the finite statistics,
can be attributed to  residual structure in the radial coordinate.
The relative acceptance,
as a function of the value of the inner radial definition cut
for both the \SwitR\  and \SwitL\ selections,
is shown 
in \fg{fig:anc_dmc9394}
for the combined 1993--1994 data sample.
The width of the shaded bands represent
the binomial errors with respect to the standard selection,
with its cut at 7.70~cm.
The relative acceptance is remarkably flat,
indicating the smoothing procedure has been properly
applied to the data.
Note that the smoothing procedure does not allow events
to move across the pad boundaries at layer 7~$\x$, i.e. at 7.45~cm, 
7.70~cm and 7.95~cm, so that the
data--Monte Carlo comparison at these points cannot be
affected by the smoothing.

The unique pad boundary at 7.70~cm in layer 7 was chosen
as the optimum point to anchor the coordinates (the primary anchor).
A large number of other pad boundaries could have been chosen,
and should give consistent results, assuming our methods are
valid.
The solid points in \fg{fig:anc_dmc9394} 
show the anchoring results for the 5-6, 6-7,
and 7-8 pad row boundaries and for layers between 2~$\x$ 
and 10~$\x$. 
The primary anchor at 7.70~cm has been required to lie at zero,
while the other points demonstrate the consistency of other
potential choices.
Even at the level of the small residual structure in the
smoothed coordinate, the alternative anchors track the 
expected acceptance variations for each of the pad rows 
and most of the depth range studied.

Looking at the anchor measurements as a function of depth, good
agreement is seen for the layers near the center of measurement.
For the deepest layers the difference between the anchors and
the measured relative acceptance changes exceeds two standard deviations.  
The character of these differences might
be expected, since the anchor measurements in the
deeper layers are much more sensitive to gain variations 
and details of shower development than those at the 
center of measurement.
The largest differences occur at the 
7-8 radial pad boundary in the left calorimeter, where
the amount of material between the detector and the
interaction point is starting to become large and the test beam
measurement of the expected bias may no longer be applicable.  
Furthermore, the non-Gaussian behaviour of the pad boundary image
at layer 10~$\x$ causes the anchor in this layer to
be sensitive to the fitting method used.
Presumably a more sophisticated treatment of the deeper layers
would give better results, but the anchors in
layers 2--9~$\x$ are more than sufficient for a complete
understanding of the radial coordinate.

We have also applied the anchoring method to determine the
bias in the radial coordinate at the outer edge of the
acceptance.  The acceptance of the luminosity
selections is approximately a factor of four less sensitive to
the outer edge than it is to the inner edge, making the result less
sensitive to biases in the smoothed coordinate in the outer region.
Since we have no test beam measurements of the pad boundary
bias for a radius of pad curvature corresponding to the outer edge
of the acceptance, our analysis of the outer anchor is also necessarily
less precise.
We assume that the pad boundary bias scales according to our
expectations (as $1/R$) but assign an additional systematic error equal
to $50\%$ of the expected bias to account for possible deviations from
this behavior.
We also treat the inner and outer anchor errors as being uncorrelated,
the most conservative assumption.

The outer anchor acceptance correction is then calculated according to
equation~\ref{eq:router}.
Tables~\ref{tab:final_anc_rout} and~\ref{tab:final_anc_lout}
give the anchoring results for the right and left sides of the detector,
respectively.
The average of the right and left outer anchor acceptance corrections
and their uncertainties are entered in tables~\ref{tab:cor_all}
and~\ref{tab:syssum_all} in section~\ref{sec:acceptance}.
The outer anchor errors are taken as mostly
correlated between the two sides, as they were for the inner anchor.

\begin{table}
\begin{center}
\begin{tabular}{|l||r|c|r|c|}
\hline
 Sample &  \multicolumn{1}{|c|}{$R_{\rm off}$}  & Estimated $w$ &
\multicolumn{1}{c}{Net coordinate bias} &
Acceptance Correction \\
        &  \multicolumn{1}{|c|}{($\mu$m)}    &  ($\mu$m)          &
\multicolumn{1}{|c|}{($\mu$m)}    & ($\times 10^{-4}$)\\
\hline\hline
  93 $-2$&$ -1.6\pm  1.8$&$ 232.\pm  62.$&$  6.7\pm  3.7$&$             $\\
  93 pk  &$ -4.7\pm  1.8$&$ 232.\pm  58.$&$  3.6\pm  3.8$&$             $\\
  93 $+2$&$ -4.8\pm  1.8$&$ 229.\pm  62.$&$  3.3\pm  3.7$&$             $\\
  94 a   &$ -4.1\pm  1.8$&$ 228.\pm  57.$&$  4.0\pm  3.7$&$             $\\
  94 b   &$ -5.0\pm  0.8$&$ 237.\pm  64.$&$  3.5\pm  3.5$&$             $\\
  94 c   &$ -4.4\pm  3.1$&$ 222.\pm  60.$&$  3.5\pm  4.4$&$             $\\
\hline
  93--94 &$ -4.6\pm  0.6$&$ 237.\pm  64.$&$  4.0\pm  3.4$&$  1.6\pm  1.4$\\
\hline
  95 $-2$&$  3.0\pm  2.0$&$ 263.\pm  80.$&$ 12.7\pm  4.1$&$             $\\
  95 pk  &$  1.7\pm  2.5$&$ 255.\pm  77.$&$ 11.0\pm  4.3$&$             $\\
  95 $+2$&$  3.7\pm  1.9$&$ 257.\pm  76.$&$ 13.1\pm  4.0$&$             $\\
\hline
  95     &$  3.5\pm  1.2$&$ 265.\pm  82.$&$ 13.2\pm  3.8$&$  5.3\pm  1.5$\\
\hline

\end{tabular}
\end{center}

\caption[Inner anchor values (right)]
{Values of the right inner anchors at 7.70~cm in layer 7
for the nine data samples.  
The errors on $R_{\rm off}$ are statistical only.
The errors on the net coordinate bias
include the systematic error from
the test beam measurement and are
largely correlated between the data samples.
The acceptance corrections due to the net
coordinate bias are calculated for the 93-94 and 95 samples separately.
\label{tab:final_anc_r} }

\end{table}
\begin{table}
\begin{center}
\begin{tabular}{|l||r|c|r|c|}
\hline
 Sample &  \multicolumn{1}{|c|}{$R_{\rm off}$}  & Estimated $w$ &
\multicolumn{1}{|c|}{Net coordinate bias} &
Acceptance Correction \\
        &  \multicolumn{1}{|c|}{($\mu$m)}    &  ($\mu$m)          &
\multicolumn{1}{|c|}{($\mu$m)}    & ($\times 10^{-4}$)\\
\hline\hline
  93 $-2$&$ -0.4\pm  1.9$&$ 257.\pm  57.$&$  9.0\pm  4.0$&$             $\\
  93 pk  &$ -0.3\pm  1.9$&$ 250.\pm  55.$&$  8.8\pm  3.9$&$             $\\
  93 $+2$&$ -1.7\pm  1.9$&$ 243.\pm  54.$&$  7.1\pm  3.9$&$             $\\
  94 a   &$ -8.3\pm  1.8$&$ 249.\pm  55.$&$  0.7\pm  3.9$&$             $\\
  94 b   &$ -2.1\pm  0.9$&$ 258.\pm  57.$&$  7.3\pm  3.7$&$             $\\
  94 c   &$ -3.4\pm  3.0$&$ 215.\pm  47.$&$  4.2\pm  4.4$&$             $\\
\hline
  93--94 &$ -2.4\pm  0.6$&$ 257.\pm  56.$&$  6.9\pm  3.6$&$  2.8\pm  1.4$\\
\hline
  95 $-2$&$  1.5\pm  1.9$&$ 262.\pm  58.$&$ 11.1\pm  4.1$&$             $\\
  95 pk  &$ -1.9\pm  2.5$&$ 248.\pm  55.$&$  7.1\pm  4.3$&$             $\\
  95 $+2$&$  0.9\pm  1.8$&$ 244.\pm  54.$&$  9.8\pm  3.9$&$             $\\
\hline
  95     &$  0.6\pm  1.2$&$ 258.\pm  57.$&$ 10.0\pm  3.7$&$  4.0\pm  1.5$\\
\hline

\end{tabular}
\end{center}

\caption[Inner anchor values (left)]
{Values of the left inner anchors at 7.70~cm in layer 7
for the nine data samples.  
The errors
on $R_{\rm off}$ are statistical only.  
The errors on the net coordinate bias
include the systematic error from
the test beam measurement and are
largely correlated between the data samples.
\label{tab:final_anc_l} 
The acceptance corrections due to the net
coordinate bias are calculated for the 93-94 and 95 samples separately.
}
\end{table}
\begin{table}
\begin{center}
\begin{tabular}{|l||r|c|r|c|}
\hline
 Sample &  \multicolumn{1}{|c|}{$R_{\rm off}$}  & Estimated $w$ &
\multicolumn{1}{|c|}{Net coordinate bias} &
Acceptance Correction \\
        &  \multicolumn{1}{|c|}{($\mu$m)}    &  ($\mu$m)          &
\multicolumn{1}{|c|}{($\mu$m)}    & ($\times 10^{-4}$)\\
\hline\hline
  93 $-2$&$  7.3\pm  4.1$&$ 279.\pm  62.$&$ 12.9\pm  6.0$&$             $\\
  93 pk  &$ 24.3\pm  4.1$&$ 262.\pm  58.$&$ 29.4\pm  5.9$&$             $\\
  93 $+2$&$ 18.0\pm  4.3$&$ 279.\pm  62.$&$ 23.7\pm  6.1$&$             $\\
  94 a   &$ 16.2\pm  4.0$&$ 259.\pm  57.$&$ 21.3\pm  5.8$&$             $\\
  94 b   &$ 12.7\pm  2.0$&$ 289.\pm  64.$&$ 18.7\pm  4.9$&$             $\\
  94 c   &$ 10.2\pm  7.4$&$ 266.\pm  60.$&$ 15.5\pm  8.5$&$             $\\
\hline
  93--94 &$ 13.8\pm  1.4$&$ 290.\pm  64.$&$ 19.7\pm  4.8$&$ -1.8\pm  0.4$\\
\hline
  95 $-2$&$  9.2\pm  5.0$&$ 360.\pm  80.$&$ 17.1\pm  7.5$&$             $\\
  95 pk  &$ 15.5\pm  6.4$&$ 345.\pm  77.$&$ 23.0\pm  8.4$&$             $\\
  95 $+2$&$  5.8\pm  4.7$&$ 345.\pm  76.$&$ 13.2\pm  7.1$&$             $\\
\hline
  95     &$ 10.6\pm  3.1$&$ 374.\pm  82.$&$ 18.8\pm  6.6$&$ -1.7\pm  0.6$\\
\hline

\end{tabular}
\end{center}

\caption[Outer anchor values (right)]
{Values of the right outer anchors at 12.70~cm in layer 7
for the nine data samples.  
The errors on $R_{\rm off}$ are statistical only.
The errors on the net coordinate bias
include the systematic error from
the test beam measurement and are
largely correlated between the data samples.
The acceptance corrections due to the net
coordinate bias are calculated for the 93-94 and 95 samples separately.
\label{tab:final_anc_rout} }

\end{table}
\begin{table}
\begin{center}
\begin{tabular}{|l||r|c|r|c|}
\hline
 Sample &  \multicolumn{1}{|c|}{$R_{\rm off}$}  & Estimated $w$ &
\multicolumn{1}{|c|}{Net coordinate bias} &
Acceptance Correction \\
        &  \multicolumn{1}{|c|}{($\mu$m)}    &  ($\mu$m)          &
\multicolumn{1}{|c|}{($\mu$m)}    & ($\times 10^{-4}$)\\
\hline\hline
  93 $-2$&$  8.5\pm  4.8$&$ 376.\pm  83.$&$ 16.8\pm  7.7$&$             $\\
  93 pk  &$ 18.0\pm  4.8$&$ 357.\pm  79.$&$ 25.7\pm  7.4$&$             $\\
  93 $+2$&$ 16.0\pm  4.7$&$ 339.\pm  75.$&$ 23.3\pm  7.1$&$             $\\
  94 a   &$  6.4\pm  4.7$&$ 362.\pm  80.$&$ 14.3\pm  7.4$&$             $\\
  94 b   &$ 10.8\pm  2.3$&$ 402.\pm  89.$&$ 19.9\pm  6.8$&$             $\\
  94 c   &$ 17.2\pm  7.0$&$ 257.\pm  58.$&$ 22.3\pm  8.1$&$             $\\
\hline
  93--94 &$ 11.2\pm  1.7$&$ 399.\pm  88.$&$ 20.2\pm  6.6$&$ -1.8\pm  0.6$\\
\hline
  95 $-2$&$ -5.5\pm  5.0$&$ 367.\pm  81.$&$  2.5\pm  7.6$&$             $\\
  95 pk  &$  3.8\pm  6.4$&$ 352.\pm  78.$&$ 11.5\pm  8.5$&$             $\\
  95 $+2$&$ -0.3\pm  4.8$&$ 353.\pm  78.$&$  7.4\pm  7.3$&$             $\\
\hline
  95     &$ -0.2\pm  3.1$&$ 386.\pm  85.$&$  8.4\pm  6.9$&$ -0.8\pm  0.6$\\
\hline

\end{tabular}
\end{center}

\caption[Outer anchor values (left)]
{Values of the left outer anchors at 12.70~cm in layer 7
for the nine data samples.  
The errors on $R_{\rm off}$ are statistical only.
The errors on the net coordinate bias
include the systematic error from
the test beam measurement and are
largely correlated between the data samples.
The acceptance corrections due to the net
coordinate bias are calculated for the 93-94 and 95 samples separately.
\label{tab:final_anc_lout} }

\end{table}

\begin{table}[tbh!]
\begin{center}
\begin{tabular}{|l||c|c|}
\hline
Effect  &  Inner   &  Outer   \\
        & ($\mu$m) & ($\mu$m) \\
\hline\hline
Measurement of $R_{\mathrm{off}}$                 &  0.6&  1.6\\
Test beam measurement of pad boundary bias        &  2.0&  4.0\\
$w_{\mathrm{a}}$ to $w$ conversion                &  1.9&  1.6\\
Fit method                                        &  1.5&  1.5\\
Azimuthal variations                              &  1.5&  2.9\\
\hline\hline
Total                                             &  3.5&  5.6\\
\hline
\end{tabular}
\end{center}
\caption[Anchor sytematic errors]
{Typical values of the errors
for the inner and outer anchors
near the longitudinal center of measurement. 
The error on the pad boundary bias and the effect of azimuthal variations
track the shower size, which grows with the greater amount of preshowering
material at the outer radius.
We treat the inner and outer errors as being completely uncorrelated,
the most conservative asumption.
\label{tab:anc_syse} 
}
\end{table}

\begin{figure}[tbh!]
 \mbox{\epsfxsize16cm\epsffile{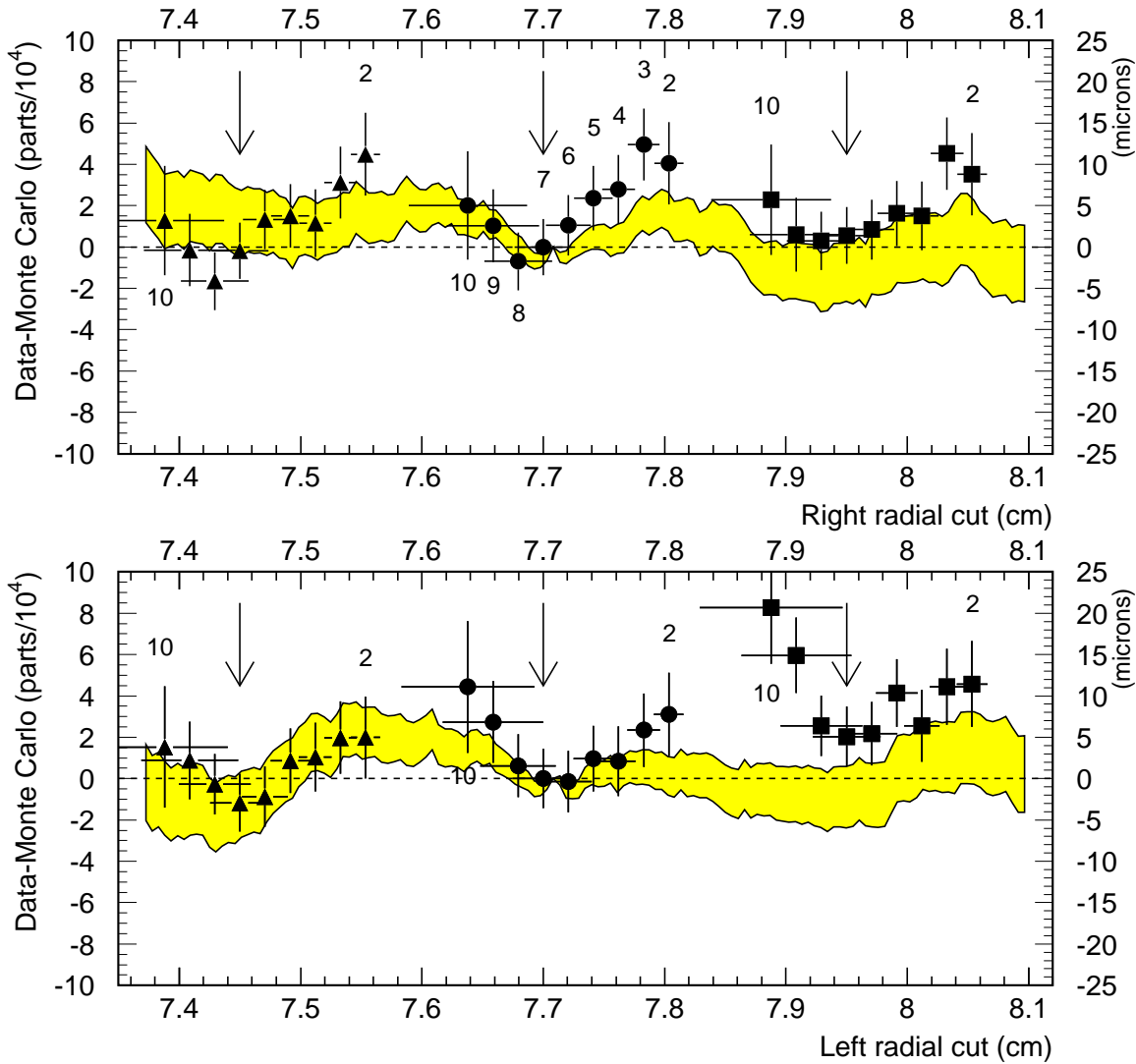}} 
 \caption[Data - Monte Carlo comparison ]{
The shaded band shows the
relative change in acceptance between data and Monte Carlo
as a function of the definition cut on the inner
radius for the \SwitR\ and \SwitL\ selections for the 
combined 1993--1994 data sample.
Departures from zero beyond the highly correlated one-sigma statistical
errors which correspond to the width of the band are due to the small
residual structure in the reconstructed radial coordinate.
The solid points show the coordinate bias measured by the anchoring
procedure.
The primary anchor at 7.7~cm is fixed to lie at zero.
The deviations of the other anchors from the data/Monte Carlo band show
the acceptance shifts which would result if any of the alternative pad
boundaries had been chosen as the reference.
Triangles, circles and squares indicate the 5--6, 6--7 and 7--8 pad
row boundaries respectively.  The numbers indicate the
layer depth in radiation lengths.
The arrows show the location of the radial pad boundaries at
the center of measurement (7~$\x$).

\label{fig:anc_dmc9394}
}
\end{figure}

\subsection{Azimuthal reconstruction}
\label{sec:azimuth}
%
%
%
%
%
The determination of the azimuthal coordinate is based on a
technique similar to the one used for the determination of
the radial layer coordinates.  
Because of the relatively
coarse azimuthal segmentation of the detector\footnote{Recall that
each layer of the detector is divided into 32 azimuthal segments, 
giving a pitch of $11.25^\circ$ 
which corresponds to a width of 1.2 to 2.7~cm.}
the azimuthal coordinate
relies primarily on the deeper layers of the calorimeter.
Another important difference between the azimuthal coordinate
and the radial coordinate is that the azimuthal towers are
projective.  Thus, the first step in the determination of
the $\phi$ coordinate is to sum the pads associated with a given 
cluster longitudinally and radially. Typically the shower
extends over three or four of the 32~$\phi$ segments.
In the summation, the deeper layers are given a larger weight.

The three most energetic contiguous $\phi$ segments are then
used to calculate the quantity $D$ given by equation~\ref{eq:coord_D}.
The distribution of the $D$ values is then fit to the
sum of two Lorentzian distributions, one with a narrow width
and the other with a wider width.  By integrating the differential
distributions, the linearization function can be extracted. 
Because
of the changing pad size, the parameters of the linearization
were determined separately in bins of radial coordinate.

After applying the linearization, a reasonably
flat $\phi$ distribution is obtained.  
Near the inner edge of the acceptance the
resolution is approximately 10~mrad
corresponding to 1/50 of a pad width.
The resolution of the $\phi$ 
coordinate depends on radius and increases by approximately 20\%
at the outer edge of the detector.

The uncertainties in the parameterization of the azimuthal coordinate 
result in a negligible systematic error on the acceptance of
the \SwitR\ and \SwitL\ selections.
The complete azimuthal coverage of the detector allows the
$\phi$--coordinates to be ignored except in determining the acoplanarity.
The azimuthal alignment of the right and left calorimeters is
verified by measuring a mean acoplanarity for Bhabhas of 6.2~mrad, in excellent
agreement with the value of 6.34~mrad expected from the magnetic deflection
of the Bhabha electrons in the OPAL magnetic field.
The azimuthal symmetry of both the detector and the Bhabha distribution
ensures that there can be no integral scale error in the acoplanarity
measurement.
Any local scale distortions will cancel in the integral over $\phi$, appearing
only as a broadening of the effective resolution.
The azimuthal resolution is determined from the width of
the beam--centered acoplanarity distribution of full--energy Bhabhas.
It is found to vary from 10--12~mrad
as a function of radius, with some evidence for non--Gaussian features.

\begin{figure}[htb!]
 \mbox{\epsfxsize17cm\epsffile{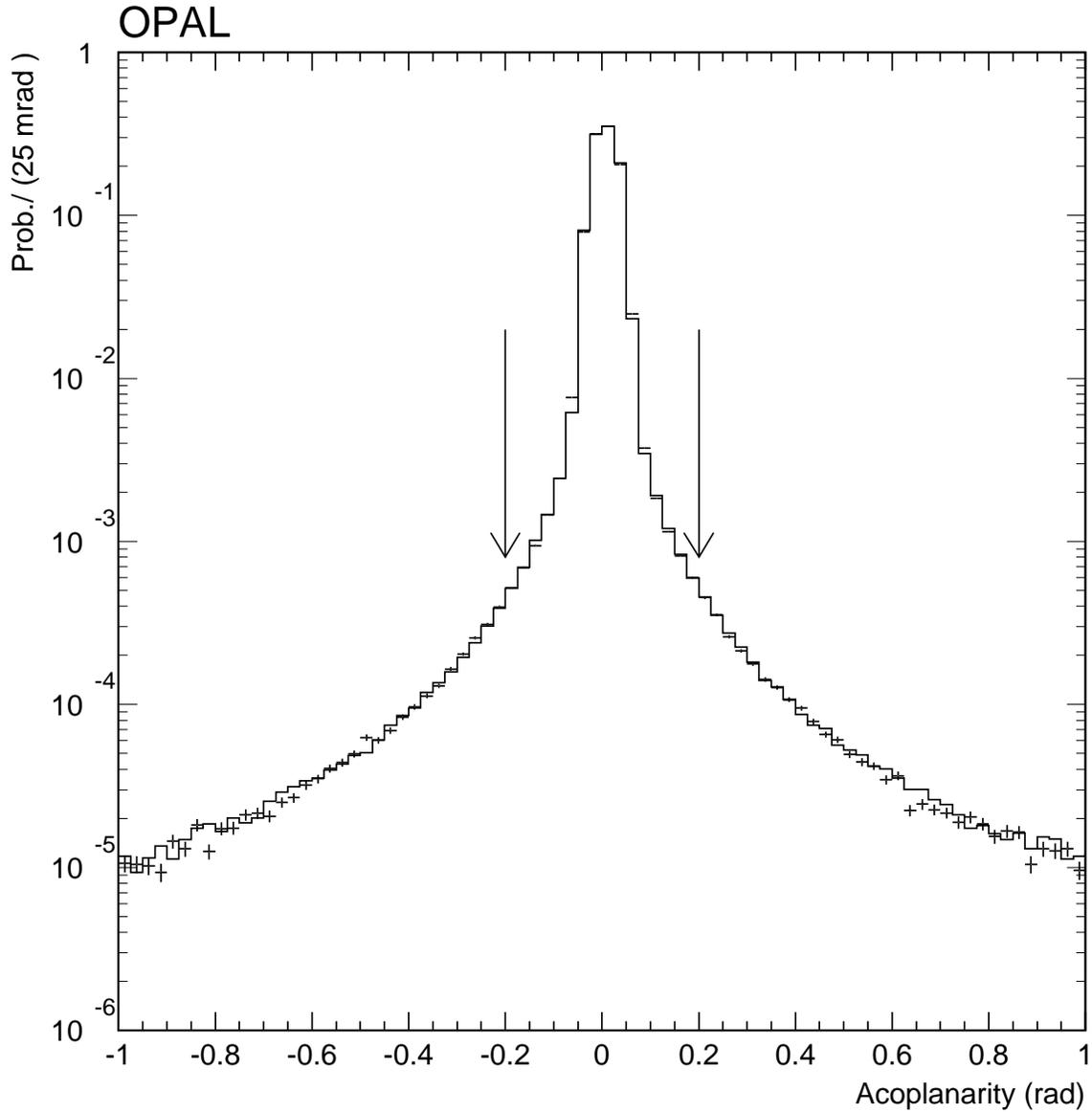}} 
    \caption[Acoplanarity]{
    The distribution of the acoplanarity
    ($(\phi_{R} - \phi_{L) - \pi)}$)
    after all the \SwitA\ selection cuts have been applied,
    except for the acoplanarity cut.
    Data are shown as points 
    and Monte~Carlo as
    the solid histogram.
    For this comparison, the background from accidental coincidences
    of off-momentum beam particles has been subtracted from the data.
    The acoplanarity cut is indicated by the two vertical
    arrows.
    The mean acoplanarity is offset from zero by
    6.2~mrad averaged over all fills.
    This can be compared with the expected effect due to the OPAL
    magnetic field, which is 6.34~mrad.
    \label{fig:sel_dphi}
    }
\end{figure}

Since the acoplanarity cut is made on the steeply falling tails of
the distribution (see section~\ref{sec:ev_samp}, figure~\ref{fig:sel_dphi})
the resolution flow is
\begin{equation}
    \frac{\Delta A}{A}
    \approx -
    \left(\frac{\sigma_{\Delta\phi}}{66 ~\mathrm{mrad} }\right)^2 \times 10^{-3}.
\label{eq:resdphi}
\end{equation}
\noindent
For the measured $11\times\sqrt{2}$~mrad resolution in $\Delta\phi$,
the resolution flow in acoplanarity
is $-0.5\times10^{-4}$.  Since the parametrized detector simulation does not
account for the non--Gaussian features of the true detector resolution,
$10\%$ of this calculated resolution flow is taken as a systematic error
in the acceptance.

\subsection{Energy reconstruction}
\label{sec:en}
%
%
%
%
%
%
%
%
\newcommand{\eraw}   {E_{\mathrm{raw}}}
\newcommand{\emain}  {E_{\mathrm{main}}}
\newcommand{\epre}   {E_{\mathrm{pre}}}
\newcommand{\fpre}   {F_{\mathrm{pre}}}
%
%
%

%
%
The reconstructed energy used in the luminosity analysis depends
on both the energy response of the \SW\ calorimeters to individual
showers and on the selection of the clusters which are used
in the final energy sum.  In this section we first discuss
the energy characteristics of Bhabha events and describe 
how the energy cuts were chosen.  
We then describe the procedure
used to correct the energy of individual clusters for
the  energy
loss in dead material and describe how the detector
energy response is extracted from the data and
parameterized.
Using a detector simulation based on this parameterization
we are able to propagate the
uncertainties in the measured energy response of
the detector to the systematic error in the luminosity
measurement.  Uncertainties in the energy resolution tails and
detector nonlinearity are found to be the dominant systematic
effects.
The effects of cluster selection
on the energy measurement are also discussed and a summary of the 
energy-related systematic errors is given.

The distribution of the summed energy in the left and right
calorimeters (after all other cuts) 
is shown in \fg{fig:sel_er_vs_elab}.  
The large accidental
coincidence background, visible at small $\er$, 
$\el$ requires that relatively tight
cuts be applied. 
Also apparent in the figure are the
radiative tails of the distribution from events which 
have lost energy due to a single initial state 
photon emitted along the beam axis.  
For these events transverse momentum conservation 
implies
\begin{equation}
    \frac{\er}{\el} = \frac{\rl}{\rr},
\label{eq:en_ratio}
\end{equation}
    so that an acollinearity cut effectively limits the energy
    lost to initial state radiation.
    The approximately 10~mrad acollinearity cut 
    raises the minimum kinematically 
    possible cluster energy for events radiating
    a single hard photon along the beam axis from $0.49  \ebeam$,
    given by the ratio of the 
    inner to the outer acceptance angles, to $0.75 \ebeam$, 
    safely above 
    the explicit energy cuts, thus reducing the
    intensity of events near the cut boundary by about a factor
    of four (see \fg{fig:sel_er_vs_elab}).

\begin{figure}[htb!]
    \begin{center}
    \mbox{\epsfxsize16cm\epsffile{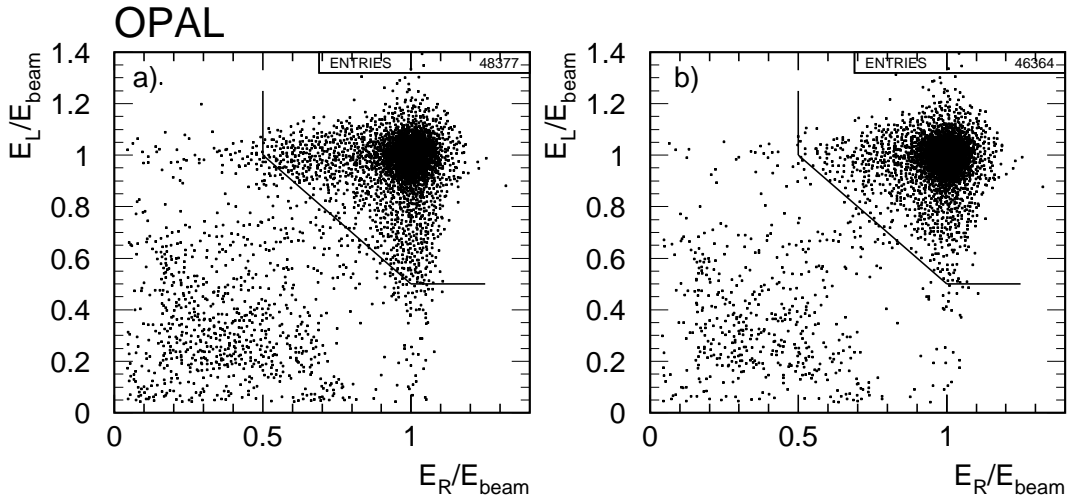}} 
    \end{center}
    \caption[$\el$~vs~$\er$]{
    The distribution of 
    $\el$~vs~$\er$,
    after all the $\SwitA$ selection cuts have been applied,
    except for the energy cuts and the acollinearity cut
    (a).
    The effect of the acollinearity cut on the
    radiative events and background is clearly visible in (b).
    The lines in the $\el$~vs~$\er$ plane indicate the cuts applied to 
    the energies.  The data shown here are from a typical run
    with approximately 40,000 $\SwitA$ events.
    \label{fig:sel_er_vs_elab}
    }
\end{figure}

    After the acollinearity cut,
    a change in the effective average 
    energy response of the calorimeters
    results in a change in the fiducial acceptance of
 \[
    \frac{\Delta A}{A} 
    \approx 
    \left( \frac{0.013  \Delta E}{E}\right)
\]
     for a range of a few percent in $\Delta E / E$.
     As the average energy response of the detector relative to
     the beam energy  can easily be adjusted to be the same in data
     and detector simulation, 
     no signficant systematic error results from our knowledge of
     the overall energy scale.
     A sample of collinear events has been used to 
    check the stability of the energy measurement.
    The average energy is stable
    at better than the percent level, leading to 
    variations in the acceptance at the level of $\sim 10^{-4}$.
    Because of
    the small size of the variations, and because the method used
    to determine the
    energy response of the 
    calorimeter automatically compensates for the
    variations in time, 
    these variations in energy scale do not cause any additional
    systematic error to the $\SwitR$, $\SwitL$ and $\SwitA$ acceptances.

    The most important source of systematic error results from 
    uncertainties associated with the
    treatment  of fluctuations in shower development due
    to energy loss in upstream material and from longitudinal
    and lateral leakage. 
    The upstream material  (mostly from
    cables and support structures) 
    has been kept to a minimum in the regions close to 
    the inner and outer edges of the acceptance.  
    However, in the middle of the acceptance the amount of material 
    traversed by the incident electrons exceeds 2 radiation lengths
    (see \fg{fig:det_mat}).  
    The energy lost due to leakage and preshowering has been parameterized
    as a function of both the radial coordinate of the
    shower and the ratio
    \begin{equation}
       \fpre = \frac{\epre} {\emain}
    \end{equation}
    where $\epre$ is the energy deposited in the first four
    layers of the calorimeter (0 to $3~\x$) 
    and $\emain$ 
    is the energy
    deposited in the remaining layers, for a given shower.
%
%
    A sample of beam energy electrons
    can be obtained by imposing a very tight
    collinearity cut 
    ($|R_{\mathrm{R}} - R_{\mathrm{L}} | < 1~\mathrm{mm}$)
    and by requiring that the measured energy on the 
    opposite side be close to the
    beam energy.\footnote{For the 
    studies of the calorimeter energy response discussed
    here, the coordinates of the scattered electron and positron
    were transformed to a system centered with respect to the LEP
    beam spot, so that the true collinearity angle is used as the 
    reference.}
    Note that except for a negligible fraction of double
    radiative events, the tight acollinearity requirement also 
    selects events with a single cluster. 
    Using this sample, the correction for dead material, in
    each of 60 radial bins, 
    is found
    from a fit of the $\fpre$ versus $\eraw$ distribution as
    illustrated in \fg{fig:en_pre}.  
    Near the inner and outer edges of the detector, where 
    the upstream material is small, the energy response is
    dominated by lateral leakage and a simple scaling, independent
    of $\fpre$ is
    applied to correct the raw energy.  The raw and corrected
    energy for the left calorimeter as a function of radius
    are illustrated in \fg{fig:en_raw}.

\begin{figure}[tbh!]
\begin{center}
 \mbox{\epsfxsize16cm\epsffile{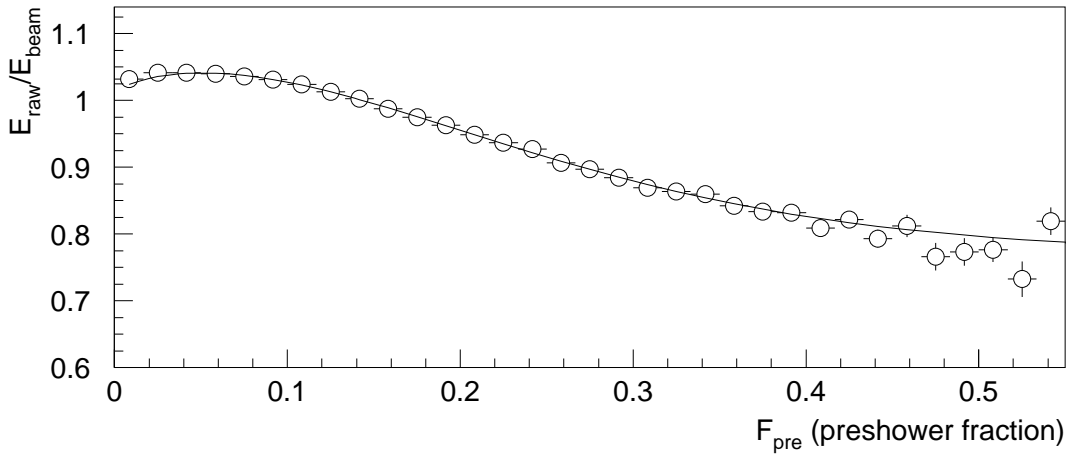}} 
\end{center}
 \caption[Mean raw energy]{
Mean raw energy, $\eraw$, measured in the calorimeter as a function
of the preshower fraction $\fpre$, for showers near the center of
the acceptance in the left calorimeter.  The solid line shows the functional
form used for the energy correction.
\label{fig:en_pre}
}
\end{figure}
\begin{figure}[tbh!]
\begin{center}
\mbox{\epsfxsize16cm\epsffile{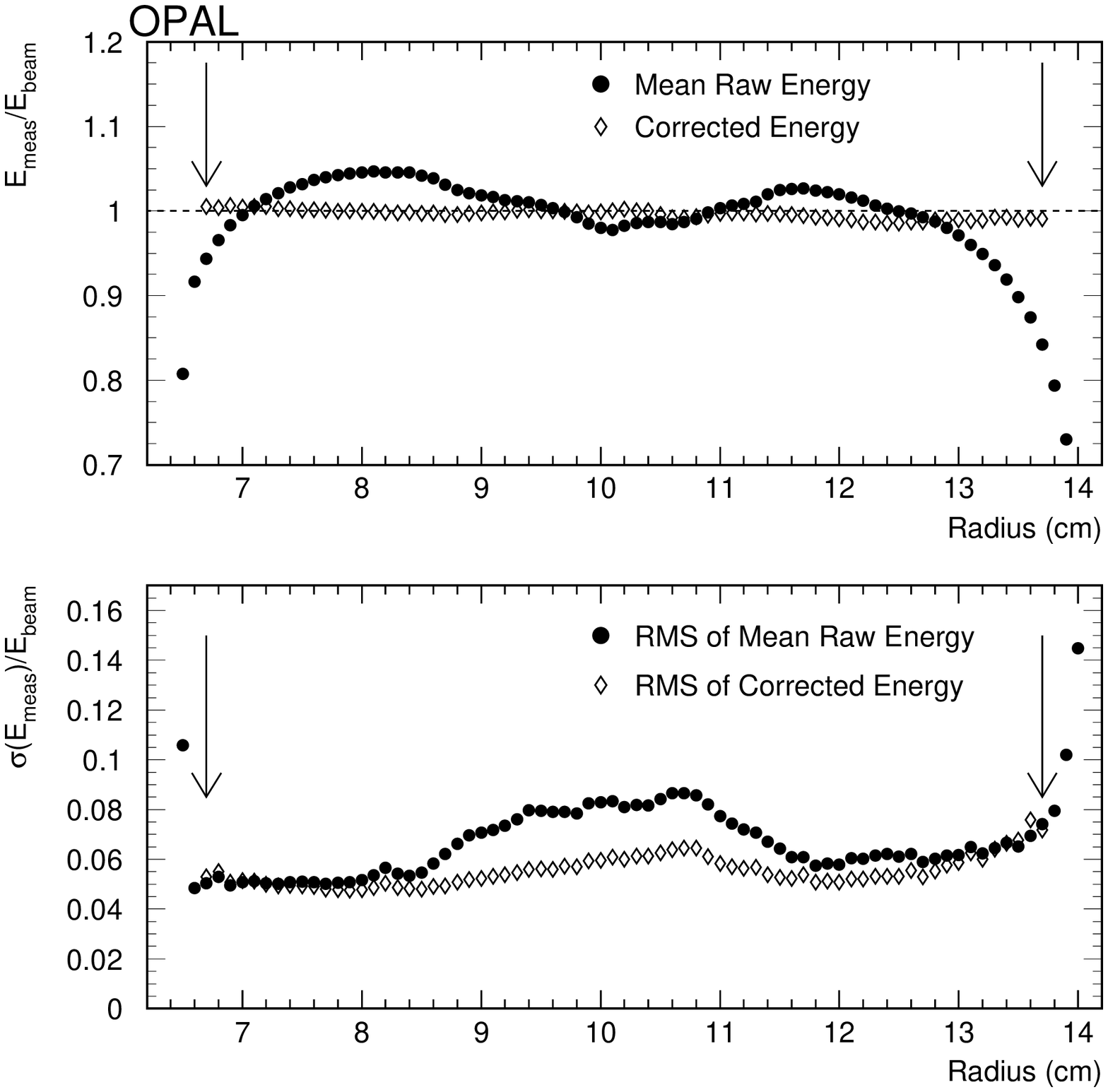}} 
\end{center}
 \caption[Corrected and raw energy]
{ Mean raw measured energy and its RMS for a sample of collinear
Bhabha electrons (solid dots),
as well as the mean corrected energy and its RMS for the same sample
(open diamonds).  All quantities are shown as
a function of radius for the left calorimeter.
The correction algorithm ensures that the energy response of the calorimeter
is not a function of radius, and recovers almost all of the resolution lost
to preshowering in the region between 9 and 11~cm, where the amount of
upstream material exceeds $2\x$.
The vertical arrows indicate the position of the loose radial cuts.
No clusters beyond these limits are used in the luminosity measurement.
\label{fig:en_raw}
}
\end{figure}

    \Fg{fig:en_eshape_def} shows the distribution of corrected 
    energies
    for collinear events in the central region
    of the SiW acceptance.
    On the right side, the energy resolution is
    approximately
    $\Delta E / E = 25 \% / \sqrt{E}$
    in resonable agreement with the expectation of 
    $\Delta E / E \sim 20 \% \sqrt{\chi_s} / \sqrt{E}$,
    where $\chi_s$ is the sampling in radiation lengths and
    $E$ is given in GeV\@.  
    On the left side the energy resolution is degraded by the
    presence of cables from the silicon microvertex detector.  
    In the region with the most material (near the middle
    of the radial acceptance) the energy resolution is
    approximately $\Delta E / E = 30 \% / \sqrt{E}$.  

    In order to describe the detector energy response, 
    the data have been divided into 60 radial bins
    and parameterized in each bin using two Gaussians of different widths
    and two Gaussians
    folded with exponentials with different exponents.
    The low exponential tail is due to events which shower 
    very late in the detector, events which are not fully contained,
    and events with electrons or positrons 
    that scatter off upstream material.

To optimise the energy response parameters, the simulated response
function was applied to four--vectors generated by the BHLUMI 4.04
Monte Carlo simulation of small-angle Bhabha scattering~\cite{bib-BHLUMI}.
The data and Monte Carlo energy distributions for events satisfying
a tight acollinearity
cut were compared as shown in \fg{fig:en_eshape_def}, and the energy
response parameters were adjusted to give the best agreement.
In selecting samples of essentially beam-energy scattered electrons, the
tight acollinearity cut prevented the manner in which photon radiation is
implemented in the Monte Carlo from influencing the parametrization of
the energy response function.
The energy spectrum of the bare Monte Carlo four-vectors, also shown in
\fg{fig:en_eshape_def}, falls two orders of magnitude below the simulated
detector response, showing that its shape cannot significantly influence the
detector response tuning.
If the response parameters had been optimised using an input delta function
rather than the BHLUMI four--vectors, the result would have been functionally
equivalent.

The most crucial feature of the energy response function is its integral
below the single-side energy cut at $E/\ebeam = 0.5$.
This controls the fraction of full-energy Bhabhas lost due to energy
resolution effects.
    In \fg{fig:en_eshape_def} it can be seen that the response function
describes this aspect of the data very well on both sides.
For the left calorimeter the simulated response function describes
the data everywhere extremely well, but for the right calorimeter, near $E/\ebeam = 0.8$, the
data are somewhat below the model.
This region of the response curve primarily affects the calculated
acceptance for single radiative events with a low energy, well measured electron
on the left, but a poorly measured full energy electron on the right.
Such events are not numerous, and we have calculated that the visible descrepancy
accounts for a systematic acceptance shift $<4\times10^{-5}$, which we
take as an energy response parametrization error in table~\ref{tab:en_sys_e_response}.

    The energy response of the detector at energies below the
    beam energy can be determined
    by using acollinear events
    which have energy in one calorimeter consistent with the
    beam energy.
    Using the ratio of radii of the showers in the
    right and left calorimeter (see equation~\ref{eq:en_ratio})
    it is possible
    to predict the energy ratio of the
    two sides with a resolution of approximately 1\%,
    resulting from the finite size and
    the divergence of the incoming beams.
    
    In principle, some nonlinearity is 
    introduced by the dead material in 
    front of the calorimeter. 
    Studies 
    of the energy correction in the acollinear events show
    that, in the energy range of 
    interest $(\ebeam \gt E \gt \ebeam/2)$,
    the preshowering correction increases as the raw energy of
    the cluster decreases, compensating for the 
    increasing fraction of energy lost in the dead material.
    Other possible effects are
    nonlinearities in the front-end electronics gain, 
    imperfectly
    determined online electronics pedestals,
    imperfectly applied cross talk corrections,
    the \mbox{${\sim 4}$~MeV}/channel threshold 
    applied to the energy sum associated
    with a cluster,
    and
    leakage of energy near the edges of the detector
    acceptance.\footnote{
    In test beam studies with a bare
    calorimeter, the energy reponse of contained
    showers was found to be linear within the
    sensitivity of the measurement (approximately 1\%).
    However, these studies can not be directly applied
    to the OPAL running because differences in
    the \hold\ timing and the trigger rate
    required special cross talk and pedestal corrections
    be applied to the test beam data.}

    Using the sample of acollinear events, the residual
    nonlinearity has been measured on the right and left
    sides separately.  The largest deviation seen
    between the expected and observed energy is 
    approximately 0.4~GeV.  
    \Fg{fig:en_nonlin} shows the difference between the observed
    and expected energy as a function of the expected
    energy for both uncorrected and corrected data.  
    The expected energy is calculated from the beam energy
    and the ratio of the radii of the two showers as described above.
    Also shown in \fg{fig:en_nonlin}
    is the expected value of this difference
    from the detector simulation (see section~\ref{sec:detsim}).
    The apparent nonlinearity visible in the figure in both the
    data and the Monte Carlo arises from effects of
    the finite size of the LEP beams and their divergence 
    convoluted with the rapidly falling
    spectrum of initial state radiation.
    Studies of data taken well above the 
    $\Zzero$ resonance show that a
    cancellation occurs between the slightly positive
    contribution from the imperfect cross-talk correction
    and the negative contribution from the pad threshold,
    giving a quite linear response for energies below 50~GeV.

    No evidence was found of any
    dependence of the nonlinearity on radius, indicating that
    nonlinearity due to shower leakage is not important for the
    event samples used to measure the luminosity.

    The nonlinearity was separately measured for the 1993--1994
    data and the 1995 data because of the 
    additional microvertex
    cables installed on the right side in the 1995 run.  
    The small sample of data taken in 1995 
    at center-of-mass energies near 133~GeV
    was also used to constrain the 1995 nonlinearity
    correction. 
    No differences, within the $0.45~\mbox{GeV}/\ebeam$
    error, between the expected and observed energies were seen.

    The parameters describing the energy response in the detector
    simulation have been varied
    within  bounds consistent with the precision with
    which they have been derived,
    allowing the
    systematic error due to the cuts on
    energy to be assessed.
    These variations included changing the contribution from
    the exponential tails by 50\%,  scaling the widths
    of the Gaussians  by $\pm 10\%$, and changing the
    method used to extrapolate the energy resolution to
    lower energies. 
    The allowed values of the
    nonlinearity were conservatively 
    assigned on the basis of a statistics-limited
    check with the 1995 high energy data (at 133~GeV).
    Any possible nonlinearity was shown to distort the
    energy scale by less than
    $\pm 0.01\ebeam$ at $E = \ebeam/2$.
    The effect of these parameter variations on the predicted 
    energy distribution is also shown in
    \fg{fig:en_eshape_def}.
    The band of variation is obtained by adding the variations
    for each of the four effects in quadrature.

\begin{table}[htbp]
\begin{center}
\begin{tabular}{|l||c|c|}
\hline
   & \multicolumn{2}{|c|} {Systematic error ($\times 10^{-4}$)} \\
\cline{2-3}
Effect & With acoll. Cut & Without acoll. cut \\          
\hline
\hline
Change tail $\pm 50\%$                              & 1.2  & 2.1 \\
Nonlinearity                                        & 1.2  & 3.2 \\
~~~~~~$(\pm 0.01\ebeam$ at $E = \ebeam/2$)          &      &     \\
Scale Gaussian $\pm 10\%$                           & 0.4  & 1.6 \\
Low energy resolution                               & 0.4  & 2.4 \\
Right side parametrization error                    & 0.4  & 0.4 \\
\hline
\hline 
Total                                               & 1.8  & 4.8 \\
\hline
\end{tabular}
\end{center}
\caption[Energy systematic errors]{
Summary of the systematic error in the luminosity
due to uncertainties in the detector energy response.
These errors were
estimated by varying the detector energy 
response parameters within the precision with which they 
could be determined from the data.
\label{tab:en_sys_e_response}
}
\end{table}

    \Tb{tab:en_sys_e_response} 
    summarizes the systematic errors assigned
    to the uncertainty in each aspect of 
    the calorimeter energy response
    discussed above.
    The total systematic uncertainty on the 
    luminosity due to uncertainty
    in the energy response is $1.8 \times 10^{-4}$.

    \Tb{tab:en_sys_e_response} also demonstrates 
    that the acollinearity cut reduces
    the acceptance uncertainty due to the energy cut
    by more than a factor of two.
    We have explicitly compared the results of the 
    luminosity determination
    with and without the acollinearity cut.
    The data and Monte~Carlo simulation track each other to within
    $1\times10^{-4}$,
    even though the density of events near the energy cuts increases by
    a factor of approximately four.

    Another potential systematic error in the energy measurement
    can arise from the method used to select clusters to include in the
    energy sum.  If only the most energetic cluster in each calorimeter
    were used in the energy sum, our analysis would be sensitive to the
    implementation of final state radiation in the physics Monte Carlo
    and to the modeling of two-cluster separation in the detector 
    simulation.  By summing the energy over the fiducial acceptance
    of the detector, our analysis is insensitive to the detailed
    implementation of final state radiation and to the modeling of
    two-cluster separation.
The kinematic cluster selection algorithm discussed in
section~\ref{sec:kine} efficiently includes the clusters which belong to
the Bhabha scattering event while rejecting the background clusters
generated by the off--momentum electrons discussed in section~\ref{sec:back}.
 
    We have also varied both of the energy cuts used in the analysis and 
    compared data and Monte Carlo.
    The minimum $(E_{\mathrm{min}})$ and 
average $(E_{\mathrm{ave}})$ energy cuts were 
    simultaneously varied so that
    the minimum energy requirement for a single radiative event remained
    coincident with the average energy cut, 
    i.e. $E_{\mathrm{ave}} = ( 1 + E_{\mathrm{min}} )/ 2$,
    as in the standard set of cuts shown in 
    \fg{fig:sel_er_vs_elab}.
    The comparison  for the 1993 and 1994 data 
    is shown in \fg{fig:en_scaled_e}.
    For this comparison a detailed background subtraction
    based on our delayed
    coincidence trigger sample has been performed in each energy bin 
    (see section~\ref{sec:back}).
    It can be seen that,  with the acollinearity cut applied,
    the measured luminosity changes by less than 
    $10^{-4}$ as the average energy cut is varied in the range from 
    0.7 to 0.8 of the beam energy.  
    (Also shown in \fg{fig:en_scaled_e}
    is the variation without the acollinearity cut.)
    Similar results were obtained for the 1995 data samples.

%
%
%

%
\begin{figure}[tbh!]
\mbox{\epsfxsize16cm\epsffile{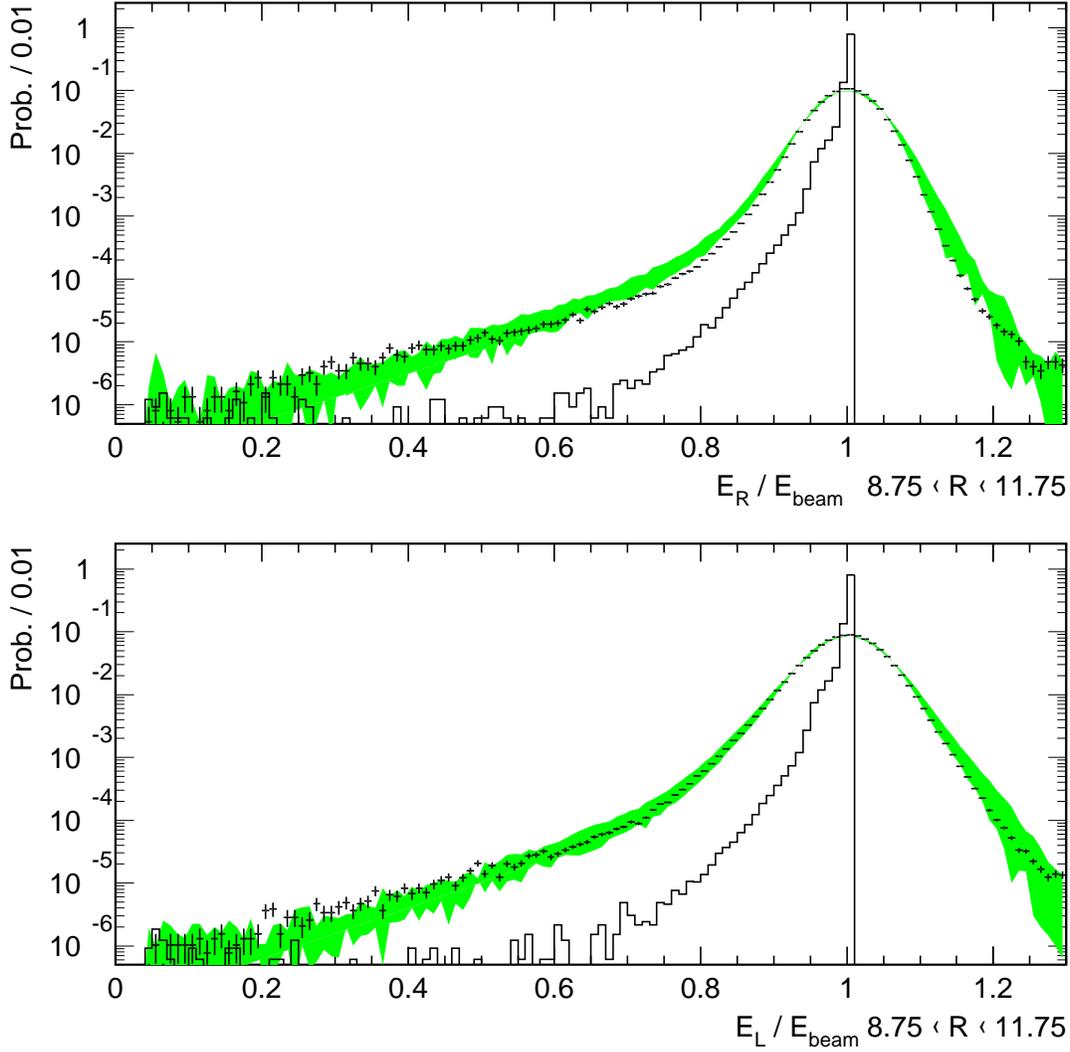}} 
\caption[Data-Monte Carlo energy comparison (fiducial region)]
{ The observed detector energy response to very collinear events,
with $|R_{\mathrm{R}} - R_{\mathrm{L}} | < 1~\mathrm{mm}$,
(points) which is used to parameterize the detector simulation
(shaded band) for showers in the radial region between 8.75 and 11.75~cm.
The width of the shaded band shows the variation in the parametrization
used in setting the systematic error.  Because of the tight acollinearity
cut applied to the samples, single radiative and double radiative
events are highly suppressed, and the observed distribution is
dominated by resolution effects, as illustrated by the narrow histogram
of the bare BHLUMI four--vectors, whose tail falls two orders of magnitude
below the simulated detector response.
\label{fig:en_eshape_def}
}
\end{figure}

\begin{figure}[tbh!]
  \mbox{\epsfxsize16cm\epsffile{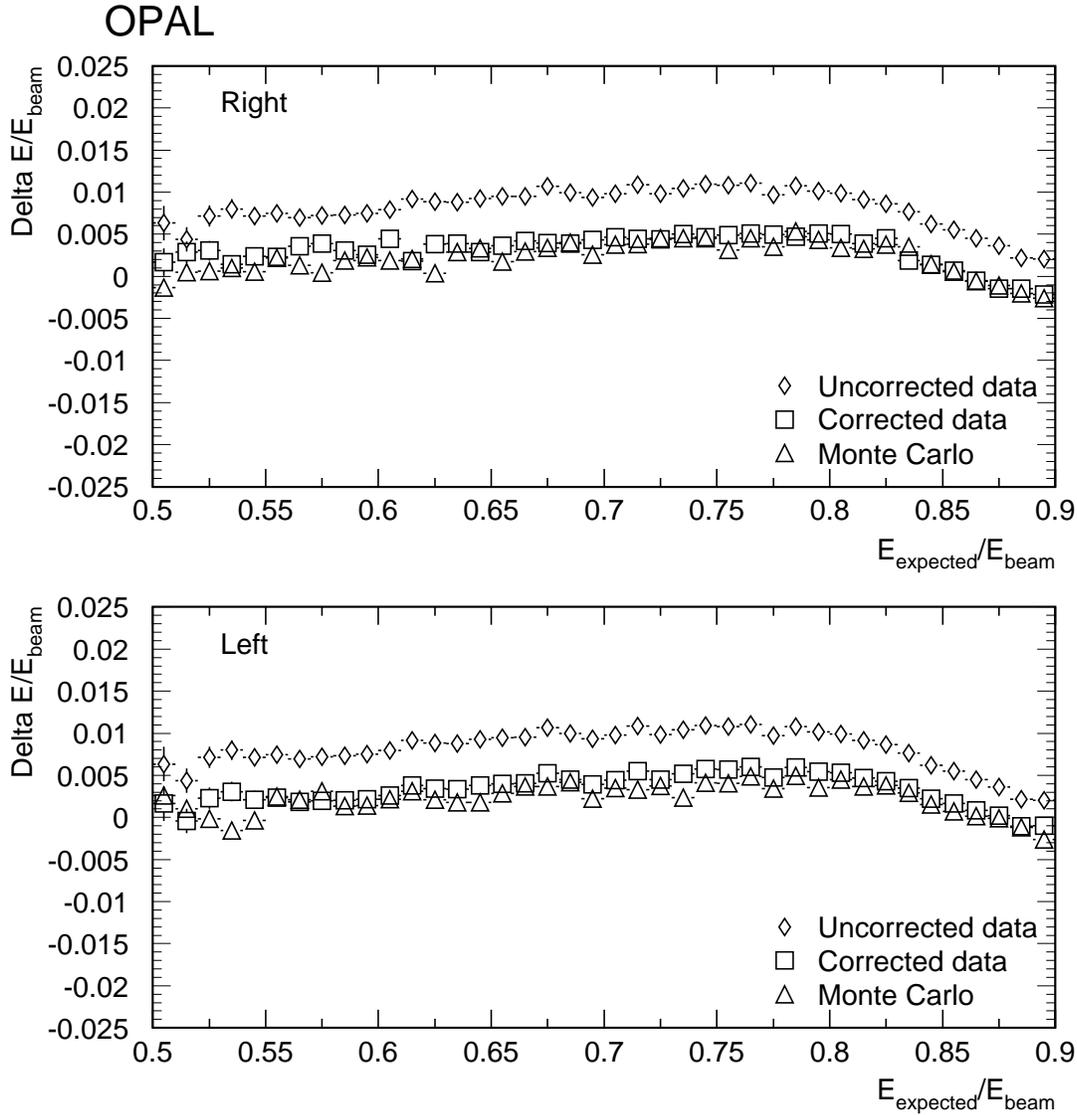}}
\caption[Energy nonlinearity]
{ Fractional difference between the expected energy 
(calculated 
from the radii of the scattered electron and positron) 
and both the corrected and uncorrected 
measured energy for the sample of acollinear
events.   Also shown is the difference expected from
the detector simulation.
    The apparent nonlinearity visible in both the
    data and the Monte Carlo above a fractional energy of 0.8
    arises from effects of
    the finite size of the LEP beams and their divergence 
    convoluted with the rapidly falling
    spectrum of initial state radiation.
\label{fig:en_nonlin}
}
\end{figure}

\begin{figure}[tbh!]
  \mbox{\epsfxsize16cm\epsffile{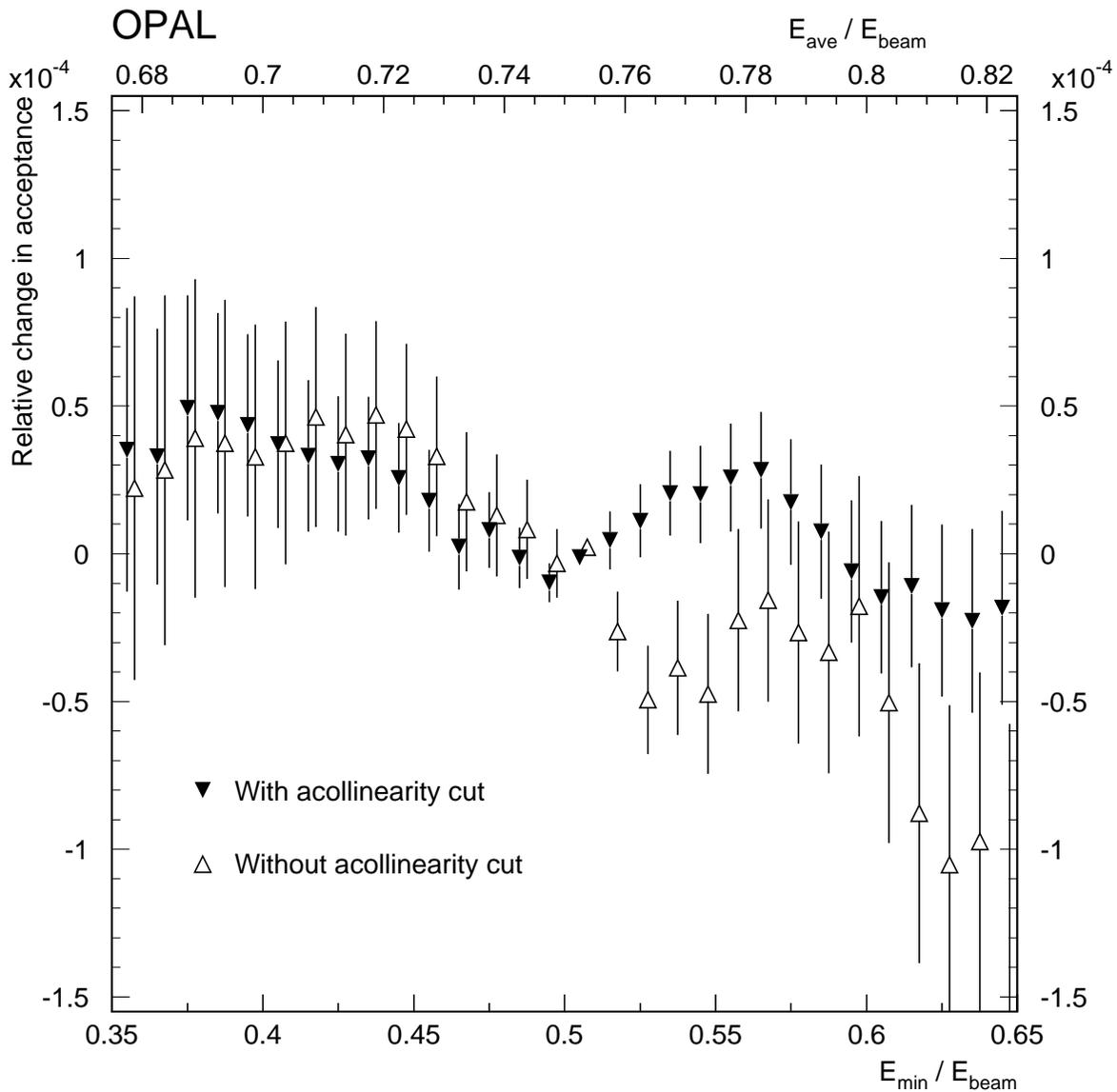}}
\caption[Change in acceptance (1994)]
{ The relative change in acceptance between the
Monte Carlo and the 1993--1994 data as a function of
scaling the energy cut.  
The observed deviations are well within
the systematic error assigned to detector energy response.
The error bars show the binomial uncertainties 
with respect to the standard cut values.
\label{fig:en_scaled_e}
}
\end{figure}
%


\clearpage
\section{Background and accidentals}
\label{sec:back}
%
%
%

The primary source of background to the luminosity measurement
is from 
off-momentum electrons and positrons generated by beam-gas
scattering in the straight RF sections on either
side of the experiment which
are deflected by the mini-beta quadrupoles into the luminosity
monitor.  The size of this background varies with time, and
depends on the quality of the vacuum in the straight
sections on either side of the OPAL interaction region and
on the settings of the LEP collimators.  
The occupancies of
the left and right calorimeters for 
clusters with $E > 1.0$~GeV, 
averaged over the 1993, 1994 and 1995 runs, are given in 
table~\ref{tab:acc_bxr}.
The 1993 and 1994 values are from the occupancy of random
triggers.  In the 1995 bunch train running, random triggers can 
only be used to measure the occupancy of the last bunchlet.
Therefore, in 1995 the occupancy was estimated from the rate of
extraneous clusters in non-radiative Bhabha events.

\begin{table}[bth]
\begin{center}
\begin{tabular}{|l|c|c|c|}
\hline
calorimeter      & 1993   & 1994 & 1995 \\
\hline
right &   $8.0 \times 10^{-3}$  & $5.1 \times 10^{-3}$ & $8.4 \times 10^{-3}$\\
left  &   $5.1 \times 10^{-3}$  & $3.7 \times 10^{-3}$ & $6.2 \times 10^{-3}$ \\
\hline
\end{tabular}
\end{center}
\caption[Occupancy]
{Probability for a cluster of 
$E > 1.0$~GeV to be found in the right or left calorimeters 
for the 1993, 1994 and 1995 runs.  
}
\label{tab:acc_bxr}
\end{table}


The background from off-momentum particles
affects the luminosity measurement in two ways.
Accidental coincidences between background clusters in the right and
left calorimeters can produce events which will potentially
pass the $\SwitR$ and $\SwitL$ selection. 
The accidental overlap of a background cluster with a Bhabha
event can also change the values of
reconstructed quantities, modifying the acceptance.

The rate of random coincidences 
between background in the left and right calorimeters 
is approximately given by the product of the rates and
is of order $5 \times 10^{-5}$ per beam crossing.  
This rate is comparable to the rate expected for true Bhabha events.
Fortunately the background events are primarily at low energy and 
can be isolated from Bhahba events as was shown in
section~\ref{sec:en},
\fg{fig:sel_er_vs_elab}.

In the 1993 and 1994 data,
this background can be studied using 
delayed coincidence triggers. 
Because the trigger signals from every
beam crossing are digitized and recorded, it
is possible to form a trigger 
which requires a coincidence when one side is
delayed by one full revolution of LEP
with respect to the other.
In these events, full detector readout is only 
available from the current beam crossing.
For the delayed beam crossing, 
only trigger information
is available.  
The calorimeter which exceeds the trigger threshold
in the current beam crossing is called the
in-time side.  The opposite calorimeter
is referred to as the
out-of-time side.
The available information
allows the energy and the azimuthal coordinate
of the out-of-time side to be reconstructed, but not the
radius.  In order to apply our standard selection procedure to
the delayed-coincidence events, 
the radial coordinate of the out-of-time
side is randomly generated as a function of azimuth and energy
from the three-dimensional distribution of $R$,$E$, and $\phi$
seen in that calorimeter when it 
was the in-time side in the sample of the delayed-coincidence triggers.
This procedure assumes that the correlations among
energy, radius and azimuth are relatively constant for 
a given year.

\Fg{fig:back_en94} shows the lower portion of 
the 1994 total energy spectrum
with all cuts applied, except that on the total energy.
The background, as calculated using the method described above,
agrees with the observed event rate at low energies to better
than 10\%.
The out-of-time background events above the cut on the total
energy of $(\er+\el)/(2 \ebeam) > 0.75$ are used to determine
the background in the selected luminosity sample.
The background fractions for 1993 and 1994
were between 0.1 and 0.6 $\times 10^{-4}$.
The small number of
accidental triggers which passed all cuts gave a statistical
error of about 30\% 
on the background which is much larger than the 10\%
systematic uncertainty.

\begin{figure}[tbh!]
\begin{center}
  \mbox{\epsfxsize17cm\epsffile{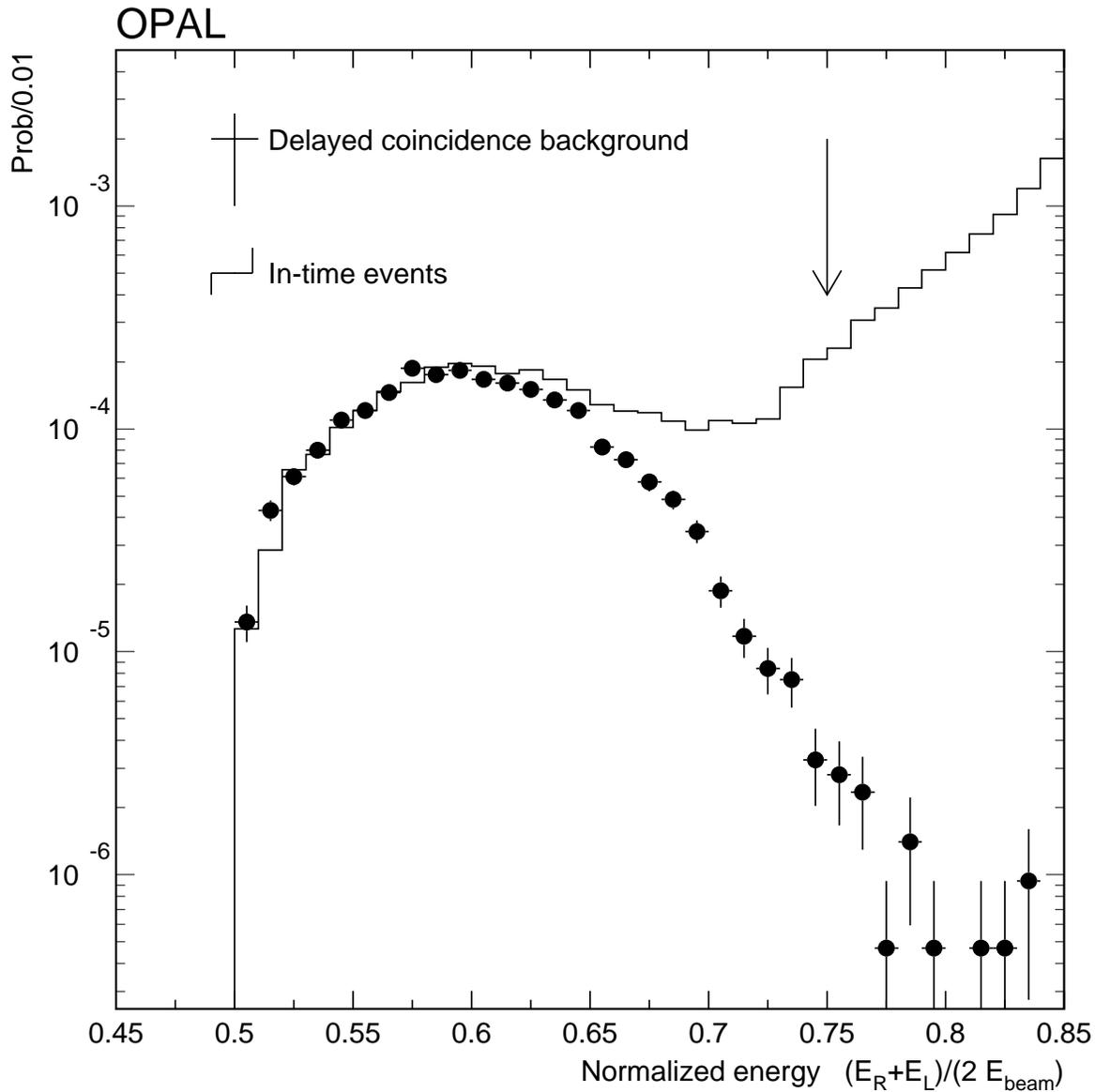}}
\end{center}
\caption[Low engery tail (1994)]
{ 
The low energy tail of the
total energy spectrum normalized to the number 
of accepted luminosity events for the 1994 data (histogram).  The
points with errors show the background spectrum derived from the 
delayed-coincidence triggers.
The arrow indicates the position of the selection cut made on the average
energy.
\label{fig:back_en94}
}
\end{figure}

The determination of the coincidence background in 1995 is somewhat
more involved. For almost all of the 1995 running, the \SW\ ``wagon
tagger'' was configured to require a coincidence between the 
energy deposition in the left and right calorimeters, 
before issuing a \hold\ signal to the front end electronics.
If no coincidence was found, the \hold\ signal was asserted on
the last bunchlet.
This means that the delayed-coincidence trigger
only operated properly for the last bunchlet of
each train.
For earlier bunchlets, the measured energy was less than the 
true energy.  
For example,
a \hold\ issued on the last bunchlet measured only 15\% of energy
deposited on the first bunchlet and 50\% of the energy deposited
on the second bunchlet.

A second consequence of the ``AND mode'' 
of operation 
was that only
coincidences between the same positron and electron bunchlets 
can produce background which is energetic enough to 
pollute the signal region.  
If the off-momentum background
for the three bunchlets were equal, the delayed coincidence trigger
would simply measure one third of the background.   
The characteristics of the background in 
the three bunchlets was studied by using
the background (``extra'') clusters in full-energy back-to-back
Bhabha events.
The overall energy spectrum 
of the background in each of the
bunches was found to be similar, but the
rate depended on the bunchlet.
Using the shape of the background as measured in the last
bunchlet and normalizing the energy spectra 
of the predicted background to the low energy tail
of the in--time data,
the background was determined separately for each of the
three energy points as
illustrated in \fg{fig:back_en95}.
Note that the cuts on the right and left
calorimeter energies 
(~$\er > 0.5 \cdot \ebeam$ and $\el > 0.5 \cdot \ebeam$~)
have already been applied, leaving only the high energy tail of the 
background.
The resulting corrections for the 1995 accidental coincidence background are
between 0.1 and $0.15\times10^{-4}$, as shown in table~\ref{tab:cor_all}.

\begin{figure}[tbh!]
\begin{center}
  \mbox{\epsfxsize17cm\epsffile{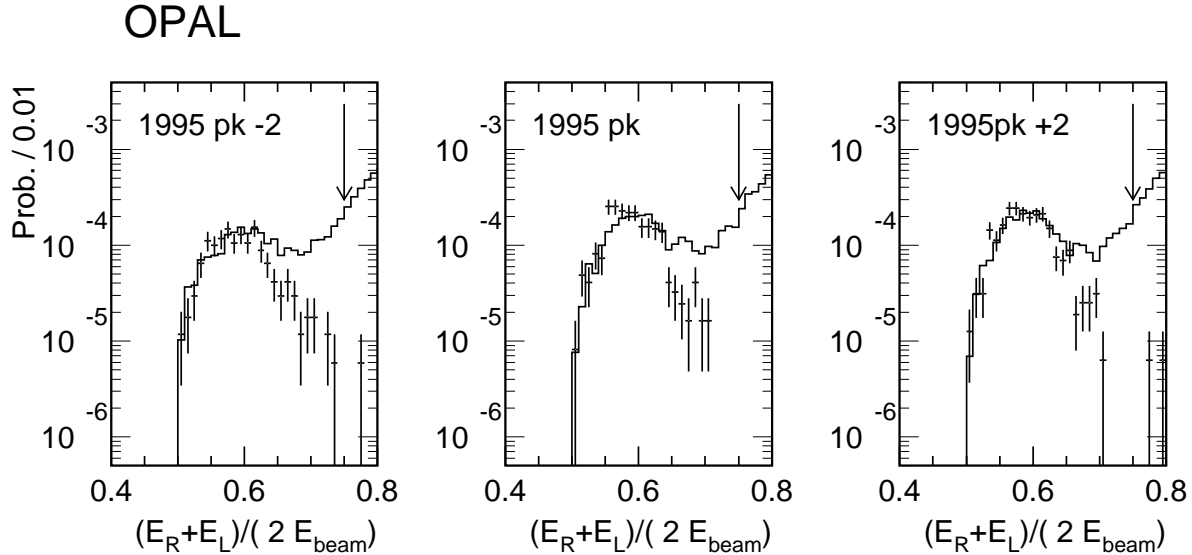}} 
\end{center}
\caption[Low energy tail (1995)]
{
The low energy tail of the 
total energy spectrum normalized to the number 
of accepted luminosity events for the three 1995 energy points.  
The data points show the accidental background as derived from the delayed
coincidence accidental triggers and rescaled to account for the background
distribution among bunchlets.
The arrows indicate the position of the selection cut made on the average
energy.
\label{fig:back_en95}
}
\end{figure}

Off-momentum electrons can also strike the detector in coincidence with
a true Bhabha event.
Such {\em overlap background}
can cause Bhabha events which would have otherwise failed the
selection cuts to pass them. 
It can also cause Bhabha events which would have otherwise passed the
selection cuts to fail them.  These effects have been evaluated 
by incorporating the measured background into the detector
simulation as described in section~\ref{sec:detsim}, and produce
an increase in the acceptance of less than $10^{-4}$.
This effect has also been evaluated by convoluting
the measured background with data events.
In this case,
slightly larger values for these effects were found.  
We assign a systematic error
of  $1.0 \times 10^{-4}$, divided equally into components correlated
and uncorrelated among the different data analysis samples,
which covers the difference 
between the two methods and allows for the uncertainty
associated with the implementation of
two-cluster separation in the detector simulation.
This uncertainty dominates the total uncertainty assigned to background
effects, as shown in table~\ref{tab:syssum_all}.

A particularly striking form of overlap ``background'' occurs
when two Bhabha scatterings are recorded in the same
bunch crossing.  At typical LEP luminosities of 
$3 \times 10^{30}$ sec$^{-1}$cm$^{-2}$, a few $\times 10^{-6}$
of all \SW\ Bhabha events can be expected to be overlaid with
a second Bhabha, consistent with observations.  We count
such double-Bhabhas as one, and make no correction for such 
overlaps.  
%
\section{Trigger efficiency}
\label{sec:trig}
%
%
%
%
%
%
%
%
%
%
%
%
%

\SW\ luminosity events are triggered by two signals based on hardware sums
of the energy deposited in the calorimeters: \SWSEG, which requires
deposition of energy in back-to-back segments in the two calorimeters, and
\SWHILO, which requires a large total energy deposition in one of the
calorimeters and a smaller total energy deposition in the other. 

Since the nominal hardware thresholds lie well below the minimun energies
required to accept fiducial Bhabha events, demonstrating that the trigger
system provides adequate efficiency primarily requires studying the
mechanisms which can generate low energy tails in the trigger response. 
One such mechanism involves electronic effects which cause small, but
coherent, shifts in the pedestals of the large number of individual
channels which need to be summed. 
Such shifts occur when the steady rhythm of track/hold signals is
disturbed, either by the readout of a triggered event, or by a rejected
pretrigger. 
These shifts after a rejected pretrigger can be as large as 10~GeV for
the total energy sum used in
\SWHILO, but less than 2~GeV for the segment sums used in \SWSEG. 

Another mechanism involves the energy absorbed in material upstream of the
calorimeters. 
The corrections applied to the measured cluster energies on the basis of
the longitudinal shower profile allow such absorption to be largely
compensated in the luminosity selection (see section~\ref{sec:en}), 
but the hardware
trigger signals are sensitive only to the energy actually deposited in the
detector. 

The digitized values of the raw trigger signals stored with each read out
event allow the relations between the fully corrected cluster energies and
the energies seen by the trigger to be studied in detail. 
We use the large sample of full-energy Bhabhas to measure the extreme tails
of the distribution of differences between the corrected cluster energy and
the trigger energy all the way down to trigger threshold. 
We then transport the difference spectrum away from the peak to estimate
the number of radiative Bhabhas which might have been lost due to
unfavorable trigger energy fluctuations. 
These studies carefully preserve the temporal and spatial pattern of
correlations in the trigger signal fluctuations induced by the two chief
known mechanisms mentioned above. 

Since the event sample used in these studies requires a \SW\ Bhabha to have
successfully triggered, a truely catastrophic failure mode of the trigger
system might remain masked. 
To uncover such possibilities, we also studied a sample of over 600,000 
events
independently triggered by the OPAL forward-detector calorimeter 
with shared energy in \SW\@. 
No catastrophic \SW\ trigger failures were observed, giving a lower limit of
$0.37\times10^{-5}$ on the \SW\ trigger system efficiency at 90\% CL. 

\Fg{fig:trig_cont} shows how the calculated trigger efficiency 
contours are well
adapted to the distribution of energies in fiducial \SwitA\ events. 
\Tb{tab:trig_inef} 
gives the calculated trigger inefficiency integrated over the
Bhabha event distribution for each year of data. 

\begin{figure}[htb!]
\begin{center}
\mbox{\epsfxsize17cm\epsffile{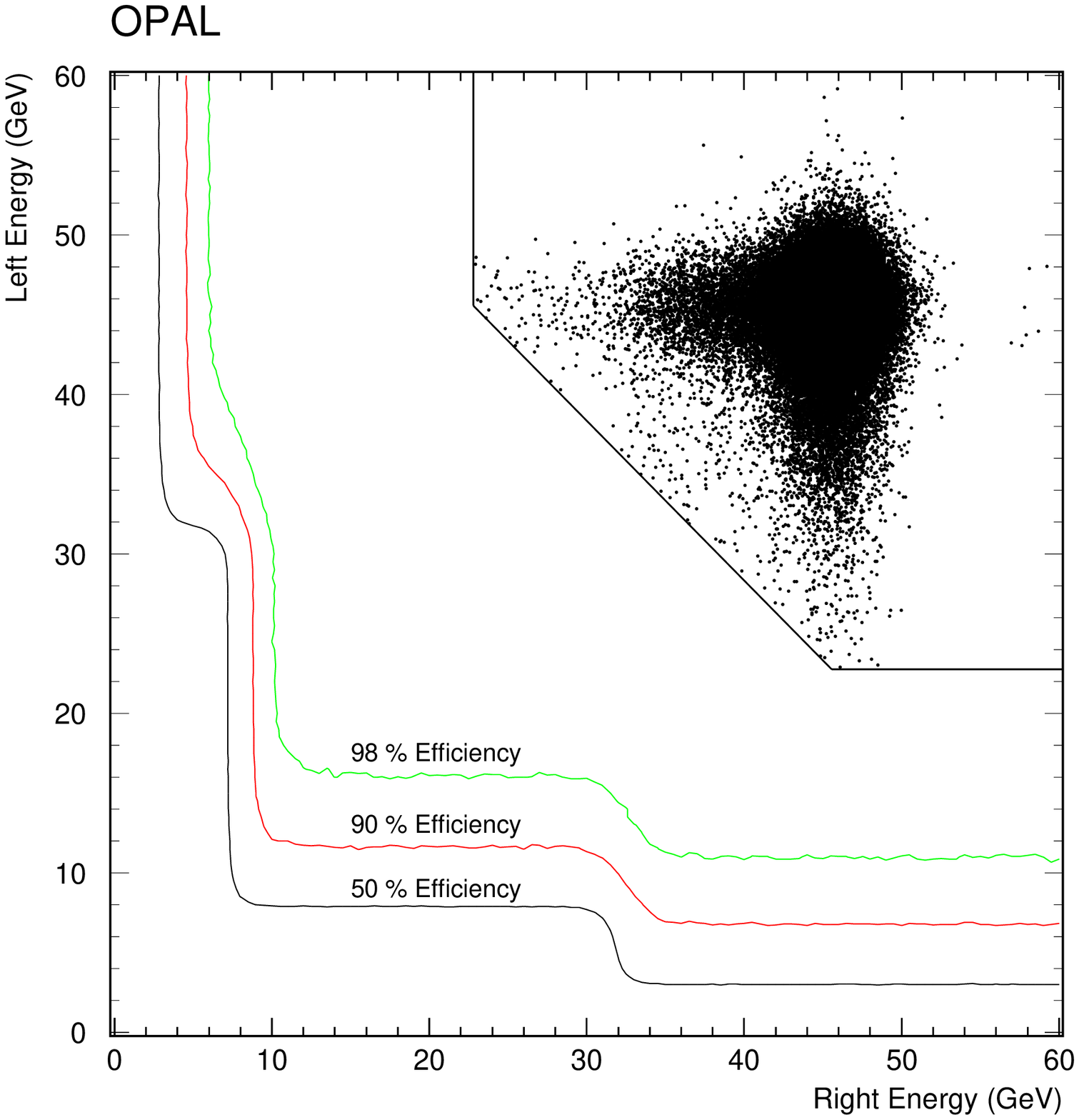}}
\end{center}
\caption[Efficiency countors]
{Left versus right summed energy for $10^{5}$ \SwitA\ selected 
events in 1994 with calculated 50, 90, and 98 \% 
efficiency contours for the combined \SWSEG, \SWHILO\ trigger signals
for the threshold settings. 
The straight lines indicate the energy cuts imposed
in selecting the \SwitA\ sample.
}
\label{fig:trig_cont}
\end{figure}

\begin{table}[htbp!]
\begin{center}
\begin{tabular}{|c||c|c|c|}
\hline
\multicolumn{4}{|c|}{ 1993 estimated \SW\ trigger inefficiencies}\\
\hline
Beam energy  & \SWSEG\   & \SWHILO\   &  Total\\
(GeV)     &$\times10^{-5}$&$\times10^{-5}$&$\times10^{-5}$\\
\hline
\hline
43.57& 3.5 $\pm$ 1.3 $\pm$ 0.9 & 692 $\pm$ 7 $\pm$ 14 & 0.14 $\pm$ 0.38 $\pm$ 0.18 \\
44.57& 2.7 $\pm$ 1.2 $\pm$ 0.7 & 538 $\pm$ 7 $\pm$ 12 & 0.12 $\pm$ 0.38 $\pm$ 0.17 \\
45.57& 2.1 $\pm$ 1.2 $\pm$ 0.6 & 380 $\pm$ 6 $\pm$ 9  & 0.09 $\pm$ 0.37 $\pm$ 0.16 \\
46.57& 1.6 $\pm$ 1.2 $\pm$ 0.5 & 240 $\pm$ 4 $\pm$ 7  & 0.07 $\pm$ 0.37 $\pm$ 0.14 \\
47.57& 1.2 $\pm$ 1.2 $\pm$ 0.4 & 137 $\pm$ 3 $\pm$ 4  & 0.06 $\pm$ 0.37 $\pm$ 0.14 \\
\hline
\multicolumn{4}{c}{ }\\
\hline
\multicolumn{4}{|c|}{ 1994 estimated \SW\ trigger inefficiencies}\\
\hline
Beam energy  & \SWSEG\   & \SWHILO\   &  Total\\
(GeV)     &$\times10^{-5}$&$\times10^{-5}$&$\times10^{-5}$\\
\hline
\hline
45.57& 0.02 $\pm$ 1.2 $\pm$ 0.01 & 21 $\pm$ 1.1 $\pm$ 0.8 & 0.001 $\pm$ 0.37 $\pm$ 0.001\\
\hline
\multicolumn{4}{c}{ }\\
\hline
\multicolumn{4}{|c|}{ 1995 estimated \SW\ trigger inefficiencies}\\
\hline
Beam energy  & \SWSEG\   & \SWHILO\   &  Total\\
(GeV)     &$\times10^{-5}$&$\times10^{-5}$&$\times10^{-5}$\\
\hline
\hline
43.57& 0.10 $\pm$ 1.18 $\pm$ 0.01 & 36 $\pm$ 2.0 $\pm$ 1.3 & 0.014 $\pm$ 0.37 $\pm$ 0.005 \\
44.57& 0.07 $\pm$ 1.18 $\pm$ 0.01 & 19 $\pm$ 1.4 $\pm$ 0.8 & 0.010 $\pm$ 0.37 $\pm$ 0.004 \\
45.57& 0.05 $\pm$ 1.18 $\pm$ 0.01 & 10 $\pm$ 1.1 $\pm$ 0.6 & 0.007 $\pm$ 0.37 $\pm$ 0.003 \\
46.57& 0.04 $\pm$ 1.18 $\pm$ 0.01 &  5 $\pm$ 0.9 $\pm$ 0.4 & 0.005 $\pm$ 0.37 $\pm$ 0.002 \\
47.57& 0.03 $\pm$ 1.18 $\pm$ 0.01 &  3 $\pm$ 0.8 $\pm$ 0.3 & 0.004 $\pm$ 0.37 $\pm$ 0.002 \\
\hline
\end{tabular}
\end{center}
\caption[Trigger efficiencies]
{Table of estimated trigger inefficiencies for \SwitA\ events
for an energy range twice as broad as the scan, in 
1993 and 1995, and at the peak energy in 1994.
The first error gives the uncertainty due to the limited
statistics of the OPAL forward detector sample
used to check for catastrophic \SW\ trigger system failures.
The second error covers the uncertainty in convoluting the tails of
the trigger response distribution over the energy spectrum of 
fiducial Bhabhas.} 
\label{tab:trig_inef}
\end{table}

\begin{figure}[htb!]
\begin{center}
\mbox{\epsfxsize17cm\epsffile{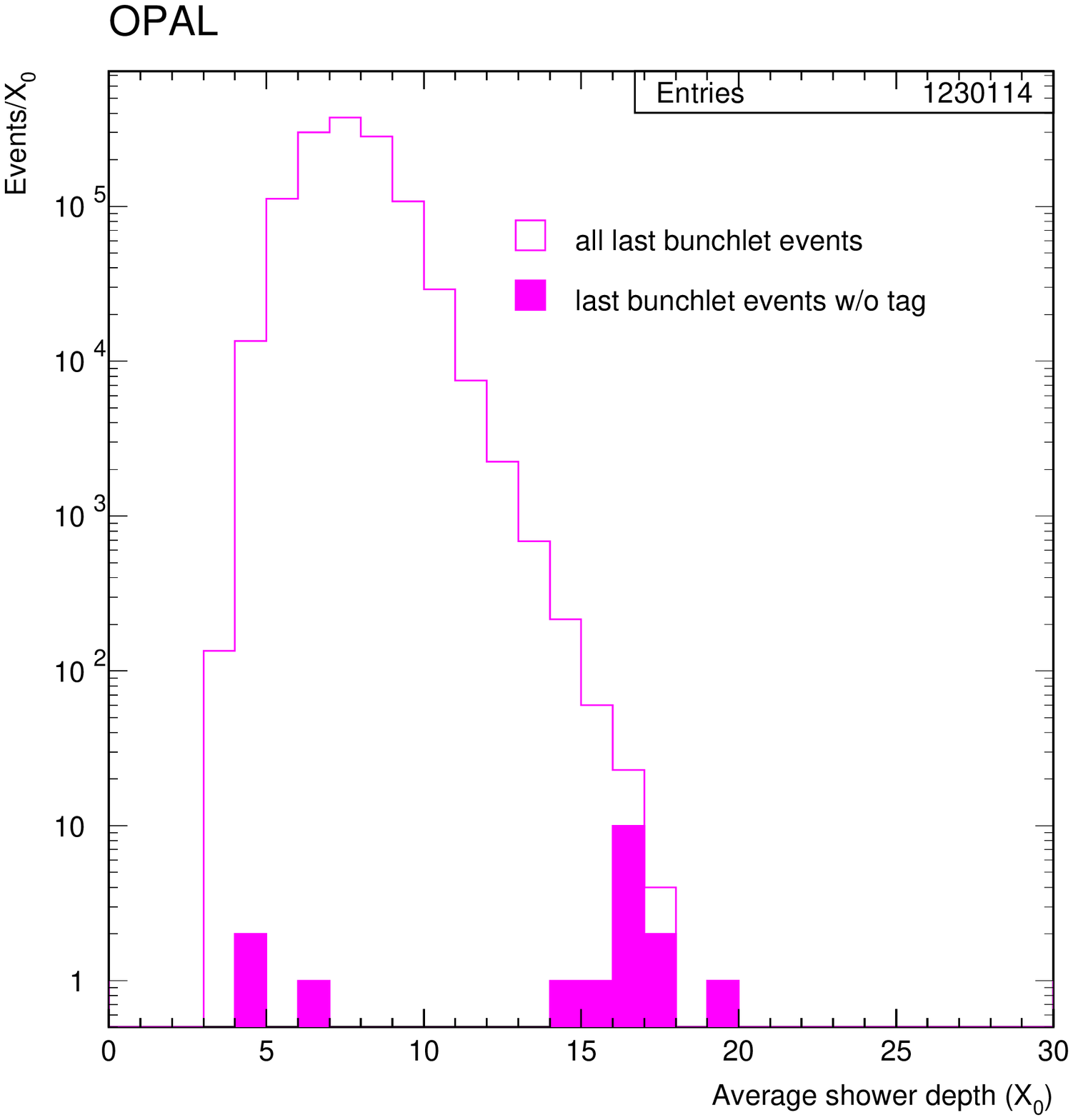}}
\end{center}
\caption[Wagon tagger failures]
{   The efficiency of the \SW\ wagon tagger as measured on the last bunchlet, 
where \hold\ is always issued, independent of any tag.  For the sample of 
accepted Bhabha events held on the last bunchlet, the average shower depth 
parameter is plotted for all showers (open) and showers lacking a tag 
(shaded).  Showers with anomalous development profiles, either deeper or 
shallower than the wagon tagger layers, are seen to be the dominant cause of 
tagger inefficiency.
}
\label{fig:trig_wt}
\end{figure}

In 1995 the proper functioning of the wagon tagger was also necessary for
valid luminosity measurements. 
The fact that the \SW\ \hold\ was issued on the last bunchlet of each
bunchtrain, even if the tagger did not fire, allowed us to measure the
tagger efficiency on last-bunchlet fiducial Bhabhas in a straight-forward
manner. 
In the entire sample of 1,230,114 such events, 18 failures were observed, as
shown in figure~\ref{fig:trig_wt}. 
All the tagger failures are due to showers with extremely late or early
development profiles which
deposited essentially no energy in the wagon tagger layers. 
Considering all bunchlets, we estimate the fraction of fiducial
Bhabhas lost due to wagon tagger inefficiency is $1 \times 10^{-5}$,
with an uncertainty of $2 \times 10^{-6}$.

\section{LEP beam parameters}
\label{sec:beam}
%
%
%
%
In this section we consider the effects of the LEP beam geometry on
the luminosity measurement.  
Although the Bhabha
selection procedure is chosen to minimize these effects, the luminometer
acceptance is affected in second order due to the fact that the LEP beams
depart from the ideal geometry of beams free of angular divergence
passing on-axis through the center of the bores of the two calorimeters.

We utilize the considerable power of the SiW luminometer itself, as well
as measurements using the OPAL tracking detectors and constraints from
operational parameters of the accelerator to determine the offset and tilt
of the beams with respect to the calorimeter bores, as well as the angular
divergence of the beams and the physical size of the intersection region.
Measurements of the above beam parameters, many of them determined
fill-by-fill, enter the Monte-Carlo simulation
of the detector described in section~\ref{sec:detsim}, and result in acceptance
corrections of typically $3\times10^{-4}$, which we know with a precision of
about $1\times10^{-4}$.
The two most significant acceptance corrections come from the transverse beam
offsets and the angular beam divergence.

The principle systematic errors in these corrections are due to discrepancy
between the beam divergence measured from the acollinearity of SiW Bhabhas and
that expected from accelerator parameters, as well as a conservative allowance
for rapid variations in the tilt of the beam axis, which can only be roughly
constrained by our measurements.

The offset and tilts of the LEP beams, as well as the beam 
divergence, all modulate or smear the
$1/\theta^3$ Bhabha spectrum leading to a net
increase in acceptance.
It is convenient to sort the effects of the beam parameters
into the two categories of (i) beam offsets
and tilts and (ii) beam size and divergence,
as they affect the acceptance in slightly
different ways.  In practice the separation can
not be made absolutely and some effects of
rapid variations in beam offsets and tilts 
are included in the beam size and divergence category.
We first consider the effects of transverse beam offsets
(section~\ref{sec:beam_trans}) and
then longitudinal offsets (section~\ref{sec:beam_long}).
Finally, we consider the effects of beam size and
divergence (section~\ref{sec:beam_sz}).

\subsection{Transverse beam offsets and tilts}
\label{sec:beam_trans}

Beam offsets
transverse to the beam axis produce a first order modulation
in the \SwitR\ and \SwitL\ acceptances 
as a function of the azimuthal angle.  After
integrating over all azimuth, a second-order
correction remains.  
The importance of this correction can be seen by considering the 
effect of a beam offset or tilt on the acceptance of
a single side of the calorimeter due to
the {\em definition} cuts.  
An analytical calculation
assuming a $1/\theta^3$ differential distribution
gives an azimuthal modulation of the luminosity of
\begin{equation}
\frac{ \mathrm{d} \Lrl} { \mathrm{d} \phi} \propto
 1 + 
\frac{8}{3} \frac{\ec}{r_1} \frac{ 1 - \left(\frac{r_1}{r_2}\right)^3 }
                                 { 1 - \left(\frac{r_1}{r_2}\right)^2 } 
\cos{(\phi-\phi_\ec)} +
{\left( \frac{\ec}{r_1} \right)}^2 \frac{ 1 - \left(\frac{r_1}{r_2}\right)^4 }
                        { 1 - \left(\frac{r_1}{r_2}\right)^2 } 
(6 \cos^2{(\phi-\phi_\ec)} - 1) 
\label{eq:ecc_az}
\end{equation}
where $r_1$ and $r_2$ are the inner and outer radii of the acceptance
and $\ec$ is the magnitude of the 
eccentricity which corresponds to the offset
of the outgoing beam
at the reference plane of the calorimeter.
The $x$ and $y$ components of the eccentricity are given by
\[
\begin{array}{ccl}
\ec_x &=& \ec \cos\phi_\ec \\
\ec_y &=& \ec \sin\phi_\ec \ . \\ 
\end{array} 
\]
The amplitude of a typical first-order azimuthal Bhabha intensity
modulation is $\sim5\%$, but
only the much smaller second-order term contributes to the average
over azimuthal angle.
We are therefore in the favorable situation of being able to use a sizeable
first-order effect to determine a small second-order correction.
The increase in  acceptance of the
\SwitR\ or \SwitL\ selection due to the
eccentricity can be obtained from equation~\ref{eq:ecc_az} and
is given by 
\begin{equation}
    \frac{\Delta A_{\mathrm{RL}}}{A_{\mathrm{RL}}} 
    \approx 
     +\left(\frac{\ec}{1.5\ {\mathrm  {mm}}}\right)^2 \times 10^{-3},
\label{eq:ecc}
\end{equation}
where $\ec$ is the magnitude of the 
eccentricity of the beam on the relevant side.
Typical values of the eccentricity of the LEP beam 
in the OPAL interaction region are 1 to 2 mm, making this a
significant effect.  
However, the effect of
the eccentricities on the \SwitR\ and \SwitL\ selections 
is reduced by more than
50\% when the {\em isolation} cuts are considered.
The reduction is due 
to the loss of radiative events whose detector centered 
value of
acollinearity or acoplanarity is increased by the
eccentricities.

One of the concepts motivating the \SwitA\ luminosity selection is
the reduction of acceptance dependence on the position of the beam 
spot.
If the true Cartesian half-difference between the shower impact points in
the two calorimeters is used in determining the acceptance for
back-to-back Bhabhas, then the acceptance is effectively 
defined in the
beam-centered frame, and is therefore independent of any beam offset. 
In determining \SwitA, however, we actually approximate the Cartesian 
coordinate half-difference by forming the average radius in polar 
coordinates, calculating
\begin{equation}
          R_A = \frac{1}{2} (R_R + R_L)
\end{equation}
which differs from the exact Cartesian expression
\begin{equation}
    R_A = \sqrt{ \left(\frac{x_R - x_L}{2}\right)^2 + 
                 \left(\frac{y_R - y_L}{2}\right)^2 }
\end{equation}
by a small second-order term in the beam offset.
An analytic calculation of the resulting azimuthal modulation of the \SwitA\
luminosity, similar to the calculation made for \SwitR\ and \SwitL\ in
equation~\ref{eq:ecc_az}, considering only the effect of the
definition cuts, gives 
\begin{equation}
\frac{ \mathrm{d} \La } { \mathrm{d} \phi} \propto
 1 + 
{\left( \frac{\ec}{r_1} \right)}^2 \frac{ 1 - \left(\frac{r_1}{r_2}\right)^4 }
                        { 1 - \left(\frac{r_1}{r_2}\right)^2 } 
\sin^2{(\phi-\phi_\ec)} 
\label{eq:ecc_swita}
\end{equation}
where $\ec$ is here the average eccentricity of the beams in the two 
calorimeters, equal to the transverse offset of the beamspot.
Considering the definition cuts alone, there is no first-order 
azimuthal modulation of the acceptance, and the second-order increase 
in total acceptance is $\frac{5}{16}$ as large as the corresponding 
increase for \SwitR\ and \SwitL.
The effect of the isolation cuts, however, is similar, and
therefore \SwitA\ suffers a net {\em decrease} in acceptance about half
the size of the net increase suffered by \SwitR\ and \SwitL.

Beam tilts with respect to the calorimeters result in different
right and left eccentricities.
Thus, it is necessary to determine the  $x$ and $y$ 
components of the eccentricities associated with the
\SwitR\ selection $(\ec_{\mathrm xR},\ec_{\mathrm yR})$ 
and the \SwitL\ selection 
$(\ec_{\mathrm xL},\ec_{\mathrm yL})$ separately.
The acceptance
of the \SwitR\ and \SwitL\ definition cuts 
will be independent of the eccentricity on 
the opposite side,
while the
acceptance of the isolation cuts depends on the values
of the eccentricities on both sides of the detector. 
The \SwitA\ acceptance shares a similar dependence on the
isolation cuts due to beam tilts, and also picks up a
second-order increase in acceptance through the definition
cut.

In order to calculate the effect of the
mean beam offset and tilt, these quantities are first determined
from the resulting azimuthal modulation of the Bhabha rate for each OPAL
run.\footnote{An OPAL run typically corresponds to one
LEP fill.  In the data considered here, the average number
of accepted Bhabha events in an OPAL run was approximately
25,000.}
The average of the eccentricities gives the beam offset, i.e,
\[
\begin{array}{cc}
x_{\mathrm {off}} = \frac{1}{2} (\ec_{\mathrm xR} + \ec_{\mathrm xL}) &
y_{\mathrm {off}} = \frac{1}{2} (\ec_{\mathrm yR} + \ec_{\mathrm yL}) \\
\end{array}
\]
The beam tilt is defined as the difference of the left and 
right eccentricities, i.e,
%
%
\[
\begin{array}{cc}
x_{\mathrm {tilt}} =  \frac{1}{2} (\ec_{\mathrm {xR}} - \ec_{\mathrm{ xL}}) &
y_{\mathrm {tilt}} =  \frac{1}{2} (\ec_{\mathrm {yR}} - \ec_{\mathrm{ yL}}) \\
\end{array}
\]

As with any second-order correction a lack of precision
in determing the offsets will result in a net bias in the 
calculated corrections.
Due to the limited statistical precision of the offsets
determined from the eccentricities ( $ \sim 250\,\mu$m/run), 
a second method is used to monitor the
beam offset at the interaction point.
This method uses the radial and azimuthal coordinates of Bhabha events
with beam energy electrons in both calorimeters.
These events will have little or no initial state radiation and
the line joining each such pair of back-to-back showers therefore passes
through the beam spot.
For each OPAL run we determine the average beam spot position by fitting
for the point which minimizes the
distances of closest approach over the ensemble of such lines.

\begin{figure}[htb!]
  \mbox{\epsfxsize18cm\epsffile{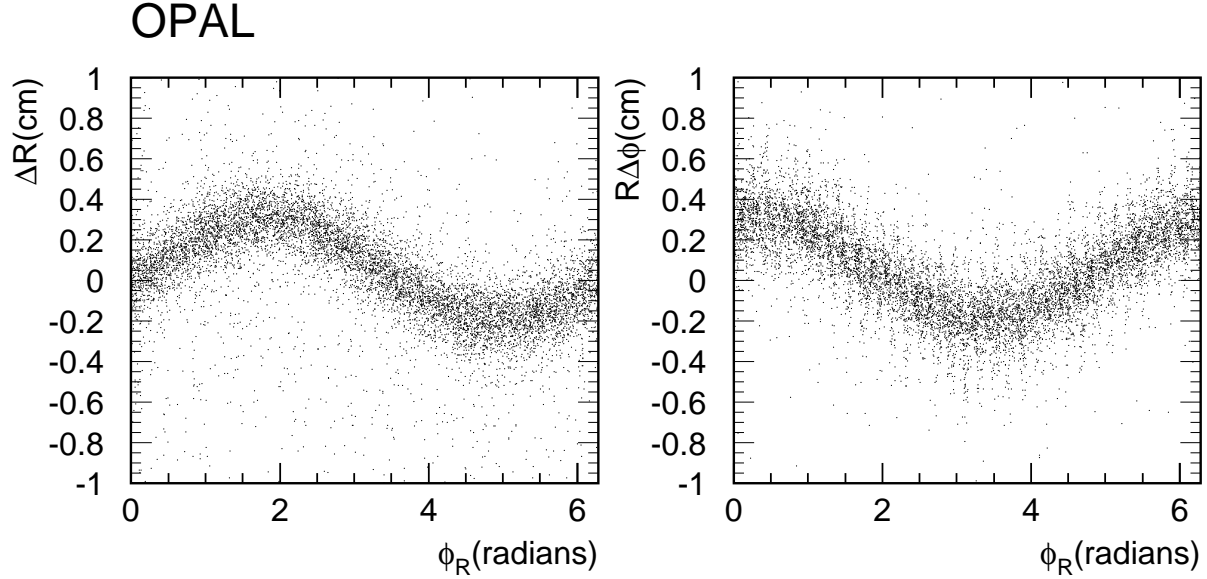}}
    \caption[Beam spot determination from shower coordinates]
{  The beam spot position can be determined with great precision from the
measured coordinates of back-to-back Bhabha clusters.
The plot on the left shows the acollinearity of non-radiative Bhabhas
in the detector frame as a function of $\phir$.
The plot on the right shows $R$ times acoplanarity for the same sample. 
    \label{fig:beam_spot}
}
\end{figure}

The resolution of this method is illustrated by the
apparent acollinearity or acoplanarity observed in such non-radiative
events in the {\em detector frame}.
A single event measures two of the three spatial coordinates
of the interaction region with a resolution limited only by the inherent
position resolution of the detectors, the angular divergence of the beams,
and, of course, the size of the luminous region itself.
A plot of $\Delta R = (\rr - \rl)$ versus $\phir$ for an entire run, as shown in
figure~\ref{fig:beam_spot}, allows the beam spot to be determined in three
dimensions, and with greatly enhanced statistical precision.
The $\cos$ and $\sin$ amplitudes of the azimuthal modulation give the
$x-$ and $y-$components, while the constant offset gives the
the $z-$component.
Similarly, a plot of $R \Delta\phi$ versus $\phir$, also shown in
figure~\ref{fig:beam_spot}, reveals the $x-$ and $y-$components of
the mean beam spot, with a $90^{\circ}$ phase shift,
while the constant offset of about 0.6~mm yields no additional
beam spot position information, but agrees with the offset expected from
the curvature of the Bhabha tracks in the OPAL magnetic field.
The RMS width of the scatter about the mean sinusoids in these plots
is about 0.9~mm, dominated by two essentially equal components, one from
the inherent resolution of the detector and the other from the angular
divergence of the beams.
The apparent structure in the $R \Delta\phi$ versus $\phir$ plot is due to the
strong variation of the $\phi$ resolution across each pad.
The statistical precision of this method is high; the beam spots
are determined with an accuracy of better than 10~$\mu$m 
for all but a few runs.

In order to calculate the acceptance corrections due to
beam offsets and tilts, the
measured run-by-run distributions of offsets and tilts for a given
period are used as input to the detector simulation, as 
described in section~\ref{sec:detsim}.
Because the offsets are determined from the fitting procedure described
above, they do not suffer any bias from the isolation cuts or
any significant error from finite statistical errors.

The beam spot determined through shower coordiates
agrees well with the spot determined through Bhabha intensity modulation,
taken as the average of the right and left eccentricites.
The coordinate method, however, yields no information on the tilt
of the beam, which we continue to extract from the difference of the right
and left eccentricites.
The statistical precision of $\sim250\mu$m on the measured tilt in a typical
run is adequate to limit the non-linear bias in the acceptance correction to
a manageable $3\times10^{-5}$, but
it is not possible to usefully probe the tilts on time scales shorter than an
individual run of approximately 10 hours.
Rapid variations of the tilts could produce 
a correction to the acceptance
which is not monitored through the azimuthal intensity modulation.\footnote{
Note that random variations in the angles of the electron and
positron beams will appear as additional contributions to
the beam divergence which is evaluated at the end of this
section.  In order for the tilts to have an effect on
the acceptance which is not included in the divergence
correction, the trajectories of the incoming
positron and electron beams must 
change in a correlated manner.}
Because the typical
beam offsets of 1--2~mm are appreciably larger than
than the typical tilts of $\sim300\,\mu$m for the
data analyzed here, 
the main effect of these variations would be
to increase either the \SwitR\ or \SwitL\ selection at the expense
of the other, resulting in no change to the measured luminosity.  
However, a small additional correction of order
\begin{equation}
    \frac{\Delta A_{\mathrm{RL}}}{A_{\mathrm{RL}}} 
    \approx 
     +\left(\frac{\Delta \ec_{\mathrm{rms} }}
                 {1.5\ {\mathrm{mm} }}\right)^2 \times 10^{-3}
\label{eq:ecc_rms}
\end{equation}
will remain due to the quadratic dependence
of the acceptance on the eccentricity (see equation~\ref{eq:ecc}). 
Studies of the data from the LEP beam orbit monitors show
no evidence for rapid motion of the beams, and such
movements can be safely assumed to be
smaller than the average tilts given by equation~\ref{eq:ecc_az}.
The acceptance correction for a typical 
tilt of $\ec = 300 \mu$m
would be $0.4 \times 10^{-4}$.  
It is possible that the isolation cuts
introduce a correction of opposite sign.
To evaluate this effect,
the change in the $\Lrl$  average was measured 
in the Monte Carlo with and without simulated tilts.  
For the 1994b sample, which has a typical tilt and offset, 
turning on the tilt gave an acceptance change 
of  $+0.4 \pm 0.3 \times 10^{-4}$,
in agreement with the estimate from equation~\ref{eq:ecc_rms}.
We take 100\% of this effect as the possible systematic
error from variations of the tilt within a run.

\subsection{Longitudinal offsets}
\label{sec:beam_long}

 Since the luminosity is calculated from the average of \SwitR\ and 
\SwitL, 
offsets in the position of the interaction region along the 
beam direction almost cancel in the average.
Possible second order effects are taken into account by
including the measured longitudinal beam offset in the
detector simulation. 
In order to check the expected effects of the longitudinal beam
position on the acceptance of the \SwitR\ and \SwitL\
selections, we plot  
\[
  \Dlr 
 = \frac{ (N_{\SwitL} - N_{\SwitR}) }
 { \frac{1}{2} ( N_{\SwitL} + N_{\SwitR}) }
\]
for the nine data samples and compare it with 
analytical and Monte Carlo expectations.
In \fg{fig:beam_long} it can be seen that $\Dlr$
depends linearly on the $z$-offset with the expected slope.  
The 
scatter of the data with respect to a straight line 
is due to beam tilts, which also effect $\Dlr$. 
These tilts have been included in the detector simulation and there
is reasonable  agreement between the data and Monte Carlo prediction
for $\Dlr$, with
a $\chi^2$/d.o.f of 15/9 (C.L. = 0.09). However,
perfect agreement between the detector simulation and the data is not
expected, because the detector simulation does not include correlations
between the longitudinal beam position and changes in the tilts.  
These correlations will affect the $\Dlr$ ratio, but will
not change the measured luminosity.

\subsection{Beam size and divergence}
\label{sec:beam_sz}

Random movements of the colliding beams on a short time scale,
as well as the finite beam sizes and divergences 
at the OPAL interaction region, cannot be monitored using
the measured eccentricities.
The largest of these effects is the beam divergence 
which affects the acceptance in the same
way as the radial resolution, i.e.,
\begin{equation}
    \frac{\Delta A}{A} 
    \approx 
+\left( \frac{\sigma_{\theta}}{450 \ {\mathrm {\mu rad}} } \right)^2 
         \times 10^{-3} 
\label{eq:beam_div}
\end{equation}
giving a correction to the luminosity of approximately 
$1 \times 10^{-4}$ for the typical LEP beam divergence
of $150 \mu$rad.  
The effects of finite beam size 
have a slightly different effect on the luminosity,
and give rise to a slightly smaller correction.
The most visible effect of the beam size and divergence
is to broaden the acollinearity distribution.  
In the
following, we compare the beam divergences extracted
from the acollinearity distribution with
those predicted from the LEP machine optics 
and measured beam size.  
The difference between these estimates is then used to 
evaluate the systematic error on the divergence.

By using measurements of the longitudinal and transverse
size of the luminous region from
the OPAL central detector and micro-vertex
detector, it is possible to derive a number of
constraints on the beam size and divergence.
Tracks from the process ${\Zzero \rightarrow \mu^+ \mu^-}$
were used to determine the horizontal size of the luminous region, 
$\sigma_x$.
The beam divergence in the horizontal plane is related to
the horizontal size of the luminous region by
$$ \sigma_x^\prime = \sqrt{2} \frac{\sigma_x}{\beta_x^*} $$
where $\beta_x^*$  is the $\beta$ function at the interaction region
and the factor $\sqrt{2}$ accounts for the difference between the
beam size and the size of the luminous region.  
The horizontal beam divergence determined from this method
is given in \tb{tab:beam_hor}.
(Note that for period 1994c the microvertex
detector was not installed and an adequately precise value of the
horizontal beam size is not available.)

\begin{table}[htbp!]
\begin{center}
\begin{tabular}{|l||c|c|}
\hline
 Sample & $\sigma_x^\prime$ (a)    & $\sigma_x^\prime$ (b) \\
        & (from acollinearity) & (from beam spot) \\
\hline\hline
93 $-2$ & $ 110 \pm 15\,\mu$rad         & $ 90 \pm 10\,\mu$rad \\  
93 pk   & $ 110 \pm 15\,\mu$rad         & $ 90 \pm 10\,\mu$rad \\ 
93 $+2$ & $ 100 \pm 15\,\mu$rad         & $ 90 \pm 10\,\mu$rad \\
94 a    & $ 120 \pm 15\,\mu$rad         & $ 85 \pm 10\,\mu$rad \\
94 b    & $ 110 \pm 15\,\mu$rad         & $ 85 \pm 10\,\mu$rad \\
95 $-2$ & $ 145 \pm 15\,\mu$rad         & $ 80 \pm 10\,\mu$rad \\
95 pk   & $ 135 \pm 15\,\mu$rad         & $ 80 \pm 10\,\mu$rad \\
95 $+2$ & $ 155 \pm 15\,\mu$rad         & $ 80 \pm 10\,\mu$rad \\
\hline
\end{tabular}
\end{center}
\caption[Horizontal beam divergence]{
Values of the horizontal beam divergence,  $\sigma_x^\prime$, 
as determined from (a) the acollinearity distribution and
(b) the horizontal beam size as measured by the
OPAL central detector.  
The discrepancy between the two values is used as the
basis for the systematic error due to the uncertainty
in the beam divergence.  If this discrepancy
is due to random changes in the directions
of the incoming beams during runs, then using
beam divergence as determined from the acollinearity will
give the correct acceptance.
\label{tab:beam_hor}
}
\end{table}

The horizontal 
beam divergence may also be estimated by
comparing the acollinarity distribution of Bhabha events in the luminometer,
corrected for the beam offset,
with that of simulated events.
The simulated events 
incorporated the radial 
detector resolution, as measured in the test beam,  and
the measured longitudinal size of the luminous region,
as measured by muon pairs.  
To reduce the tails 
in the acollinearity distribution from radiative events,
the events were required to satisfy ${\mathrm{ E > 0.9 \cdot E_{beam}}}$
on both sides of the detector. 
Comparing the width 
of the acollinearity distribution of Bhabha events
in the horizontal plane with simulated
events we obtain the values given in \tb{tab:beam_hor}.
The values of the beam divergence 
estimated from the Bhabha acollinearity 
distribution are systematically larger than those
inferred from the size of the luminous region
and the expected value for $\beta$.
This effect is especially pronounced for the 1995  
data.
These differences may 
be due to an incorrect estimate of the value of
$\beta_x^*$. However, a more likely explanation is that
they are due to random changes in the angles of the incoming
beams over time.  
These changes would be expected to be larger for
the 1995 bunch train running.
In the following, we use the value of divergence 
estimated
from the acollinearity distribution and 
take 100\% of the difference, in quadrature, 
between the two methods as a systematic error. 
This gives a $\sim 100$~$\mu$rad
uncertainty for 1993 and 1994 and 
110 to 125~$\mu$rad for 1995.
These results are summarized in \tb{tab:beam_hor}.

In the vertical plane, the beam size is much smaller,
(${\mathrm{\sim 5~\mu m}}$) and is not easily measured with
tracks from the central detector and microvertex detector.
For the effective 
beam size in the vertical direction we take 25 $\mu$m which
includes typical motions of the beam in the vertical plane
over the course of a run.

The beam divergence in the vertical plane can not be directly
determined from measurements of the central detector alone. 
For the nominal LEP optics the vertical and horizontal divergences
are equal, but in practice the vertical divergence
is expected to be somewhat larger than the horizontal divergence.  
Values of $\sigma_y^\prime$ between
125~$\mu$rad  and 185~$\mu$rad are obtained
for the nine data samples by
using Bhahba
events in the vertical plane and comparing their 
acollinearity with that of simulated events.
These values, which are used in calculating the acceptance for
each period,  are given in \tb{tab:beam_off}.

The effects of the finite beam size and beam divergence have
been incorporated into the simulation used to extract the
final acceptance.  
To estimate the systematic error due to uncertainties
in these beam parameters equation~\ref{eq:beam_div} 
is used.
The largest single effect is from the beam divergence.
The uncertainty in the
divergences is parameterized by a 
term of $\pm 100 \mu {\mathrm {rad}}$
which adds in quadrature to the divergence. 
The resulting systematic error on the luminosity is 
$0.4 \times 10^{-4}$.  
Since this uncertainty could be
fully correlated, or fully uncorrelated between the samples,
we conservatively assume both a correlated and uncorrelated error
of $0.4 \times 10^{-4}$.  The uncorrelated error allows for different
systematic effects at the three energy points of each scan.  We
also allow for a systematic difference between the 1993-1994 and 
1995 running which reflects the increase in the difference
between the measured and calculated divergence for the 1995 run.
The systematic
error from uncertainties on the values of the transverse beam size and
radial resolution is negligible.
These effects are summarized in \tb{tab:beam_sys}.

The effect of the finite longitudinal beam 
size on the luminosity measurement differs slightly from that
of the transverse size. Assuming a $1/\theta^3$ distribution,
it can be parameterized as
\begin{equation}
    \frac{\Delta A}{A} 
    \approx 
   +\left( \frac{ \sigma_z}{{\mathrm {8.0~cm}}} \right)^2 \times 10^{-3}.
\end{equation}
The
longitudinal beam size, $\sigma_z$ , 
averaged over the 1993, 1994 and 1995 running,
is known to better than  0.02~cm, giving negligible systematic 
error.\footnote{Large, $\sim 30\%$, 
variation in the longitudinal beam size within
a run can occur due to the use of wigglers to limit the beam-beam interactions.
However, the average behavior in each of the nine samples was nearly
the same.}
\begin{table}[htbp!]
\begin{center}
\begin{tabular}{|l||c|c|c|c|c|}
\hline
Sample & $\sigma_x$ & $\sigma_y$ & $\sigma_z$ & 
$\sigma_x^\prime$  & $\sigma_y^\prime$ \\ 
\hline\hline
&&&&&\\
93 $-2$&   156 $\mu$m     &  25 $\mu$m    &  0.70 cm  & 
 110   $\mu$rad  &   140 $\mu$rad  \\
93 pk  &   156 $\mu$m     &  25 $\mu$m    &  0.70 cm  & 
 110   $\mu$rad  &   140 $\mu$rad  \\
93 $+2$&    140 $\mu$m     &  25 $\mu$m    &  0.70 cm  & 
  100   $\mu$rad  &  125 $\mu$rad  \\
94 a &   120 $\mu$m     &  25 $\mu$m    &  0.70 cm  &
  120  $\mu$rad   &   155 $\mu$rad  \\
94 b &   120 $\mu$m     &  25 $\mu$m    &  0.70 cm  &
  110  $\mu$rad   &   150 $\mu$rad  \\
94 c &   150 $\mu$m     &  25 $\mu$m    &  0.70 cm  &
  135  $\mu$rad   &   185 $\mu$rad  \\
95 $-2$& 140 $\mu$m & 25 $\mu$m         & 0.70 cm &
  145 $\mu$rad  & 170 $\mu$rad \\
95 pk  & 140 $\mu$m & 25 $\mu$m         & 0.70 cm &
  135 $\mu$rad  & 165 $\mu$rad \\
95 $+2$& 140 $\mu$m & 25 $\mu$m         & 0.70 cm &
  155 $\mu$rad  & 180 $\mu$rad \\
\hline
\end{tabular}
\end{center}
\caption[Beam parameter values ]{
Values of the beam parameters used in the
detector simulation.  The dimensions of the 
luminous region, $\sigma_x$, $\sigma_y$ and $\sigma_z$,
include the estimated rms motion of the beam
offsets over the course of a typical run.
\label{tab:beam_off}
}
\end{table}

\begin{table}[htbp!]
\begin{center}
\begin{tabular}{|l||c|c|c|}
\hline
Effect  & Systematic error &  
          Systematic error & 
          Additional error    \\
        & (correlated)    &                 
          (uncorrelated)  &
          (1995 only)     \\
        & $\times 10^{-4}$ & 
          $\times 10^{-4}$ & 
          $\times 10^{-4}$ \\
\hline\hline
&&&\\
beam divergence         &    0.40  & 0.40    &  0.50    \\
transverse beam offsets &    0.01  & 0.01    &  $<0.01$ \\
average beam tilts      &    0.01  & 0.01    &  $<0.01$ \\
variation in beam tilts &    0.40  & 0.40    &  $<0.01$ \\
longitudinal beam size  &    0.02  & 0.02    &  $<0.01$ \\
transverse   beam size  &  $<0.01$ & $<0.01$ &  0.20    \\
&&&\\
\hline
\hline
Total                   &   0.57   & 0.57   &   0.50 \\
\hline                 
\end{tabular}

\end{center}
\caption[Beam parameter systematic errors]
{Systematic errors from incomplete knowledge of the
LEP beam parameters.
\label{tab:beam_sys}
} 
\end{table}

\begin{figure}[htb!]
  \mbox{\epsfxsize16cm\epsffile{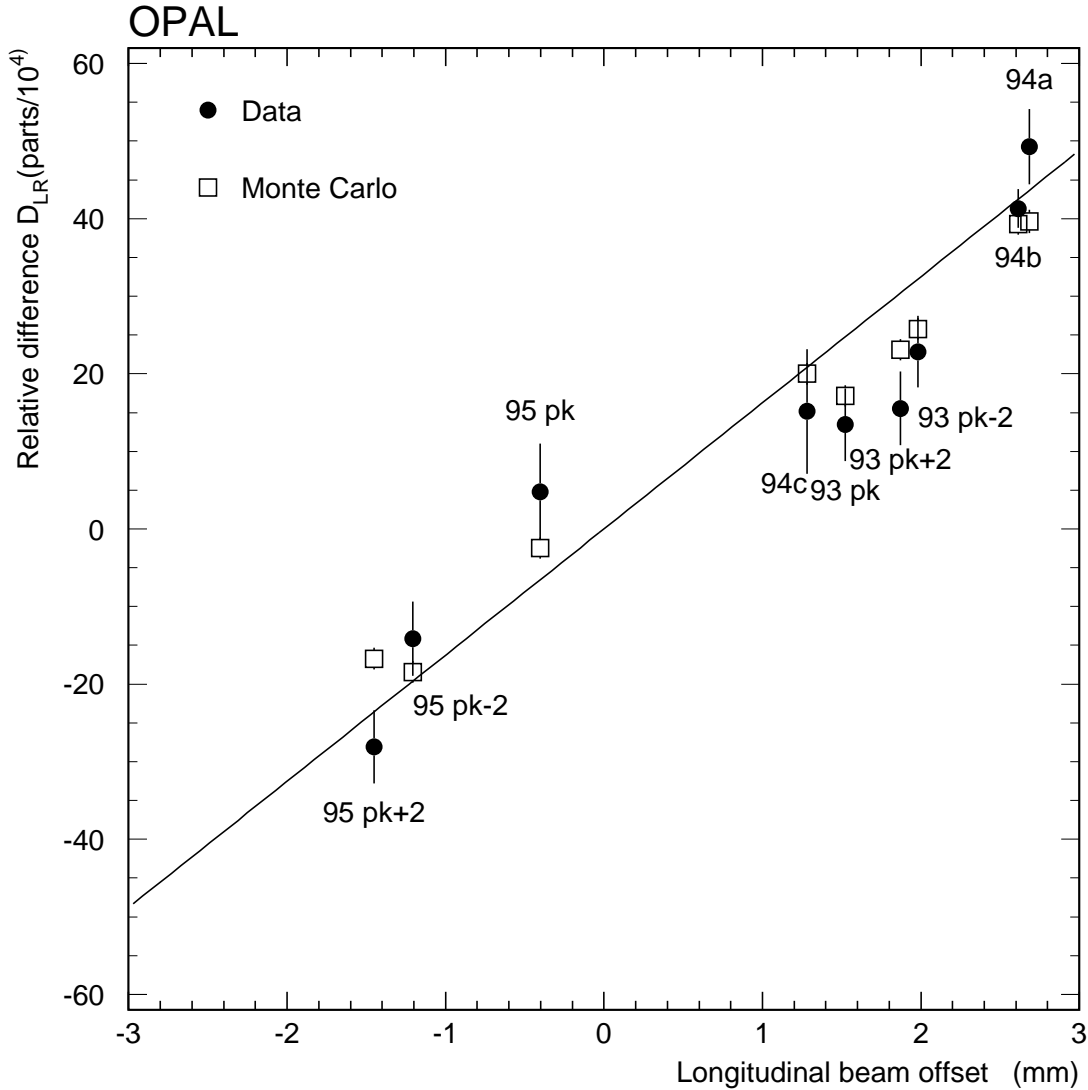}}
    \caption[Relative right-left luminosity difference ($\Dlr$)]
{  $\Dlr$,
the difference between \SwitL\ and \SwitR\
selections, divided by their average, 
as a function
of the mean measured longitudinal offset of the beam spot.
The points show the value for each of the nine data taking periods
for 
both the data, and the detector simulation.  
The detector simulation  
uses the reconstructed longitudinal
beam position and the transverse offsets and tilts as inputs.
The solid
line shows the expected ratio if there were no effects from
energy resolution and from differences in the eccentricities of beam
in the left and right calorimeter.
    \label{fig:beam_long}
}
\end{figure}

\subsection{Beam energy error and spread}
\label{sec:beam_energy}

The errors in the luminometer acceptance due to the uncertainty of the LEP
energy at each energy point~\cite{bib-hildreth} are not included in the
OPAL luminosity analysis. These highly correlated uncertainties of order
$10^{-4}$ are, however, taken into account when fitting for the $\Zzero$
resonance parameters. 
In propagating the LEP energy uncertainties to the measured cross sections,
the derivative of the luminosity with respect to energy plays a small role
compared to the derivatives of the cross sections themselves, and
ultimately has a negligible effect on the final $\Zzero$ parameters and
errors. 

The rms spread of the LEP central energy of about 55~MeV generates a
correction of $\sim10^{-3}$ to the OPAL peak cross sections
due to the curvature of the $\Zzero$ lineshape.
The off--peak corrections are substantially less.
The corresponding correction to the luminosity is everywhere
$2\times10^{-6}$, and is neglected. 

\section{Detector simulation}
\label{sec:detsim}
%
%
%
%
%
The OPAL \SW\ detector 
simulation 
does not rely on a detailed physical simulation of 
electromagnetic showers in the detector.  
Instead, we utilize the experimentally determined
response of the detector in energy, radius and
azimuth to generate the measured quantities corresponding
to the clusters of section~\ref{sec:ev_recon}
directly from the four-vectors of the physics Monte Carlo.
This approach gives a much more reliable description of the tails
of the detector response functions, which are primarily due to extreme
fluctuations in shower development, than we could obtain using any existing
code which attempts to simulate the basic interactions of photons and
electrons in matter.
We also choose to simulate an {\em ideal} detector, which differs from the
actual detector in having nominal dimensions, a perfectly efficient trigger
and wagon tagger, as well as being free of accidental coincidence (although
not overlap) backgrounds.
We treat the differences between this ideal detector and the actual detector
by applying what we term {\em external} corrections
(see section~\ref{sec:acceptance}).

The simulation consists of five steps.  First,
the four-vectors are used to generate particle trajectories
according to the beam parameters.  
Second,
the energy of each of the four-vectors is smeared by
an energy resolution algorithm, producing ``showers''.
Third,
background (as measured from random beam crossings) is
added.  
Fourth, showers are grouped together, to produce the
equivalent of the clusters (see section~\ref{sec:ev_recon}). 
Fifth, radial and azimuthal
resolutions are simulated.  In the following, these
steps are described in somewhat more detail.


The beam spot offset for each event is
chosen randomly from
the distribution of measured $x$, $y$ and $z$ run-averaged
beam spot positions.  
This run-averaged beam spot is
further smeared by the measured beam sizes averaged
over a given data sample (see section~\ref{sec:beam}).
Each event is then randomly boosted and rotated so as
to simulate the effects of finite beam divergence
while keeping the center-of-mass energy of each event
fixed at its generated value.

Each of the rotated and boosted four-vectors is then extrapolated
to the detector.
The energy of each four-vector is smeared
using the parameterization described in section~\ref{sec:en}.


Accidental overlap background is taken from a library based on random
beam crossings.  Separate libraries are available for 1993,
1994 and 1995 runs to reflect different background conditions.  
Systematic errors from these effects were evaluated in section~\ref{sec:back}.

At the next step, showers are merged into at most three
clusters per side which closely 
approximate those found in the data.
The clusters are sorted according to energy and any clusters
outside of the detector 
(that is, clusters with radii outside 6.20 and 14.20~cm) 
are discarded.  
(The relatively tight isolation cuts of the
selections ensure that the calculated acceptance is
insensitive to the treatment of showers close to
the edge of the detector.)
The most energetic remaining shower is chosen as the
shower core.  
If any other remaining shower is within the
measured two-cluster separation of the
main shower core in azimuth and radius, 
it is merged with the shower core.
The probability to merge two showers has been parameterized 
by using the data: the actual pad signals from showers due to photons 
in events with final state radiation 
were overlaid at variable positions on the signals from the electron
showers, and the resulting
cluster merging probabilities were measured (see section~\ref{sec:ev_recon})
as a function of shower separation.
The merging probability reaches 50\% at
a separation of 1.0 cm, or 4 pad widths.  When two
showers are merged, their energies are summed,
and the radial position of the most energetic
cluster is used unless the radial position of the
two four-vectors fall within a window,
in which case the position is taken as the energy weighted mean of both
clusters.
The scale of this window is set by the three pads
used in the radial position algorithm and
varies as
a function of energy from
two pad widths (0.5~cm) at beam energy to four pad widths 
(1.0~cm) at low energies.
Studies were made varying the size of the window used
for forming clusters, resulting in
changes in acceptance below the assigned systematic error
of $1 \times 10^{-4}$. 

In the case of closely spaced, but resolvable clusters,
no attempt has been made to simulate possible
distortions in the energy
assignment made by the cluster finding 
algorithm.  
This introduces a negligible 
correction to
the luminosity acceptance because the
energy cut is based on the sum of energy
in the fiducial region of the detector. It may,
however, affect the comparison of final state
radiative events in the data and Monte Carlo.


The radial and azimuthal positions of the remaining clusters are further
smeared using resolution measurements from the test beam and from the data.  
The radial and azimuthal response functions are taken as Gaussians whose
mean values reproduce the input four-vectors and whose resolutions
are assumed to scale as $1/\sqrt{E}$, where $E$ is the cluster energy. 
The detector simulation allows the evolution of resolution
with energy to be changed to assess the systematic
errors due to these aspects of the detector response.
The test beam
measurement is used to set the radial resolution of $130\mu$m at pad boundaries
in the clean acceptance near the inner edge of
the detector for beam energy electrons and positrons.  
The effects of the increased material at greater radii are estimated
from changes in the acollinearity and acoplanarity distributions as a function
of radius which are not accounted for by the known beam sizes and divergences.
These studies indicate that the apparent resolution at the outer edge
and in the central portion of
the detector, behind the bulk of the pre--showering material, is
degraded by a factor of 2 to 2.5.

The effect of this degradation in resolution is especially important for
the acceptance of the acollinearity cut,
since in many of the events having an acollinearity
near 2.5~cm, one of the electrons will have an energy significantly
less than $E_{beam}$, and lie in a
region of the detector where the effects
of the pre--showering material are most severe.
Because the apparent degradation in radial resolution could be partially
caused by an underestimate in the length of the beamspot,
we assign the full calculated effect of this degradation
as an uncertainty of $2\times10^{-5}$
in the acceptance due to uncertainties in the radial resolution.
Similarly, the radial coordinate of such lower--energy electrons in the
region behind the bulk of the pre--showering material may develop a net
bias which we estimate could be as large as $15\,\mu$m.
According to equation~\ref{eq:biasacoll} we therefore assign a systematic
error of $3.6\times10^{-5}$ due to possible bias in the measured
acollinearity.

The azimuthal resolution is taken as $1/\sqrt{2}$ times the width of
the beam--centered acoplanarity distribution of full--energy Bhabhas,
and varies from 10--12~mrad as a function of radius.
An acceptance uncertainty of $4\times10^{-6}$, equal to $10\%$ of the
resolution flow calculated according to
equation~\ref{eq:resdphi}, accounts for the non--Gaussian
features of the actual detector resolution, which are not reproduced
by the detector simulation.

The final clusters include values of $E$, $R$, and $\phi$ and are
passed through the same analysis chain as the data.  
%
%
%

%
\section{Calculation of the acceptance}
\label{sec:acceptance}
%
%
%
%
%

We are now in position to calculate the critical parameters mentioned in
section~\ref{sec:sel}, the acceptances for the $\SwitX$
luminosity selections, $A_{\mathrm{RL}}$ and $A_{\mathrm{A}}$.
We proceed to make the calculation in four steps:
\begin{itemize}
\item We use the best available Monte-Carlo simulation of small--angle
Bhabha scattering (BHLUMI) to calculate, at the level of the generated
four--vectors, a {\em theoretical reference acceptance}
($A_{\mathrm{RL}}^{\mathrm{ref}}$)
which is very close to the actual experimental acceptance.
We make this calculation at a single reference energy of 91.100~GeV, and
generate a very large sample of events to reduce the statistical error
on this absolute accepted cross section to a level well below any of our
other errors.
\item We use another Monte-Carlo (BHAGEN), and an analytic calculation
(ALIBABA), to calculate the energy evolution of the acceptance through
the small relevant range of energies about the $\Zzero$.
At this point we also consider the contribution of the processes
$\ee \rightarrow \gamma \gamma$ and $\ee \rightarrow \ee X$.
After applying these corrections ($\epsilon_{\mathrm{theory}}$) we arrive at the
{\em ideal theoretical acceptance}.
\item We pass a more limited sample of four--vectors from the BHLUMI
Monte-Carlo through our parametrized detector response simulation described in
section~\ref{sec:detsim} to calculate the small binomial differences
($\epsilon_{\mathrm{sim}}$) between
the ideal theoretical acceptance and the {\em ideal experimental acceptance}. 
\item Finally, we apply the {\em external} corrections
($\epsilon_{\mathrm{ext}}$)
to convert the ideal experimental acceptance into the
{\em actual experimental acceptance}, accounting for the small departures
of the actual detector from its nominal geometry revealed by the metrology
and coordinate anchoring studies.
The external corrections also account for any trigger and wagon tagger
inefficiencies, as well as
accidental coincidences from beam--related backgrounds.
\end{itemize}
The final actual luminometer acceptance, $A_{\mathrm{RL}}$, is therefore
calculated as:
\begin{equation}
A_{\mathrm{RL}} = A_{\mathrm{RL}}^{\mathrm{ref}}
( 1 + \epsilon_{\mathrm{theory}} + 
\epsilon_{\mathrm{sim}} +\epsilon_{\mathrm{ext}} ) .
\end{equation}
\noindent
The major advantage of the above organization is that our clear definition
of an ideal theoretical acceptance at four--vector level allows our
result to be
easily adjusted to incorporate a potential improvement in the theoretical
understanding of small--angle Bhabha scattering.
Only a four--vector level calculation using an improved Bhabha Monte-Carlo
needs to be undertaken.
Our decision to calculate the energy dependence of the theoretical acceptance
separately,
rather than to use the full BHLUMI Monte-Carlo everywhere, is motivated
solely by the desire
to limit the relative acceptance errors, which would
otherwise be larger due to the difficulty in running BHLUMI with
sufficient statistics at a large number of energies.
Tests have demonstrated that the energy dependence of BHAGEN and ALIBABA
agrees with that of BHLUMI within the quoted errors, as demonstrated
by the direct BHLUMI~4.04 points plotted in \fg{fig:theo_zgamma}.

\subsection{The theoretical reference acceptance}
\label{sec:theory}

This section describes
the calculation of the theoretical reference cross section, which is 
based only on four-vector quantities produced by 
the BHLUMI Monte Carlo.  

At Born level, Bhabha scattering is described by equation~\ref{eq:born}
and results in the production of exclusively collinear, beam--energy,
scattered electrons.
Considering first--order radiative corrections in leading log allows photons
collinear with either the in--coming or out--going electrons to be radiated.
In sub--leading log, these photons can be acollinear.
The consideration of progressively higher--order terms in the perturbative
expansion allows correlations in multiple--photon emission to be taken
into account.
Exponentiation is a useful calculational technique which allows lower--order
terms in the perturbative expansion to be used in approximating the effect
of higher--order terms which have not been explicitly evaluated,
and allows an $O(\alpha)$ generator to simulate final states with multiple
photons.
Higher--order vertex corrections also make themselves felt
in modifying the cancelation of the infra--red divergences inherent in the
external photonic corrections.
The effect of vacuum polarization modifies the internal photonic propagators
and introduces a dependence on hadronic couplings which is most accurately
quantified through a dispersion relation to experimental measurements
of $\ee \rightarrow $ hadrons~\cite{bib-radcor}.
All these effects modify the Born--level acceptance of the luminometer
by only a few percent.

    The BHLUMI Monte~Carlo program~\cite{bib-BHLUMI}
is accepted as embodying the considerable theoretical work which
has been undertaken to calculate small--angle Bhabha scattering in
the LEP era.
    It is a multiphoton exponentiated 
    $O(\alpha)$ generator for small--angle Bhabha scattering only.
    It includes the majority of the leading logarithmic and part of 
    the next to leading logarithmic terms of $O(\alpha^{2})$ for the
    Bhabha scattering process.
    The current version, BHLUMI~4.04, has an 
    estimated precision~\cite{bib-montagna, bib-ward, bib-yr-LEP2}
    of $6.1 \times 10^{-4}$
    for use with the OPAL acceptance.

    The parameters used in generating the theoretical reference
    cross section are given in \tb{tab:theo_bhlumi}.  The range of
    the allowed t-channel transfers correspond to an inner acceptance for
    non-radiative events of 18.9~mrad 
    and an outer acceptance of 150~mrad.
    This range is sufficiently broad to cover nearly all
    radiative events which could trigger our detector.  
    The fraction of small t-channel transfer events
    not generated by the Monte Carlo, 
    but which would be accepted by the 
    $\SwitR$, $\SwitL$ or $\SwitA$ selections was determined to be 
    less than $2 \times 10^{-5}$ by varying the value of the minimum t-channel
    transfer.
With the parameter set used, our BHLUMI calculation includes the effect of
all $\gamma$ exchange processes within the scope of the program,
but specifically excludes the effects of $\Zg$ interference,
which we consider separately.

\begin{table}[htb]
\begin{center}
\begin{tabular}{|l|l|l|}
\hline
Parameter &   Value  & BHLUMI variable\\
\hline
\hline
Center-of-mass energy ($\sqrt{s}$)    & 91.100 GeV                & \\
Minimum t-channel transfer            &  0.741  GeV$^2$           & \\
Maximum t-channel transfer            & 46.6    GeV$^2$           & \\
Soft photon cutoff    ($\epsilon = \frac{2 E_{\gamma}}{\sqrt{s}})$  
                                      &  $ 1.0 \times 10^{-4}$     & \\
Vacuum Polarization                   & Method 3~\cite{bib-radcor} &
                                                    (KeyPia =3)    \\
$\Zg$ interference                     & Off & 
                                                      (KeyZet =0)\\
\hline
\end{tabular}
\end{center}
\caption[BHLUMI parameters]
{Parameters used with BHLUMI 4.04 Monte Carlo to obtain
the reference cross section.  Note that the $\Zg$ interference
correction is kept separate so that it is not necessary to
generate large statistics Monte Carlo samples at each beam
energy.
\label{tab:theo_bhlumi} }
\end{table}

The theoretical reference acceptance is defined in the Monte Carlo
calculation by taking the energy in each detector as the sum of
all four-vectors with polar angles between 27.227~mrad and
55.629~mrad.
The energy in each calorimeter is required to be greater than 
$0.50 \ebeam$, while the total energy is required to be greater than
$0.75 \roots$, where $\ebeam$ is the nominal LEP beam energy and
$\roots =2 \ebeam$.
The tight radial cuts of 7.7~cm and 12.7~cm, 
which are equivalent to polar angles of
31.288~mrad and 51.576~mrad, as well as the acoplanarity
cut of 200~mrad, and an acollinearity cut of 
10.162~mrad (which closely corresponds to 2.5~cm in radial difference), 
are imposed on the
coordinates of the single most energetic ``particle'' in each detector.


An important feature of this ideal configuration is that it
incorporates the distinction which is made in the experimental
analysis between the particles used to calculate energy (the
sum of the whole detector) and those used to calculate
position.
The definition of the most energetic four-vector
may depend on the parameters of the Monte Carlo and the 
order of the calculation.  For the results presented
here we defined a ``particle'' by clustering together all
four-vectors within $\sim 3$~mrad of each other.
In the BHLUMI 4.04 Monte Carlo this has almost no effect
on the acceptance (less than 0.2 $\times 10^{-4}$).
In reference~\cite{bib-yr-LEP2} the impact of the
definition of the ``most energetic particle'' on
comparisons between different Monte Carlo generators
was evaluated and no differences were observed for acceptances
similar to the one described here.

The accepted theoretical reference cross sections for these
$L_{\mathrm{RL}}$ and $L_{\mathrm{A}}$ selections were calculated to be 
\begin{eqnarray}
A_{\mathrm{RL}}^{\mathrm{ref}}& = & 78.898 \pm 0.005 \mbox{~nb} \nonumber \\
A_{\mathrm{A}}^{\mathrm{ref}} & = & 79.413 \pm 0.005 \mbox{~nb} \nonumber
\end{eqnarray}
\noindent
at $\roots = 91.100$~GeV, where the error is statistical
only. Note that these cross sections do not include the
$\Zg$ interference, $ \ee \rightarrow \gamma \gamma$,
or the detector acceptance corrections given in \tb{tab:cor_all}.

The acceptance obtained by simply integrating the Born cross section
of equation~\ref{eq:born} over the geometrical acceptance of
the detector is 81.0298~nb, 2.7\% larger than the BHLUMI~4.04 value.
Vacuum polarization reduces the Born cross section by 4.1\%,
while other radiative corrections make a net positive contribution of 1.4\%,
most of which is due to the $O(\alpha)$ terms.

\subsection{Theoretical acceptance corrections}
\label{sec:theorycorr}

In this section we describe adjustments made to the BHLUMI cross section
for $\Zg$ interference, the running of the vacuum polarization as a function
of t across the energies of the scan,
and for the processes $\ee \rightarrow \gamma \gamma$ and 
$\ee \rightarrow \ee X$.

  The energy dependence of the luminometer acceptance
  was calculated using the ALIBABA~\cite{bib-alibaba} and 
  BHAGEN~\cite{bib-bhagen} calculations.
  Excluding the fundamental $1/s$ dependence of the Bhabha cross section
  leaves the effects of $\Zg$ interference and the variation in vacuum
  polarization
  across the relevant region of the scan.
The correction factor, as shown in \fg{fig:theo_zgamma}, is calculated as the
ratio
\begin{equation}
 \frac{ \sigma_{\mathrm{bhagen}}(E)   }
      { \sigma_{\mathrm {bhagen}}^{\mathrm{no} \Zzero}(91.1\,\mathrm{GeV})  }
 \:\left(\frac{91.1\,\mathrm{GeV}}{E}\right)^2 \nonumber
\end{equation}
where $\sigma_{\mathrm{bhagen}}$ is the full BHAGEN accepted cross section
with values of the $\Zzero$ mass and width taken to be
91.187 and 2.489~GeV~\cite{bib-bhagen-par},
and $\sigma_{\mathrm {bhagen}}^{\mathrm{no} \Zzero}$ is the BHAGEN cross section
calculated with the effect of the $\Zzero$ turned off.
    The precision of the interference 
    correction calculated from the BHAGEN Monte~Carlo has been estimated 
    by comparison  with the published 
    results of ALIBABA~\cite{bib-alibaba} in the same acceptance.
    Both BHAGEN and ALIBABA include the complete $O(\alpha)$
    and $O(\alpha^{2})$ LL photonic corrections. 
    The quoted physical 
    precision 
    of the ALIBABA $\Zg$ 
    interference correction is $1.5 \times 10^{-4}$, essentially
    due to the lack of NLL $O(\alpha^{2})$ terms.
    The absolute difference between
    the interference corrections of the two programs 
    is below $6 \times 10^{-5}$
    over a wide range of center-of-mass energies. 
    Around 
    the peak the difference is
    consistent with zero within the statistical error, 
    which is $2\times10^{-5}$.
    Taking into account the $1.5\times10^{-4}$ uncertainty due to missing 
    NLL $ O(\alpha^{2})$ terms in ALIBABA, we assign to BHAGEN 
    a  precision of $1.5\times10^{-4}$ in the relative cross sections
    at the three points of the energy scan.  The error matrix associated
    with this uncertainty is given in \tb{tab:theo_int}.
 
Potential physics backgrounds from the processes 
$ \ee \rightarrow \gamma \gamma$
and $\ee \rightarrow \ee X$ have also been considered.
The correction for the former process, which our calorimetric measurement
does not distinguish from $ \ee \rightarrow \ee$, has been determined by
using a radiative $\ee \rightarrow \gamma \gamma (\gamma)$ Monte Carlo
generator~\cite{bib-berendsgg} at four--vector level.
The $\gamma\gamma$ cross section within our idealized acceptance is
found to be $(14.0\pm0.2)$~pb at 91.1~GeV, or $1.8\times10^{-4}$ times the
Bhabha cross section in the same acceptance, with negligible
uncertainty.
A simple analytic calculation at Born level agrees within 10\%

The contribution of soft pair production in the reaction
$\ee \rightarrow \ee X (X = ee,\mu\mu,...)$,
giving rise to final states in configurations which would be
accepted by the luminosity Bhabha selection,
has been calculated in our idealized acceptance according to
reference~\cite{bib-softpairs}.
Applying the calculated fractional contribution of
$(-4.4\pm1.4)\times10^{-4}$~\cite{bib-softpairsopal}
as a correction to the BHLUMI reference cross section (which does not include
such pairs) allows the BHLUMI~4.04 theoretical error of $6.1 \times 10^{-4}$
to be reduced to $5.4 \times 10^{-4}$~\cite{bib-softpairs}.

The corrections to the theoretical reference cross section, including
$\Zg$ interference and physics backgrounds($\epsilon_{\mathrm{theory}}$),
are illustrated in \tb{tab:theo_x}.
The particular energies chosen in \tb{tab:theo_x} correspond
to the OPAL multihadron cross section sample.
In order to determine the cross 
sections for leptonic and hadronic
decays of the $\Zzero$, the luminosity is counted separately for each 
channel.  
The center-of-mass energy for each channel is slightly
different because of variations in detector status requirements (e.g.
the muon chambers do not have to work to count electrons). 
These variations give the 
$\Zg$ and $1/s$ corrections a very small channel dependence. 

\begin{table}
\begin{center}
\begin{tabular}{|l||c|c|c|c|c|c|c|c|c|}
\hline           
& 93 $-2$ & 93 pk &93 $+2$& 94 a & 94 b  & 94 c &95 $-2$ & 95 & 95 $+2$\\
\hline
\hline

 93 $-2$& 1.0& 0.0&-1.0& 0.0& 0.0& 0.0& 1.0& 0.0&-1.0\\
 93 pk  & 0.0& 1.0& 0.0& 1.0& 1.0& 1.0& 0.0& 1.0& 0.0\\
 93 $+2$&-1.0& 0.0& 1.0& 0.0& 0.0& 0.0&-1.0& 0.0& 1.0\\
 94 a   & 0.0& 1.0& 0.0& 1.0& 1.0& 1.0& 0.0& 1.0& 0.0\\
 94 b   & 0.0& 1.0& 0.0& 1.0& 1.0& 1.0& 0.0& 1.0& 0.0\\
 94 c   & 0.0& 1.0& 0.0& 1.0& 1.0& 1.0& 0.0& 1.0& 0.0\\
 95 $-2$& 1.0& 0.0&-1.0& 0.0& 0.0& 0.0& 1.0& 0.0&-1.0\\
 95 pk  & 0.0& 1.0& 0.0& 1.0& 1.0& 1.0& 0.0& 1.0& 0.0\\
 95 $+2$&-1.0& 0.0& 1.0& 0.0& 0.0& 0.0&-1.0& 0.0& 1.0\\
\hline
\end{tabular}
\caption[Theory error matrix]{ The correlation matrix for
$\Zg$ interference and the change in vacuum polarization.
\label{tab:theo_int}
}
\end{center}
\end{table}

\begin{table}
\begin{center}
\begin{tabular}{|l|c|r|r|c|c|c|}
\hline
Sample                                   & 
Average $\ecm$                           & 
\multicolumn{1}{|c|}{$1/s$}                & 
\multicolumn{1}{|c|}{$\Zg$}                &
$\ee \rightarrow \gamma \gamma$          &
soft                                     &
cross section\\
 &    
 & 
\multicolumn{1}{|c|}{correction} &
\multicolumn{1}{|c|}{correction} &
background &
pairs       &
(ideal theoretical)      \\
      & GeV &  \multicolumn{1}{|c|}{$\times 10^{-4}$} &
\multicolumn{1}{|c|}{$\times 10^{-4}$} &  $\times 10^{-4}$ & $\times 10^{-4}$ &
  nb \\
\hline
 93  $-2$ &$  89.4505 $&$  372.2 $&$  18.5\pm 1.5 $&$ 1.8 $ & $ -4.4\pm1.4 $&
$  81.965\pm0.090 $ \\
 93 pk    &$  91.2063 $&$  -23.3 $&$   3.6\pm 1.5 $&$ 1.8 $ & $ -4.4\pm1.4 $&
$  78.722\pm0.087 $ \\
 93  $+2$ &$  93.0351 $&$ -411.7 $&$ -16.3\pm 1.5 $&$ 1.8 $ & $ -4.4\pm1.4 $&
$  75.507\pm0.083 $ \\
 94 a     &$  91.2358 $&$  -29.7 $&$   2.8\pm 1.5 $&$ 1.8 $ & $ -4.4\pm1.4 $&
$  78.665\pm0.087 $ \\
 94 b     &$  91.2165 $&$  -25.5 $&$   3.3\pm 1.5 $&$ 1.8 $ & $ -4.4\pm1.4 $&
$  78.702\pm0.087 $ \\
 94 c     &$  91.4287 $&$  -71.8 $&$  -2.3\pm 1.5 $&$ 1.8 $ & $ -4.4\pm1.4 $&
$  78.293\pm0.086 $ \\
 94 c'    &$  91.2195 $&$  -26.2 $&$   3.3\pm 1.5 $&$ 1.8 $ & $ -4.4\pm1.4 $&
$  78.697\pm0.087 $ \\
 95  $-2$ &$  89.4415 $&$  374.3 $&$  18.5\pm 1.5 $&$ 1.8 $ & $ -4.4\pm1.4 $&
$  81.981\pm0.090 $ \\
 95 pk    &$  91.2829 $&$  -40.0 $&$   1.5\pm 1.5 $&$ 1.8 $ & $ -4.4\pm1.4 $&
$  78.573\pm0.086 $ \\
 95  $+2$ &$  92.9715 $&$ -398.5 $&$ -16.5\pm 1.5 $&$ 1.8 $ & $ -4.4\pm1.4 $&
$  75.609\pm0.083 $ \\
\hline
\end{tabular}
\end{center}
\caption[Energy correction to luminosity]{
Summary of corrections to the theoretical reference cross section
of $78.898 \pm 0.005$~nb
calculated at $\ecm = 91.100$.  The cross section listed in
the last column
does not include any detector related effects and the error
is the theoretical error only.  
The actual luminometer
acceptance is obtained by applying the acceptance corrections
of \tb{tab:cor_all} to the ideal theoretical cross sections given in
the last column.
\label{tab:theo_x}
}
\end{table}

\begin{figure}[tbh]
\mbox{\epsfxsize16cm\epsffile{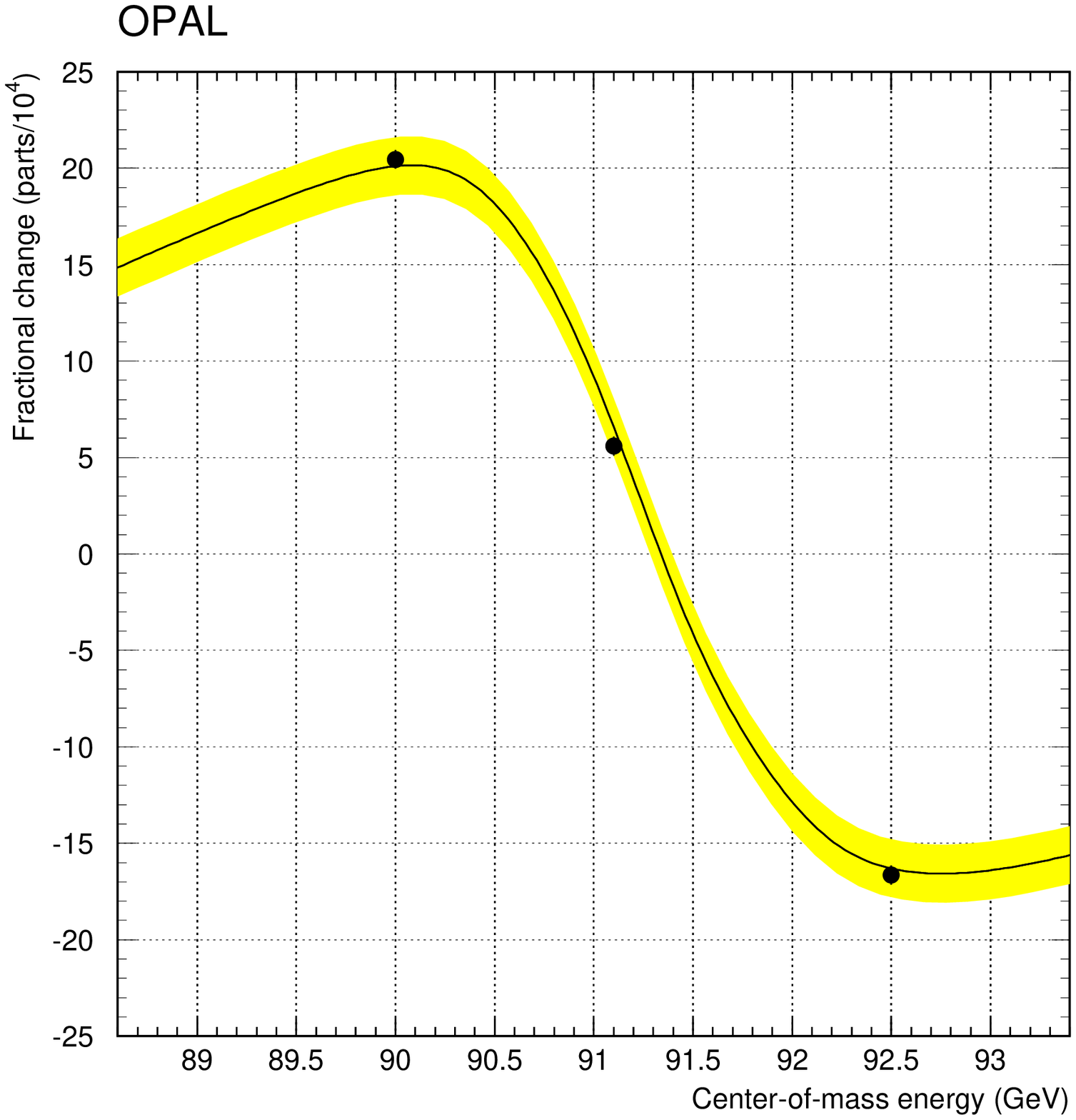}}
\caption[$\Zg$ interference]
{ Relative size of the energy dependence of the SiW acceptance as a function
of center-of-mass energy near the $\Zzero$ peak, excluding the fundamental
$1/s$ dependence of the cross section. 
The curve includes the effects of $\Zg$ interference and the variation
($6\times10^{-4}$) in vacuum polarization.
The band gives the error assigned to this correction.
The points with (statistical) error bars display the result of direct
calculations at three energy points using BHLUMI~4.04.
\label{fig:theo_zgamma}
}
\end{figure}

\subsection{Experimental acceptance corrections}
\label{sec:syssum}
%
%
%

As mentioned at the begining of this section, we divide the experimental
acceptance corrections into two categories: those evaluated using the
parameterized detector simulation ($\epsilon_{\mathrm{sim}}$), and those 
excluded from the detector simulation, and applied externally
($\epsilon_{\mathrm{ext}}$).

To evaluate $\epsilon_{\mathrm{sim}}$
a set of four--vectors from the BHLUMI~\cite{bib-BHLUMI} event
    generator has been used to convolute 
    the effects of the LEP 
    beam parameters and of the detector response, 
    determined directly from the data, 
    with the underlying distributions of Bhabha events.
This procedure relates the ideal detector acceptance, data set by data set,
to the ideal theoretical acceptance with only small binomial uncertainties.
The resulting correction factor ($\epsilon_{\mathrm{sim}}$)
for a typical data sample
is given in table~\ref{tab:detsim}, as well as its breakdown into
individual effects as estimated with the Monte Carlo.

    The largest correction, 
    $\sim (-4.1 \pm 1.8)\times 10^{-4}$, 
    is due to the energy response 
    of the detector  (section~\ref{sec:en}).
    The parameterization of the detector energy response has been studied in
    detail in the data, making extensive use of our ability to geometrically
    define the kinematics of Bhabha events.
    The systematic uncertainty quoted on the acceptance correction is based
    on varying
    the parameters describing the energy response over a range which safely
    covers that allowed by our studies.
    The corrections due to beam parameters are discussed in
    section~\ref{sec:beam}, those due to coordinate resolutions
    and clustering in section~\ref{sec:detsim}.
The corrections for resolution flow in the radial and azimuthal
coordinates agree with the analytic expressions discussed in
sections~\ref{sec:radial_intro} and~\ref{sec:azimuth}.
    The effect of finite energy
    resolution on the acceptance is nearly cancelled by an  
    increase in acceptance from the beam offsets  
    (section~\ref{sec:beam}), resulting in a very small net difference
    between the ideal theoretical and ideal experimental acceptances.
The $\Delta R - \Delta \Theta$ cut difference correction in table~\ref{tab:detsim}
results from a small technical
discrepancy between the acollinearity cut defined in terms of a fixed
difference in $\Theta$ ($\Delta\Theta = 10.162 $mrad) used for the
theoretical reference acceptance, and the acollinearity
cut defined in terms of a fixed difference in R ($\Delta R = 2.5$ cm)
used for the experimental acceptance.

\newlength{\gnat}
\settowidth{\gnat}{Typical acceptance correction}
 
\begin{table}[htbp!]
\begin{center}
\begin{tabular}{|l|p{0.33\gnat}p{0.33\gnat}p{0.33\gnat}|}
\hline
Effect  & \multicolumn{3}{|c|} {Typical acceptance correction} \\
        & \multicolumn{3}{|c|} {      ($\times 10^{-4}$)     } \\
\hline
\hline
Energy resolution                         &  $- 4.1$&$\pm 1.8 $&$\pm 0.33$\\
Beam parameters                           &  $+ 2.7$&$\pm 0.8 $&$\pm 0.56$\\
Radial resolution                         &  $- 0.4$&$\pm 0.2 $&$\pm 0.33$\\
Acollinearity bias                        &  $+ 0.0$&$\pm 0.36$&$\pm 0.33$\\
Azimuthal resolution                      &  $- 0.2$&$\pm 0.04$&$\pm 0.13$\\
Clustering                                &  $- 0.3$&$\pm 1.0 $&$\pm 0.08$\\
$\Delta R - \Delta \Theta$ cut difference &  $- 0.7$&$\pm 0.0 $&$\pm 0.03$\\
\hline
\hline
Total                                     &  $- 3.0$&$\pm 2.3 $&$\pm 0.81$\\
\hline
\end{tabular}
\end{center}
\caption[Effects evaluated with the Monte Carlo]{
Values of the acceptance corrections for data sample 94b due to effects
which are evaluated using the Monte Carlo ($\epsilon_{\mathrm{sim}}$).
The first error is systematic (see \tb{tab:syssum_all}), the second is the
statistical error of the Monte Carlo.
\label{tab:detsim}
}
\end{table}

\begin{table}[htbp!]
\begin{center}
\begin{tabular}{|l||*{9}{r|}}
\hline
 Effect           
& 
\multicolumn{1}{|c|}{ 93 $-2$}  &
\multicolumn{1}{|c|}{ 93 pk}  &
\multicolumn{1}{|c|}{ 93 $+2$}  & 
\multicolumn{1}{|c|}{ 94a}    & 
\multicolumn{1}{|c|}{ 94b}    &
\multicolumn{1}{|c|}{ 94c}    &
\multicolumn{1}{|c|}{ 95 $-2$}  &
\multicolumn{1}{|c|}{ 95   }  &
\multicolumn{1}{|c|}{ 95 $+2$} \\
\hline
\hline
%
Radial Metrology (sec ~\ref{sec:rmetro})&$ 1.55$&$ 1.55$&$ 1.55$&$ 1.55$&$ 1.55$&$ 1.55$&$ 1.55$&$ 1.55$&$ 1.55$\\
Radial Thermal (sec~\ref{sec:rmetrolep})&$-0.86$&$-0.85$&$-0.85$&$-1.13$&$-1.22$&$-1.31$&$-2.72$&$-2.69$&$-2.72$\\
Inner Anchor (sec~\ref{sec:anchor})     &$ 2.18$&$ 2.18$&$ 2.18$&$ 2.18$&$ 2.18$&$ 2.18$&$ 4.64$&$ 4.64$&$ 4.64$\\
Outer Anchor (sec~\ref{sec:anchor})     &$-1.78$&$-1.78$&$-1.78$&$-1.78$&$-1.78$&$-1.78$&$-1.19$&$-1.19$&$-1.19$\\
Z Metrology (sec~\ref{sec:zmetro})      &$ 0.50$&$ 0.59$&$ 0.50$&$ 1.16$&$ 0.72$&$ 0.08$&$-0.95$&$-0.95$&$-0.95$\\
Background  (sec~\ref{sec:back})        &$ 0.65$&$ 0.50$&$ 0.30$&$ 0.16$&$ 0.09$&$ 0.22$&$ 0.15$&$ 0.10$&$ 0.10$\\
Trigger (sec~\ref{sec:trig})            &$ 0.00$&$ 0.00$&$ 0.00$&$ 0.00$&$ 0.00$&$ 0.00$&$ 0.00$&$ 0.00$&$ 0.00$\\
Wagon Tagger (sec~\ref{sec:trig})       &$ 0.00$&$ 0.00$&$ 0.00$&$ 0.00$&$ 0.00$&$ 0.00$&$-0.10$&$-0.10$&$-0.10$\\
\hline
 Total External ($\epsilon_{\mathrm{ext}}$)&$ 2.24$&$ 2.19$&$ 1.90$&$ 2.13$&$ 1.54$&$ 0.94$&$ 1.38$&$ 1.36$&$ 1.33$\\
\hline
\hline
    Total Simulation ($\epsilon_{\mathrm{sim}}$)&$-2.13$&$-1.81$&$-1.96$&$ 0.36$&$-2.98$&$ 0.26$&$-2.33$&$-2.33$&$-2.18$\\
         Grand Total     &$ 0.11$&$ 0.38$&$-0.06$&$ 2.50$&$-1.44$&$ 1.20$&$-0.95$&$-0.97$&$-0.85$\\
\hline
\end{tabular}
\end{center}
\vspace{-0.5cm}
\caption[Correction to absolute luminosity]{
    This table summarizes the 
    detector acceptance corrections for the $\Lrl$ 
    luminosity 
    measurement for the nine data samples. 
    All corrections are in units of $10^{-4}$.
    The sum of $\epsilon_{\mathrm{ext}}$ and $\epsilon_{\mathrm{sim}}$,
    the grand total,
    has been applied to the ideal theoretical cross sections
    given in section~\ref{sec:theory}, \tb{tab:theo_x}
    to obtain the actual luminometer acceptance for each data sample.
\label{tab:cor_all}
}
\end{table}

The relatively
large variations in LEP beam parameters, as well as changes
in the energy resolution
due to the configuration of upstream material,
require that the acceptance be calculated separately for
each of the data samples, resulting in variations of
about $2\times10^{-4}$.
The total simulation correction ($\epsilon_{\mathrm{sim}}$)
for each data sample is given in table~\ref{tab:cor_all}.

    The external corrections ($\epsilon_{\mathrm{ext}}$) account for
differences between the {\em actual} detector and the {\em ideal}
detector described by the parametrized simulation.
    These corrections were discussed in 
    sections~\ref{sec:rmetro} and \ref{sec:zmetro} (metrology),
    \ref{sec:anchor} (radial anchoring),
    \ref{sec:back} (background), and 
    \ref{sec:trig} (trigger).
    Accurate metrology of the detector dimensions, and direct measurements
    of bias in the radial coordinate from test 
    beam data, enable us to control
    the uncertainty in the radial acceptance-defining cuts at better
    than the $2 \times 10^{-4}$ level.
    Careful monitoring of backgrounds and a well-designed trigger
    keep corrections and systematic errors from these 
    sources small.
In \tb{tab:cor_all} we list the values of these external corrections
($\epsilon_{\mathrm{ext}}$) for each of the nine data samples.

    In table~\ref{tab:syssum_all}
    we give the systematic uncertainties on each of 
the $\epsilon_{\mathrm{ext}}$ and $\epsilon_{\mathrm{sim}}$ correction
    factors. The grand total quadrature sum of these uncertainties gives
    the total experimental uncertainty in the absolute luminosity 
    measurement using the average of the \SwitR\ and \SwitL\ selections.
    Taking into account all sources of 
    experimental error,
    the quadrature sum of the correlated and uncorrelated systematic errors is 
    $3.3 \times 10^{-4}$, degrading slightly to
    $3.5 \times 10^{-4}$ in 1995.
    Due to the statistical manner in
    which the greater part of these systematic errors have been
    evaluated, as well as the fact that four independent sources
    of error make contributions greater than $1 \times 10^{-4}$,
    we would like to point out that these one-sigma errors can be
    extrapolated to greater levels of significance in a
    Gaussian fashion.  
    The final experimental luminosity error correlation matrix is given 
    in \tb{tab:lumi_matrix}.  
The classification of the detailed sources of error into correlated and
uncorrelated components given in \tb{tab:syssum_all} does not reveal the
complete pattern of correlations embodied in the full correlation matrix.
In the table, the detailed errors classified as correlated are fully
correlated between all
data samples, while those classified as uncorrelated are often correlated within
a given year, but uncorrelated between years.

\begin{table}[htbp]
\vspace{-2.0cm}
{\footnotesize
\begin{center}
\begin{tabular}{|l|c||*{9}{r|}}
\hline
 Uncertainty & section           
& 
\multicolumn{1}{|c|}{ 93 -2}  &
\multicolumn{1}{|c|}{ 93 pk}  &
\multicolumn{1}{|c|}{ 93 +2}  & 
\multicolumn{1}{|c|}{ 94a}    & 
\multicolumn{1}{|c|}{ 94b}    &
\multicolumn{1}{|c|}{ 94c}    &
\multicolumn{1}{|c|}{ 95 -2}  &
\multicolumn{1}{|c|}{ 95   }  &
\multicolumn{1}{|c|}{ 95 +2} \\
\hline
\hline
%
        Radial Metrology & \ref{sec:rmetro} &&&&&&&&&\\
 ~~~~~~uncorrelated&&~~ 0.00&~~ 0.00&~~ 0.00&~~ 0.00&~~ 0.00&~~ 0.00&~~ 0.00&~~ 0.00&~~ 0.00\\
 ~~~~~~~~correlated&&~~ 1.40&~~ 1.40&~~ 1.40&~~ 1.40&~~ 1.40&~~ 1.40&~~ 1.40&~~ 1.40&~~ 1.40\\
         Radial Thermal & \ref{sec:rmetrolep} &&&&&&&&&\\
 ~~~~~~uncorrelated&&~~ 0.06&~~ 0.00&~~ 0.06&~~ 0.09&~~ 0.11&~~ 0.11&~~ 0.25&~~ 0.25&~~ 0.25\\
 ~~~~~~~~correlated&&~~ 0.18&~~ 0.18&~~ 0.18&~~ 0.18&~~ 0.18&~~ 0.18&~~ 0.18&~~ 0.18&~~ 0.18\\
         Inner Anchor & \ref{sec:anchor}    &&&&&&&&&\\
 ~~~~~~uncorrelated&&~~ 0.23&~~ 0.23&~~ 0.23&~~ 0.23&~~ 0.23&~~ 0.23&~~ 0.58&~~ 0.58&~~ 0.58\\
 ~~~~~~~~correlated&&~~ 1.36&~~ 1.36&~~ 1.36&~~ 1.36&~~ 1.36&~~ 1.36&~~ 1.36&~~ 1.36&~~ 1.36\\
         Outer Anchor & \ref{sec:anchor}    &&&&&&&&&\\
 ~~~~~~uncorrelated&&~~ 0.13&~~ 0.13&~~ 0.13&~~ 0.13&~~ 0.13&~~ 0.13&~~ 0.28&~~ 0.28&~~ 0.28\\
 ~~~~~~~~correlated&&~~ 0.31&~~ 0.31&~~ 0.31&~~ 0.31&~~ 0.31&~~ 0.31&~~ 0.30&~~ 0.30&~~ 0.30\\
         Z Metrology  & \ref{sec:zmetro}   &&&&&&&&&\\
 ~~~~~~uncorrelated&&~~ 0.00&~~ 0.00&~~ 0.00&~~ 0.00&~~ 0.00&~~ 0.00&~~ 0.37&~~ 0.37&~~ 0.37\\
 ~~~~~~~~correlated&&~~ 0.41&~~ 0.41&~~ 0.41&~~ 0.41&~~ 0.41&~~ 0.41&~~ 0.41&~~ 0.41&~~ 0.41\\
         Background   & \ref{sec:back}   &&&&&&&&&\\
 ~~~~~~uncorrelated&&~~ 0.76&~~ 0.76&~~ 0.76&~~ 0.75&~~ 0.75&~~ 0.75&~~ 0.76&~~ 0.76&~~ 0.76\\
 ~~~~~~~~correlated&&~~ 0.75&~~ 0.75&~~ 0.75&~~ 0.75&~~ 0.75&~~ 0.75&~~ 0.75&~~ 0.75&~~ 0.75\\
         Trigger  & \ref{sec:trig} &&&&&&&&&\\
 ~~~~~~uncorrelated&&~~ 0.00&~~ 0.00&~~ 0.00&~~ 0.00&~~ 0.00&~~ 0.00&~~ 0.00&~~ 0.00&~~ 0.00\\
 ~~~~~~~~correlated&&~~ 0.04&~~ 0.04&~~ 0.04&~~ 0.04&~~ 0.04&~~ 0.04&~~ 0.04&~~ 0.04&~~ 0.04\\
         Wagon Tagger & \ref{sec:trig} &&&&&&&&&\\
 ~~~~~~uncorrelated&&~~ 0.00&~~ 0.00&~~ 0.00&~~ 0.00&~~ 0.00&~~ 0.00&~~ 0.02&~~ 0.02&~~ 0.02\\
 ~~~~~~~~correlated&&~~ 0.00&~~ 0.00&~~ 0.00&~~ 0.00&~~ 0.00&~~ 0.00&~~ 0.00&~~ 0.00&~~ 0.00\\
\hline
 Total External ($\Delta \epsilon_{\mathrm{ext}}$)&&&&&&&&&&\\
 ~~~~~~uncorrelated&&~~ 0.81&~~ 0.81&~~ 0.81&~~ 0.80&~~ 0.80&~~ 0.81&~~ 1.10&~~ 1.10&~~ 1.10\\
 ~~~~~~~~correlated&&~~ 2.16&~~ 2.16&~~ 2.16&~~ 2.16&~~ 2.16&~~ 2.16&~~ 2.16&~~ 2.16&~~ 2.16\\
\hline
         Energy  & \ref{sec:en}        &&&&&&&&&\\
 ~~~~~~uncorrelated&&~~ 0.10&~~ 0.10&~~ 0.10&~~ 0.10&~~ 0.10&~~ 0.10&~~ 0.10&~~ 0.10&~~ 0.10\\
 ~~~~~~~~correlated&&~~ 1.80&~~ 1.80&~~ 1.80&~~ 1.80&~~ 1.80&~~ 1.80&~~ 1.80&~~ 1.80&~~ 1.80\\
         Beam parameters  & \ref{sec:beam} &&&&&&&&&\\
 ~~~~~~uncorrelated&&~~ 0.57&~~ 0.57&~~ 0.57&~~ 0.57&~~ 0.57&~~ 0.57&~~ 0.57&~~ 0.57&~~ 0.57\\
 ~~~~~~~~correlated&&~~ 0.57&~~ 0.57&~~ 0.57&~~ 0.57&~~ 0.57&~~ 0.57&~~ 0.76&~~ 0.76&~~ 0.76\\
       Radial resolution & \ref{sec:detsim}  &&&&&&&&&\\
 ~~~~~~uncorrelated&&~~ 0.00&~~ 0.00&~~ 0.00&~~ 0.00&~~ 0.00&~~ 0.00&~~ 0.00&~~ 0.00&~~ 0.00\\
 ~~~~~~~~correlated&&~~ 0.20&~~ 0.20&~~ 0.20&~~ 0.20&~~ 0.20&~~ 0.20&~~ 0.20&~~ 0.20&~~ 0.20\\
       Acollinearity bias  & \ref{sec:detsim} &&&&&&&&&\\
 ~~~~~~uncorrelated&&~~ 0.00&~~ 0.00&~~ 0.00&~~ 0.00&~~ 0.00&~~ 0.00&~~ 0.00&~~ 0.00&~~ 0.00\\
 ~~~~~~~~correlated&&~~ 0.36&~~ 0.36&~~ 0.36&~~ 0.36&~~ 0.36&~~ 0.36&~~ 0.36&~~ 0.36&~~ 0.36\\
       Azimuthal resolution & \ref{sec:detsim}  &&&&&&&&&\\
 ~~~~~~uncorrelated&&~~ 0.00&~~ 0.00&~~ 0.00&~~ 0.00&~~ 0.00&~~ 0.00&~~ 0.00&~~ 0.00&~~ 0.00\\
 ~~~~~~~~correlated&&~~ 0.04&~~ 0.04&~~ 0.04&~~ 0.04&~~ 0.04&~~ 0.04&~~ 0.04&~~ 0.04&~~ 0.04\\
          Clustering  & \ref{sec:detsim}   &&&&&&&&&\\
 ~~~~~~uncorrelated&&~~ 0.00&~~ 0.00&~~ 0.00&~~ 0.00&~~ 0.00&~~ 0.00&~~ 0.00&~~ 0.00&~~ 0.00\\
 ~~~~~~~~correlated&&~~ 1.00&~~ 1.00&~~ 1.00&~~ 1.00&~~ 1.00&~~ 1.00&~~ 1.00&~~ 1.00&~~ 1.00\\
  $\Delta R - \Delta \Theta $ cut difference  & \ref{sec:syssum}   &&&&&&&&&\\
 ~~~~~~uncorrelated&&~~ 0.00&~~ 0.00&~~ 0.00&~~ 0.00&~~ 0.00&~~ 0.00&~~ 0.00&~~ 0.00&~~ 0.00\\
 ~~~~~~~~correlated&&~~ 0.00&~~ 0.00&~~ 0.00&~~ 0.00&~~ 0.00&~~ 0.00&~~ 0.00&~~ 0.00&~~ 0.00\\
         M.C. statistics & \ref{sec:detsim}  &&&&&&&&&\\
 ~~~~~~uncorrelated&&~~ 0.29&~~ 0.27&~~ 0.29&~~ 0.33&~~ 0.13&~~ 0.25&~~ 0.36&~~ 0.34&~~ 0.32\\
 ~~~~~~~~correlated&&~~ 0.80&~~ 0.80&~~ 0.80&~~ 0.80&~~ 0.80&~~ 0.80&~~ 0.80&~~ 0.80&~~ 0.80\\
\hline
    Total Simulation ($\Delta \epsilon_{\mathrm{sim}}$) &&&&&&&&&&\\
 ~~~~~~uncorrelated&&~~ 0.65&~~ 0.64&~~ 0.65&~~ 0.67&~~ 0.59&~~ 0.63&~~ 0.68&~~ 0.67&~~ 0.66\\
 ~~~~~~~~correlated&&~~ 2.32&~~ 2.32&~~ 2.32&~~ 2.32&~~ 2.32&~~ 2.32&~~ 2.37&~~ 2.37&~~ 2.37\\
\hline
\hline
         Grand Total     &&&&&&&&&&\\
 ~~~~~~uncorrelated&&~~ 1.04&~~ 1.03&~~ 1.04&~~ 1.04&~~ 1.00&~~ 1.03&~~ 1.29&~~ 1.28&~~ 1.28\\
 ~~~~~~~~correlated&&~~ 3.17&~~ 3.17&~~ 3.17&~~ 3.17&~~ 3.17&~~ 3.17&~~ 3.21&~~ 3.21&~~ 3.21\\
\hline
\end{tabular}
\end{center}
\vspace{-0.5cm}
\caption[Sytematic errors on absolute luminosity]{
    This table summarizes the 
    experimental systematic uncertainties on the absolute $\Lrl$ 
    luminosity 
    measurement for the nine data samples. The lines labeled correlated
    and uncorrelated refer to errors correlated and uncorrelated 
    among the samples.  
    All errors are in units of $10^{-4}$.
\label{tab:syssum_all}
}
}
\end{table} 

\begin{table}
\begin{center}
\begin{tabular}{|l||c|c|c|c|c|c|c|c|c|}
\hline
 Sample       
      & 93 $-2$& 93 pk &93 $+2$& 94 a & 94 b  & 94 c &95 $-2$ & 95 & 95 $+2$\\
\hline
\hline
 93 $-2$& 1.00& 0.91& 0.91& 0.90& 0.90& 0.90& 0.87& 0.87& 0.87\\
 93 pk  & 0.91& 1.00& 0.91& 0.90& 0.90& 0.90& 0.87& 0.87& 0.87\\
 93 $+2$& 0.91& 0.91& 1.00& 0.90& 0.90& 0.90& 0.87& 0.87& 0.87\\
 94a    & 0.90& 0.90& 0.90& 1.00& 0.90& 0.90& 0.87& 0.87& 0.87\\
 94b    & 0.90& 0.90& 0.90& 0.90& 1.00& 0.90& 0.87& 0.87& 0.87\\
 94c    & 0.90& 0.90& 0.90& 0.90& 0.90& 1.00& 0.87& 0.87& 0.87\\
 95 $-2$& 0.87& 0.87& 0.87& 0.87& 0.87& 0.87& 1.00& 0.91& 0.91\\
 95 pk  & 0.87& 0.87& 0.87& 0.87& 0.87& 0.87& 0.91& 1.00& 0.91\\
 95 $+2$& 0.87& 0.87& 0.87& 0.87& 0.87& 0.87& 0.91& 0.91& 1.00\\
\hline
\end{tabular}
\caption[Experimental correlation matrix]{ The 
experimental luminosity correlation matrix.
   \label{tab:lumi_matrix}
}
\end{center}
\end{table}

\section{Properties of the Bhabha event sample}
\label{sec:ev_samp}
%
%
 
    In the preceeding sections, 
    we have demonstrated that
    the characteristics
    of our detector, as well as the background in our event sample,
    are well understood. 
    In this section we compare distributions of the quantities
    on which the Bhabha selection is based with the predictions of
    the BHLUMI Monte Carlo~\cite{bib-BHLUMI}.  
    The agreement in every aspect is excellent,
    even though the tuning of the detector response functions
    was carried out in a manner almost completely independent of the
    BHLUMI Monte Carlo.
    We must also emphasize that we derive no part of our systematic
    error from these comparisons.  All such errors are determined directly
    from the data and the limitations on our ability to
    understand the detector response.
The data--Monte Carlo comparisons therefore represent, in great measure,
a blind test in which discrepancies can only be interpreted as a failure
of either our experimental or theoretical understanding.
In either case, such a discrepancy would necessarily invalidate the
luminosity measurement.

Furthermore, since the higher-order terms which represent the limiting
theoretical uncertainties in the Bhabha scattering calculation
are expected to leave little trace in the
experimentally observable quantities, even perfect agreement between the
measured and expected distributions can not be used to argue for a reduction
in the theoretical error.
The material of this section therefore represents a demanding test of
the luminosity measurement, in which it is impossible for the agreement to
be better than expected.
The agreement is, and must be, essentially perfect.

    The most important isolation cut concerns the energy.
    In \fg{fig:sel_er_vs_elcd}
    we compare the rate of ``single radiative'' events,
    predominately from initial--state radiation, 
    to the prediction
    from the Monte Carlo.  
    One of the electrons is required to 
    satisfy $E > 0.85 \cdot \ebeam$ and the energy of the other
    is plotted.  The energy response of the detector was determined
    using a sample of very collinear events and the kinematics
    of radiative events.  This determination was independent of
    the rate of radiative events, allowing this rate to be used
    as a test of the Monte Carlo.
    There is good agreement between the predicted and observed
    radiative tail.  This tail can best be observed when the
    acollinearity cut is suppressed, indicating that BHLUMI gives a valid
    description of the radiative energy tail over almost five orders
    of magnitude.

    The acollinearity distribution (difference between $\rr$ and $\rl$)
    is shown in \fg{fig:sel_dr}.  This plot also shows agreement
    between the Monte Carlo and the data over almost five orders of
    magnitude.  The two peak structure of the plot is due to
    the beam offset.  
    The good agreement near the peak is due to the use of the
    measured LEP beam parameters in the simulation.  
    The agreement
    in the tail provides more evidence that initial--state radiation
    is correctly implemented in the Monte Carlo.

Further features of radiative processes in small--angle Bhabha scattering are
revealed by studying accepted Bhabha events with secondary clusters
falling within the SiW acceptance.
The excellent capability of our detector for resolving even close--lying
clusters allows a sensitive comparison to be made between the observed
and expected rates of such secondary clusters.
Despite the fact that the luminosity acceptance depends only weakly on
the fidelity with which secondary clusters are treated experimentally,
faith in the accuracy of higher-order terms in the calculation of the
Bhabha scattering cross section would be undermined if the observed
distribution of secondary clusters did not conform to theoretical
expectations.

\Fg{fig:sel_2clus} shows the observed energy spectrum of comfortably
resolved secondary clusters which are identified by the kinematic cluster
selection algorithm as belonging to the Bhabha event.
To ensure that the events entering this study are kinematically complete,
we further require that one side of the event consists of a single
full--energy cluster, and that the total measured energy in the event is
within $10\%$ of the LEP center-of-mass energy.
The energy spectrum of such clusters predicted by
BHLUMI, also shown in \fg{fig:sel_2clus}, agrees well with the experimental
measurements, except for the two lowest energy bins.
   
    The acoplanarity distribution is displayed in \fg{fig:sel_dphi}.
    Good agreement between the Monte Carlo expectation and the 
    data is seen throughout most of the distribution.  At small
    values of acoplanarity some differences between Monte Carlo
    and data are observed.  These differences are due to
    the rather crude treatment of the $\phi$ resolution in
    the detector simulation,
which does not reproduce the non--Gaussian features of the true detector
response, and are not an indication of a
    problem in the physics Monte Carlo.   
A $10\%$ error in the calculated resolution flow in acoplanarity is assigned
to account for these imperfections (see section~\ref{sec:azimuth}).
    
    The distribution of the right radial coordinates 
    for the \SwitL\ sample, and similarly the left radial 
    coordinate for the \SwitR\ sample, are shown
    in \fg{fig:sel_rr_vs_rl}.  These distributions are closely related
    to the acollinearity distribution, as the tight radial cut 
    on the opposite side provides
    an effective acollinearity cut on the events near the inner and
    outer edges of the acceptance.

    The radial distributions after all cuts, except the
    tight radial definition cut, are displayed in
    \fg{fig:sel_def} and the locations of the definition cuts
    are indicated.  Good agreement
    between the data and Monte Carlo is seen except 
    for a small amount of structure near the center 
    of the acceptance on the right and left sides,
    behind the bulk of the upstream material (see \fg{fig:det_mat}).
    The radial distribution checks the 
    prediction of the Monte Carlo on the scale
    of the radial granularity of the detector (0.25~cm).
    The smoothing procedure,
    described in section~\ref{sec:smoothing}, 
    suppresses any structure within
    each pad at the 7~$\x$ reference plane and
    therefore the radial distributions 
    do not check the prediction of
    the Monte Carlo on scales smaller than a pad.

    Some global checks of the selection procedure have
    already been mentioned in sections~\ref{sec:anchor} and
    \ref{sec:beam}.  Shifting the location of
    the inner radial cut one pad boundary lower
    or higher
    leads to changes in measured
    luminosity of $ 0.4 \pm 1.3 \times 10^{-4}$ and 
    $ -0.9 \pm 1.3 \times 10^{-4}$ respectively. 
    A stronger check on the selection comes from
    the comparison of the luminosity measurement
    based on \SwitA\ with that based on \SwitR\ and \SwitL.  
    The \SwitA\ selection has additional systematic 
    errors beyond
    those of the \SwitR\ and \SwitL\ selections due 
    to differences in the resolution of the individual 
    right and left coordinates and the averaged coordinate.
    These systematic effects are estimated to be at the level
    of  $1.5 \times 10^{-4}$.  
    On the other hand the
    dependence of the \SwitA\ selection on the beam
    parameters is substantially different than that
    of the \SwitR\ and \SwitL\ selection.  Comparing
    the two selections we find that
    the selection for \SwitA\ and $\frac{1}{2}(\SwitR + \SwitL)$
    differ by $65.4 \pm 0.9 \times 10^{-4}$ 
    in the data and 
    $65.3  \pm 1.0 \times 10^{-4}$ in the Monte Carlo.  
    The residual 
    difference between
    data and Monte Carlo is much smaller than the additional
    systematic error of the \SwitA\ selection.

\begin{figure}[htb!]
    \mbox{\epsfxsize17cm\epsffile{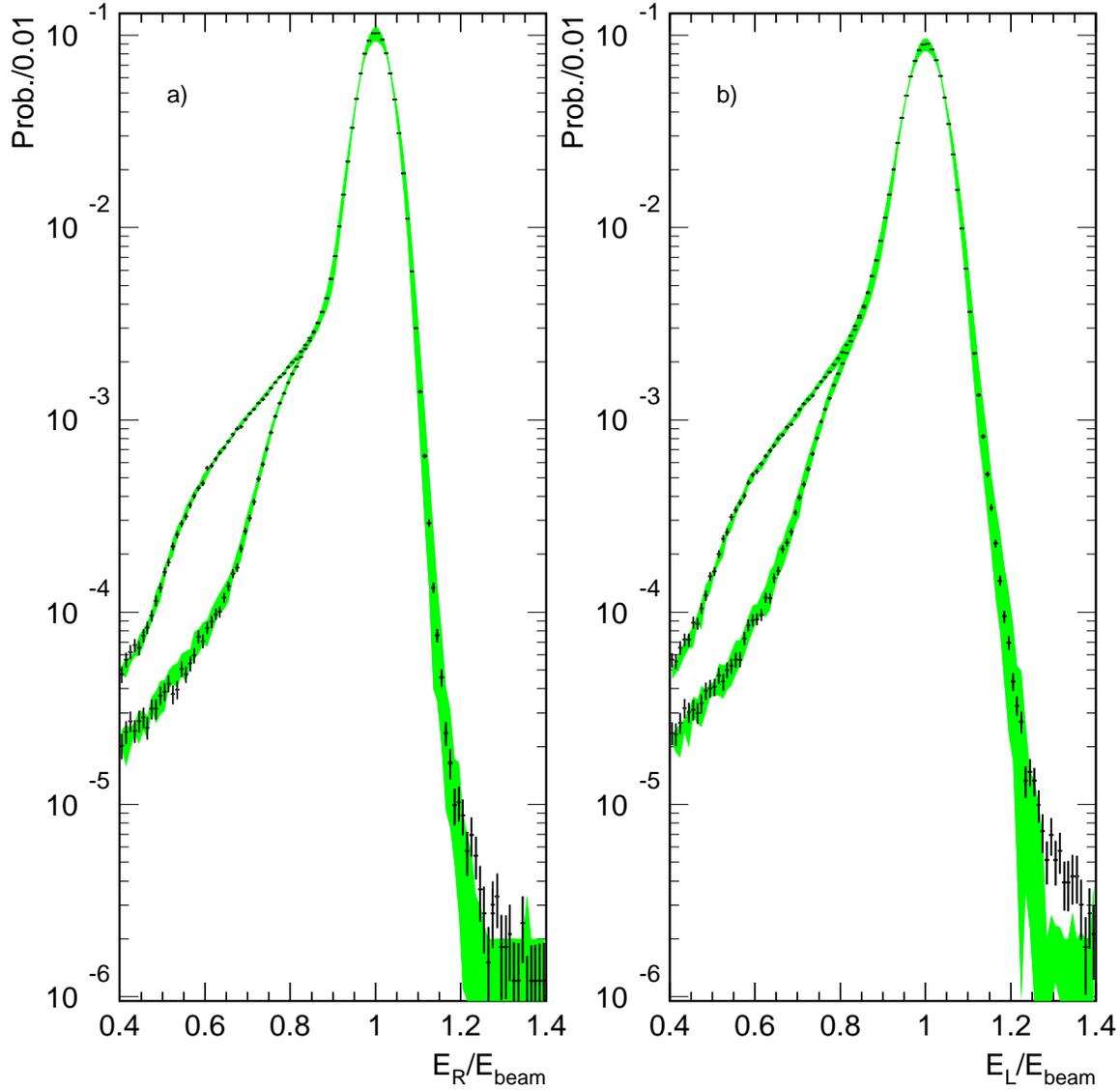}}
    \caption[$\er$ and $\el$]{
    The normalized energy distributions in the right (a)  and  left
    (b) calorimeter for 
    events passing all geometric cuts for the \SwitA\ selection and 
    which in addition 
    have more than 0.85 of the beam energy in the opposite calorimeter.
    The energy spectrum is shown respectively before and after the
    acollinearity cut has been applied in the upper and lower branches
    of the curves.
    The points with errors show the data.
    The shaded bands show the Monte--Carlo predictions for
    the range of energy parameterizations
    used in assessing the systematic error (see section~\ref{sec:en}).
    Above 0.8 the upper and lower branches of the curves are nearly
    indistinguishable.
    \label{fig:sel_er_vs_elcd}
    }
\end{figure}

\begin{figure}[htb!]
 \mbox{\epsfxsize17cm\epsffile{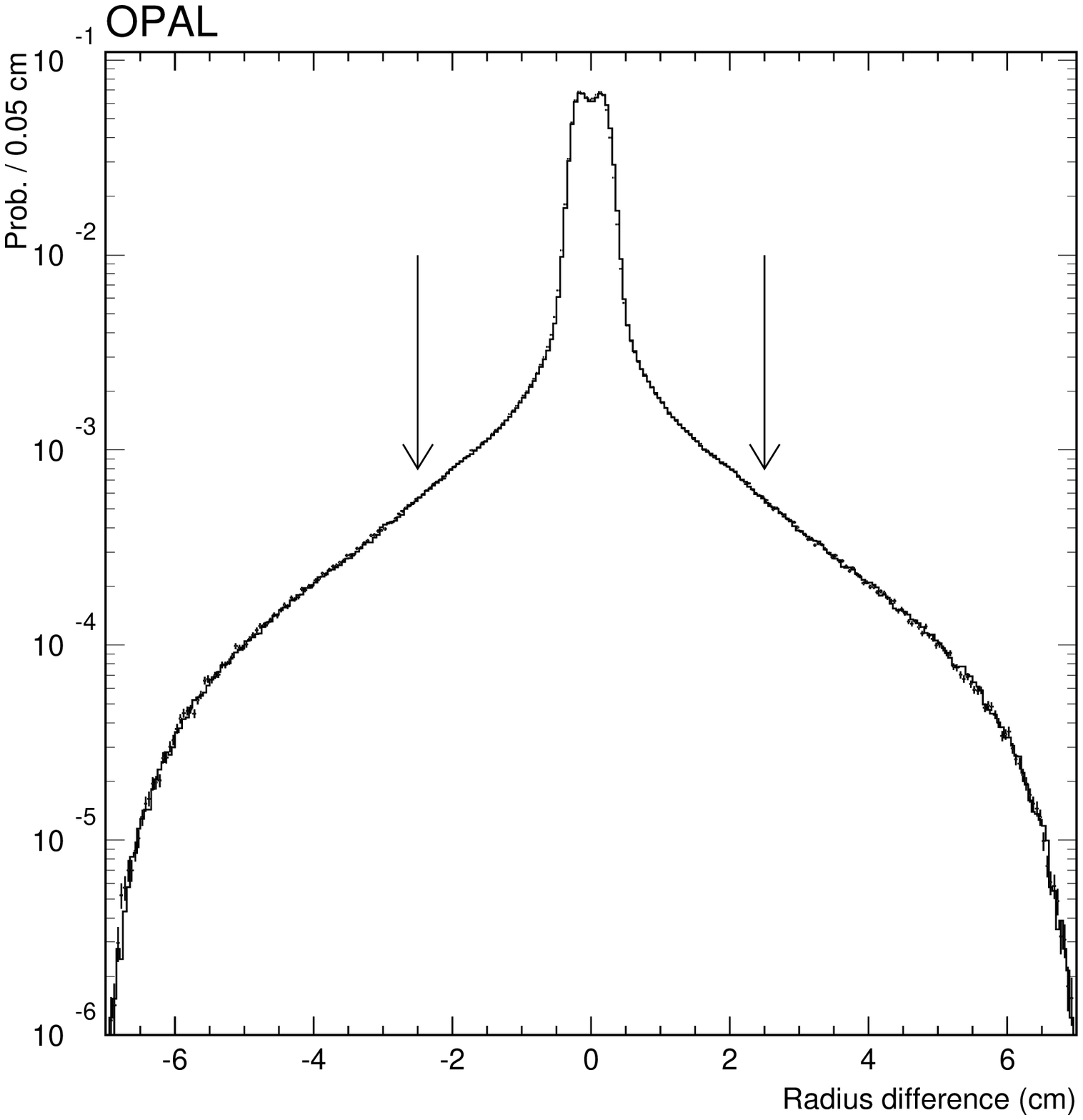}}
    \caption[Acollinearity ($\rr - \rl$)]{
    The distribution of 
    $\rr - \rl$
    after all the \SwitA\ selection cuts have been applied,
    except for the acollinearity cut.
    Data are shown as the points and Monte~Carlo
    as the solid histogram.
    For this comparison, the background from accidental coincidences
    of off-momentum beam particles has been subtracted from the data.
    The acollinearity cut is indicated by the two vertical
    arrows.
    \label{fig:sel_dr}
    }
\end{figure}

\begin{figure}[htb!]
  \begin{center}
  \mbox{\epsfxsize17cm\epsffile{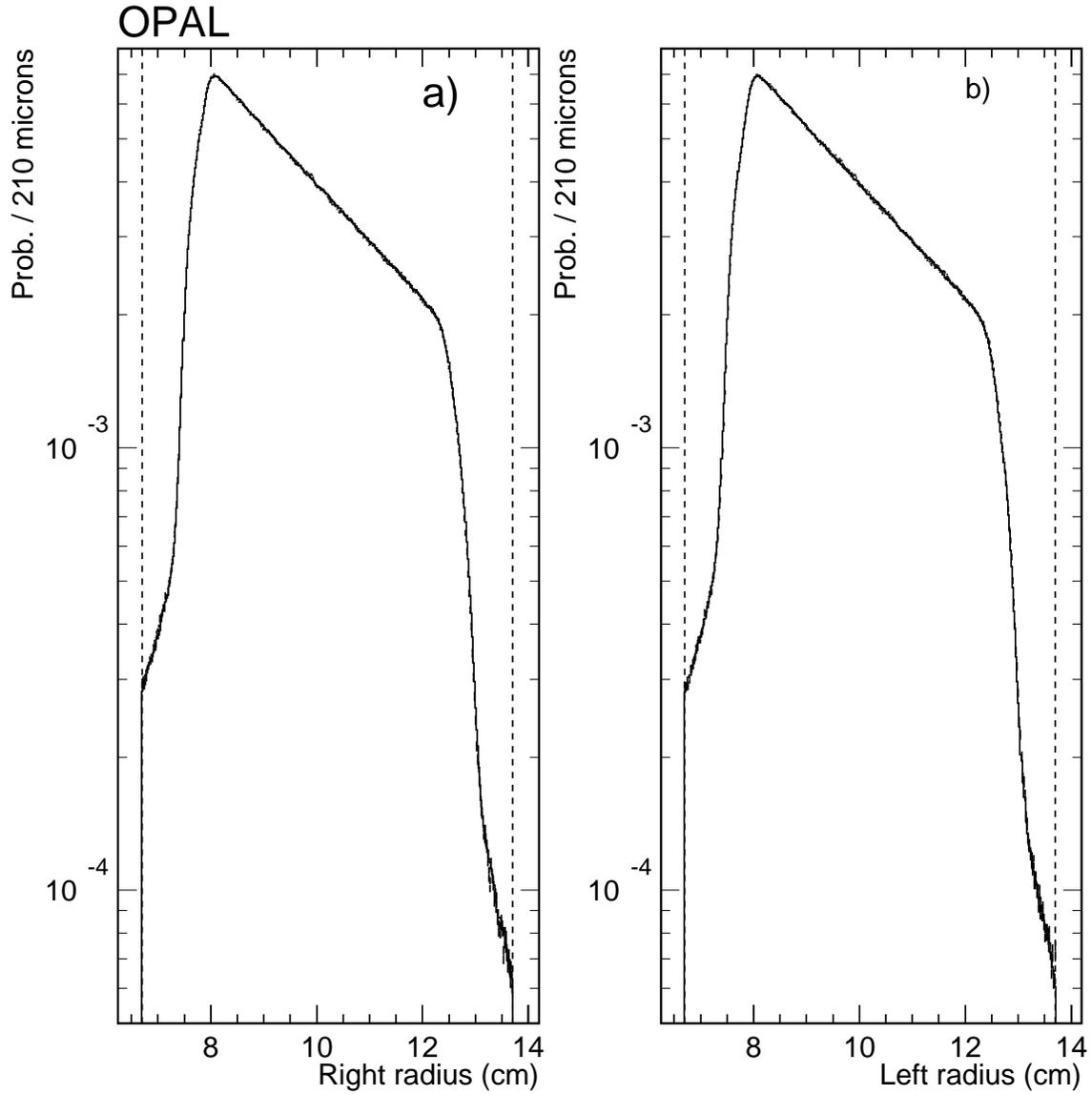}} 
  \end{center}
    \caption[Loose radial cuts]{
    The distributions of
    $\rr$ after all  \SwitL\ selection cuts have been applied (a),
    and of $\rl$ after all \SwitR\ selection cuts have been applied,
    (b).
    The points and error bars are the data and the histogram the Monte~Carlo.
    \label{fig:sel_rr_vs_rl}  The vertical dotted lines indicate the
    radial isolation cuts.
    }
\end{figure}

\begin{figure}[htb!]
  \begin{center}
  \mbox{\epsfxsize17cm\epsffile{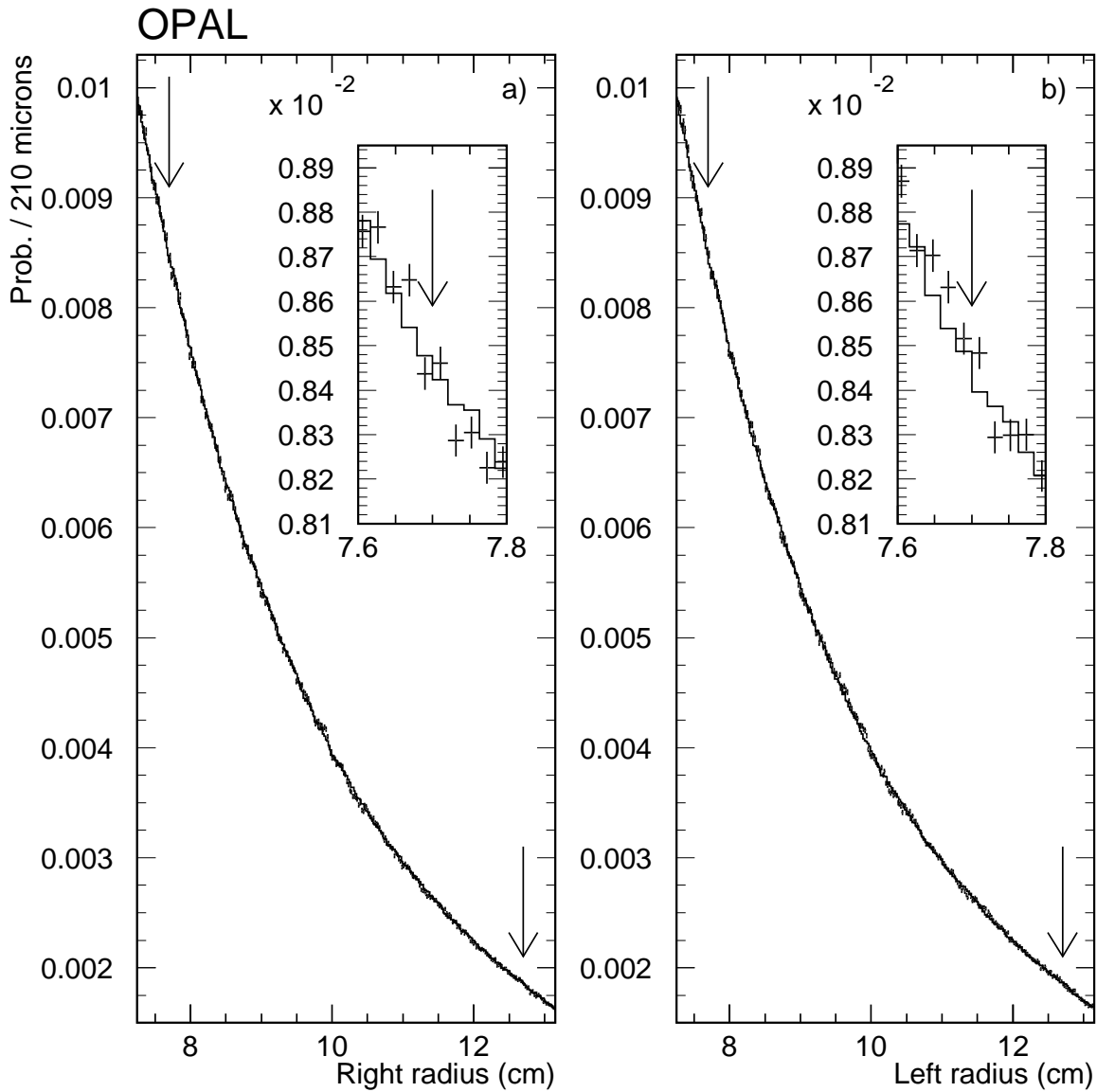}} 
  \end{center}
    \caption[Definition cuts ]{
    The distributions (after the isolation cuts
    have been applied) of 
    $\rr$ (a), and $\rl$ (b).
    The arrows indicate the location of the
    definition cuts for the \SwitR\ and \SwitL\
    selections.
    The points and error bars are the data and the 
    histogram the Monte~Carlo.  The insets show fine
    details near the inner acceptance boundaries.
    \label{fig:sel_def}
    }
\end{figure}

\clearpage
\section{Conclusion}
\label{sec:conclusion}
%
%
%
%
%

The OPAL Silicon-Tungsten Luminometer has been used to measure the 
integrated luminosity at the OPAL intersection region at LEP since its 
installation in 1993.
After stringent requirements on the condition of the detector during
data taking, it provides valid
measurements for 101~pb$^{-1}$ of OPAL data, which represents $\sim60\%$
of the entire OPAL LEP~I data sample.
The detector has achieved a fractional
experimental systematic error of \mbox{$3.4\times10^{-4}$} in defining
the acceptance within which the number of small-angle
Bhabha scattering events have been counted to determine the luminosity.
This precision represents more than an order of magnitude 
improvement with respect to previous OPAL results using the Forward 
Detector~\cite{bib-old-opal}.

As a result, the experimental precision of the luminosity measurement no
longer represents a significant limitation in the OPAL determination of the
absolute coupling strengths of the $\Zzero$. 
The statistical error inherent in the entire sample of $8\times10^6$ small
angle Bhabha events observed in the $\sim79$~nb fiducial acceptance of the
luminometer is similar in size to the experimental systematic error in the
luminosity, and also contributes negligibly to the total error.

All observable properties of the luminosity event sample agree well with the
current theoretical understanding of small--angle Bhabha scattering.
The dominant contribution to the uncertainty in the OPAL luminosity
is the current theoretical
error of \mbox{$5.4\times10^{-4}$} in the calculated acceptance.

The uncertainty in the OPAL acceptance for hadronic $\Zzero$-decays is
\mbox{$7\times10^{-4}$}~\cite{bib-ph-thesis,bib-bernd}, 
comparable to the statistical error in the total
sample of
\mbox{$2.4\times10^{6}$} hadronic $\Zzero$ decays.
The OPAL measurement of $\Ginv/\Gll$, a quantity which depends crucially
on the determination of absolute cross sections, has a total relative error of
$4.4\times10^{-3}$, of which only $1.7\times10^{-3}$ is attributable to the
luminosity measurement.
In the end, the SiW luminometer has made a dramatic
improvement in our ultimate knowledge of the $\Zzero$ couplings, quantities
whose precise measurement has been the fundamental goal of the
experimental program at LEP~I.

  The measured luminosity for the 1993-1995 multihadron  sample
is given in \tb{tab:xsec}.

\begin{table}[htbp]
\begin{center}
\begin{tabular}{|l|r|r|r|r|r|}
\hline
Sample  &  Energy & Multihadrons & $\SwitR$ & $\SwitL$ & Luminosity \\
        &  (GeV)  &              &          &          & $(\mathrm{pb}^{-1})$ \\
\hline
 93 $-2$ &$  89.4505 $&$   84710 $&$  697370 $&$  699041 $&$  8.515\pm 0.010 $ \\
 93 pk &$  91.2063 $&$  262735 $&$  689570 $&$  690487 $&$  8.761\pm 0.011 $ \\
 93 $+2$ &$  93.0351 $&$  123895 $&$  682445 $&$  683642 $&$  9.042\pm 0.011 $ \\
\hline
 94 a  &$  91.2358 $&$  267413 $&$  698423 $&$  701928 $&$  8.895\pm 0.011 $ \\
 94 b  &$  91.2165 $&$ 1238184 $&$ 3238878 $&$ 3252718 $&$ 41.229\pm 0.023 $ \\
 94 c  &$  91.4287 $&$   11135 $&$   29066 $&$   29211 $&$  0.372\pm 0.002 $ \\
 94 c' &$  91.2195 $&$   68324 $&$  177864 $&$  178154 $&$  2.261\pm 0.005 $ \\
\hline
 95 $-2$ &$  89.4415 $&$   83254 $&$  691317 $&$  690522 $&$  8.425\pm 0.010 $ \\
 95 pk &$  91.2829 $&$  139328 $&$  362725 $&$  362853 $&$  4.616\pm 0.008 $ \\
 95 $+2$ &$  92.9715 $&$  126305 $&$  677469 $&$  675479 $&$  8.944\pm 0.011 $ \\
\hline
\hline                                                                          
Total        &            &  2405283  &  7945127  &  7964035  & $101.059\pm 0.036 $ \\
\hline
\end{tabular}
\end{center}
\caption[Measured luminosity]{ The mean center-of-mass 
energy, numbers of multihadron and luminosity
events recorded in the 1993, 1994 and 1995 data samples
as well as the corresponding luminosities, with their statistical
error.
\label{tab:xsec}
} 
\end{table}

\appendix
\par
\noindent
Acknowledgements:
\par
We particularly wish to thank the SL Division for the efficient operation
of the LEP accelerator at all energies
 and for their continuing close cooperation with
our experimental group.
We thank our colleagues from CEA, DAPNIA/SPP,
CE-Saclay for their efforts over the years on the time-of-flight and trigger
systems which we continue to use.
We would also like to extend special thanks to the small community of theorists,
without whose painstaking calculation of the radiative corrections to
small--angle Bhabha scattering, the experimental precision of our result would
have been in vain.
In addition to the support staff at our own
institutions we are pleased to acknowledge the  \\
Department of Energy, USA, \\
National Science Foundation, USA, \\
Particle Physics and Astronomy Research Council, UK, \\
Natural Sciences and Engineering Research Council, Canada, \\
Israel Science Foundation, administered by the Israel
Academy of Science and Humanities, \\
Minerva Gesellschaft, \\
Benoziyo Center for High Energy Physics,\\
Japanese Ministry of Education, Science and Culture (the
Monbusho) and a grant under the Monbusho International
Science Research Program,\\
Japanese Society for the Promotion of Science (JSPS),\\
German Israeli Bi-national Science Foundation (GIF), \\
Bundesministerium f\"ur Bildung, Wissenschaft,
Forschung und Technologie, Germany, \\
National Research Council of Canada, \\
Research Corporation, USA,\\
Hungarian Foundation for Scientific Research, OTKA T-016660, 
T023793 and OTKA F-023259.\\
%


\bibliographystyle{unsrt}
\bibliography{pr289}

\end{document}